\documentclass[ba]{imsart}

\usepackage{amsthm,amsmath,amsfonts,amssymb, lineno}
\usepackage{bm, ulem, enumitem, soul, url, newtxtext}
\usepackage[subscriptcorrection]{newtxmath}
\usepackage[numbers]{natbib}
\usepackage[colorlinks,citecolor=blue,urlcolor=blue,backref=page,backref=page]{hyperref}
\usepackage{graphicx, xcolor, multirow, subcaption, tablefootnote, comment, booktabs}
\usepackage{algorithm, algpseudocode}

\pubyear{2025}
\arxiv{2412.09792}
\volume{TBA}
\issue{TBA}
\firstpage{1}
\lastpage{1}

\startlocaldefs
\theoremstyle{plain}

\newtheorem{theorem}{Theorem}[section]
\newtheorem{lemma}[theorem]{Lemma}
\newtheorem{corollary}[theorem]{Corollary}
\theoremstyle{definition}

\theoremstyle{remark}

\newtheorem{remark}{Remark}
\usepackage{xr}	
\makeatletter
\newcommand*{\addFileDependency}[1]{
  \typeout{(#1)}
  \@addtofilelist{#1}
  \IfFileExists{#1}{}{\typeout{No file #1.}}
}
\makeatother
\newcommand*{\myexternaldocument}[1]{%
    \externaldocument{#1}%
    \addFileDependency{#1.tex}%
    \addFileDependency{#1.aux}%
}
\myexternaldocument{SuppMat_R1_TH_Nov15}

\def\transp{{ \mathrm{\scriptscriptstyle T} }}

\DeclareMathOperator{\cV}{\mathcal{V}}
\DeclareMathOperator{\cF}{\mathcal{F}}
\DeclareMathOperator{\cN}{\mathcal{N}}
\DeclareMathOperator{\cG}{\mathcal{G}}

\newcommand{\by}{\mathbf{y}}
\newcommand{\bI}{\mathbf{I}}
\newcommand{\bN}{\mathbf{N}}
\newcommand{\bSigma}{\bm{\Sigma}}
\newcommand{\bPsi}{\bm{\Psi}}
\newcommand{\bLambda}{\bm{\Lambda}}
\newcommand{\bpsi}{\bm{\psi}}
\newcommand{\bphi}{\bm{\phi}}
\newcommand{\bbeta}{\bm{\beta}}
\newcommand{\btheta}{\bm{\theta}}
\newcommand{\bepsilon}{\bm{\epsilon}}
\newcommand{\nTxt}[1]{\textcolor{blue}{#1}}
\endlocaldefs

\begin{document}

\begin{frontmatter}
\title{Bayesian Nonparametric Global-local Clustering of High-dimensional Functions}
\runtitle{Bayesian Nonparametric Global-local Clustering of High-dimensional Functions}

\begin{aug}
\author[A]{\fnms{Tsung-Hung}~\snm{Yao}\ead[label=e1]{yaot@ufl.edu}\orcid{0000-0001-5342-5430}}
\and
\author[B]{\fnms{Suprateek}~\snm{Kundu}\ead[label=e2]{skundu2@mdanderson.org}}
\address[A]{Department of Biostatistics, University of Florida\printead[presep={,\ }]{e1}}
\address[B]{Department of Biostatistics, The University of Texas MD Anderson Cancer Center\printead[presep={,\ }]{e2}}
\runauthor{T. Yao and S. Kundu}
\end{aug}

\begin{abstract}
There is a rich literature on clustering functional data with applications to time-series modeling, trajectory data, and even spatio-temporal applications. Existing methods are classified into two categories of (i) global clustering that assigns an identical membership over the whole domain, and (ii) local clustering that allows multiple memberships based on the sub-domains. Classical literature has typically focused on global clustering that may struggle with degenerate clustering or produce numerous spurious clusters for high-dimensional functions. In contrast, a limited number of local clustering approaches have been proposed that are not scalable to high-dimensional functions and have limited theoretical guarantees. Focusing on basis expansions for high-dimensional functions, we propose a novel {\it global-local} framework that partitions coefficients into exclusive subsets and imposes independent Dirichlet process (DP) priors on each subset. The proposed framework reduces the total number of possible clustering configurations compared to local clustering and therefore provides greater scalability, while simultaneously providing enhanced flexibility compared to global clustering  by relaxing the assumption of identical clustering for all elements in the high-dimensional function.  Moreover, the proposed method incorporates spatially correlated errors to provide better model fitting and shows posterior consistency properties that asymptotically recover the true density of random functions. An efficient Markov chain Monte Carlo (MCMC) algorithm is developed for implementation. Extensive simulations involving images illustrate the improved scalability, clustering, and function estimation under the proposed method compared to classical approaches. We apply the proposed approach to a spatial transcriptomics application where the goal is to infer clusters of genes with distinct spatial patterns of expressions. \nTxt{This work fills an important gap in the limited literature on clustering methods for high-dimensional functions, with applications to spatial functional data and related domains.}
\end{abstract}



\end{frontmatter}

\section{Introduction}
\nTxt{Functional data analysis is a well-established and active area of research in statistics and machine learning \citep{Ramsay2005}.}
\nTxt{This is not surprising given the vast amounts of functional data collected in practice, ranging from time-series data \citep{Aghabozorgi2015}, spatial data \citep{Mateu2017}, spatio-temporal data \citep{Atluri2018}, among others. A rich literature exists for analyzing functional datasets \citep{Wang2015}, and a substantial Bayesian literature has also emerged, including both classic and more recent contributions to functional data clustering \citep[e.g.][]{James2003, Lee2013, Scarpa2014, Zhang2023, Toto2025}.}
Most Bayesian functional data clustering methods rely on parametric finite mixture models \citep{Bouveyron2019ModelClu, Chamroukhi2019} or more flexible non-parametric infinite mixtures \citep{Fruhwirth-Schnatter2019}. Existing approaches typically assign a global membership, which produce distinct groups of functions having varying cluster membership between subgroups and identical memberships within each subgroup. However, they may not capture the heterogeneity in localized features within the functions lying in the same cluster \citep{LocalClu_DiIorio2023} or address scenarios where a small subset of feature separate the different clusters \citep{Floriello2017}. 
Some examples of global clustering methods include \citep{ray2006functional, crandell2011posterior, bigelow2009bayesian, rodriguez2014functional}. 
\nTxt{We refer the readers to \citet{Wade2023BayesianClu} for a more comprehensive review of such approaches, which summarizes Bayesian model-based and loss-based clustering frameworks that primarily focus on global partitioning of functions or observations, in contrast to more recent local clustering formulations that capture within-function heterogeneity.}
The majority of Bayesian global clustering methods have focused their implementation on one-dimensional functions such as time-series curves or longitudinal trajectories. While extending these approaches to higher order functions (such as spatial functions or images) should be feasible in principle, it is not straightforward in practice. 
\nTxt{This is due to the fact that higher-order functional observations are often represented through high-dimensional parameterizations (e.g., dense spatial grids or large numbers of observed points), which pose additional computational and modeling challenges for classical global clustering methods. In this sense, we use the term high-dimensional functional data to refer to functions observed over large spatial domains or with many observed instances in the functional domain, rather than to their intrinsic functional dimensionality.}
In this context, \citet{chandra2023CurseDim} showed that classical Bayesian mixture modeling approaches produce global clustering that are susceptible to poor performance and degenerated clustering in high-dimensional settings. In particular, these approaches may potentially produce an unrealistic large number of clusters or collapse to just one cluster when the model dimension far exceeds the sample size. While these results in \citet{chandra2023CurseDim} were developed with respect to clustering of multivariate vectors, they are naturally generalizable to our functional data analysis settings of interest.

To circumvent difficulties associated with global clustering, a limited number of local clustering methods have been proposed that enable flexible clustering in different resolutions or domains \citep[e.g.][]{LPP_Dunson2009, petrone2009hybrid, Rodriguez2010}.  
These approaches produce distinct clustering allocations across various subsets of model parameters characterizing the function and thereby relax the restrictions inherent in global clustering. 
This enables one to account for greater heterogeneity and flexibility when clustering high-dimensional functions, which is desirable. Local clustering approaches can be broadly categorized into two classes. The first class of methods often directly cluster the element-wise mean parameters corresponding to the different observed instances within the function domain, while incorporating spatial dependence via these mean parameters. 
\citet{petrone2009hybrid} proposed a hybrid Dirichlet process prior that formulates the individual functions as hybrids of global functions drawn from a Gaussian process. A latent Gaussian copula is used to allow local surface selection. Related alternative approaches were proposed by \citet{Rodriguez2010}, and \citet{Nguyen2011}. The above approaches are quite useful, but can be computationally intensive due to the need to update as many labeling indicator vectors as the number of locations within MCMC sampling.  The second class of methods rely on basis expansions to smooth the mean function, while imposing suitable mixture priors on the basis coefficients.  
\citet{LPP_Dunson2009} proposed local partition processes (LPP) that enabled independent clustering across basis coefficients, while also specifying an additional shared mixture component for global clustering. \citet{suarez2016bayesian} proposed a spike and slab prior independently for wavelet basis coefficients, and proposed a post-MCMC decision-theoretic criteria to consolidate the local clusters to inform global clustering. Recently, \citet{Fan2023} proposed a hidden Markov model for local partitioning and clustering of multiple time-series data.

While local clustering approaches potentially provide greater flexibility when modeling high-dimensional functions, there are several unmet challenges: (1) intensive computational bottleneck that restricts application to higher order functions; (2) failure to incorporate correlated noise that may be restrictive; and (3) insufficient theoretical guarantees. The computational challenges are exemplified by a massive increase in the number of possible clusters under local clustering. In particular, the total number of possible number of clusters increases from $S(n,k)$ under global clustering to $\prod_{l=1}^L S(n,k_l)$ under local clustering, when clustering $n$ samples of functions \nTxt{$k_l$ as the number of possible mixture components at the lth location for $l=1,\ldots,L$ observed observations, corresponding to the number of points within the functional domain where the functional values are realized/observed.}
Here, $S(n,k)$ denotes the \nTxt{Stirling} number of the second kind.
Therefore, the increase in flexibility due to a greater number of mixture components under local clustering comes at the cost of greater computational \nTxt{cost} that must be controlled via careful modeling considerations. 
Additionally, most local clustering approaches assume independent errors in order to manage the computational burden of clustering high-dimensional functions. While this assumption may not pose serious challenges for the case of 1-D functions typically considered by existing local clustering methods \cite{LPP_Dunson2009,suarez2016bayesian,Fan2023}, it may be restrictive for our motivating spatial functions that may have clusters of neighboring voxels with correlated errors. In particular, failing to account for localized correlated errors in images may amplify local perturbations and ultimately hinder the clustering ability. 


In terms of theoretical implications of Bayesian local clustering methods, there are relatively few guarantees in literature. Using wavelet expansions, \citet{suarez2016bayesian} established asymptotic consistency in recovering the true (finite dimensional) model parameters as well as the true partition structure in the ideal case when the observed measurement error associated with the function approaches zero. Unfortunately, the assumption of vanishing noise may not be supported in most practical settings, especially for functional representations that truncate the basis expansion. 
\citet{Nguyen2011} used a finite mixture model and explored posterior consistency for density estimation in the limiting case when the number of components $k\to \infty$. While useful, they assumed the same form of the true density as the working model for their theoretical derivations that may be restrictive. Further, the working model based on finite-dimensional representations (via truncation) is deemed less flexible compared to the infinite-dimensional mixture representation that is ultimately used for their theoretical analyses. Beyond these limited findings, posterior consistency guarantees pertaining to density estimation of random functions under Bayesian nonparametric local clustering approaches are limited or essentially absent to our knowledge. 

 Our objective is to develop a novel and scalable Bayesian nonparametric global-local clustering framework that combines the advantages of global and local clustering when modeling high-dimensional functions. We utilize classical basis representations to enforce smoothing while incorporating correlated noise terms to enhance model flexibility. The basis coefficients are partitioned into exclusive subsets that are independently modeled using distinct Dirichlet process (DP) priors,  resulting in a Bayesian product mixture prior specification. Coefficients in the same subset are jointly clustered with identical memberships resulting in global clustering, while distinct parameter subsets are clustered independently leading to local clustering. This yields a global-local clustering mechanism that includes the classical global and local clustering as special cases. Computationally, by dividing the basis coefficients into smaller subsets, the proposed approach massively reduces the possible number of clustering allocations and computational \nTxt{cost} when compared to local clustering (see details in Lemma \ref{lma:cmpComplexity}). Ultimately, this allows the proposed method to scale to high-dimensional functions, which is not feasible under existing local clustering approaches. Finally, a post hoc decision-theoretic criterion is proposed to consolidate the various local clusters to report global clustering patterns that may be of potential interest. The post hoc criteria can be tuned to regulate the impact of local clusters on the consolidated global clustering, which is particularly useful when local features introduce additional noise and heterogeneity and need to be ignored when considering global clusters.

\nTxt{Although the proposed methodology is generally applicable to a broad class of functions that are represented via wavelet expansions \citep{Daubechies1993}}
our focus in this article is on clustering spatial functions that are expressed via wavelet basis representations.  We specify our approach to incorporate resolution-specific clustering that separately clusters high-level (or global) features and finer resolution (or local) features under the wavelet representation. Motivated by recent literature on measurement error models \citep{Ma2022DWT}, we incorporate correlated measurement errors that are especially suitable for spatial functions, including images. Additionally, we allow these correlations to vary across subjects and consider both unstructured and low-rank error covariances, with the latter designed for improved scalability. Theoretical properties such as posterior consistency  for density estimation are investigated under mild assumptions, which  builds upon recent developments in non-parametric Bayesian multivariate density estimation literature \citep{Canale2017, Kundu2024PDPM}. Further, we develop an efficient
MCMC algorithm for scalable implementation of the proposed approach to the case of 2-D spatial functions.  Simulation studies illustrate strong numerical advantages under the proposed method compared to global clustering approaches, in terms of clustering and function estimation. In contrast, local clustering methods are simply not scalable to the high-dimensional images considered in our applications. The proposed method is applied to spatial transcriptomics data focused on clustering spatially
varying gene expression profiles in a breast cancer study. Our analysis discovers reproducible gene clusters with solid biological justifications.  

The article is structured as follows. We develop the proposed methodology for noisy functional data in Section \ref{sec:Method} with theoretical guarantees in Section \ref{sec:thmTmp},
while Section \ref{sec:posteriorComp} outlines posterior inference strategy. 
In Section \ref{sec:sim}, we present results from extensive simulation studies. Section \ref{sec:realDf}  applies the approach to a spatial transcriptomics data for clustering gene expressions. The Discussion section contains additional insights and Supplementary Materials contain proofs of theoretical results, detailed posterior computation steps, and additional details. 

\section{Methodology}\label{sec:Method}
\subsection{Functional Data Analysis via Basis Expansion}
Consider $n$ observed functions $y_i(\cdot)$ for units $i=1,\ldots,n$, where the support of these functions corresponds to points $v\in \cV\subset \mathbb{R}^d$, and $\cV$ refers to the space of all support points where the function can be observed. Assume that each observed function $y_i(\cdot)$ depends on a corresponding underlying random function $\theta_i(\cdot)$ subject to an additive measurement error. In other words, we consider the model:
\begin{align} \label{eq:base}
y_i(v) = \theta_i(v) +\epsilon_{iv},
\end{align}
where $\epsilon_{iv}$ is an additive measurement error for unit $i$ at point $v$, \nTxt{assuming independency across units $i$ but potentially correlated across locations $v$ within each unit}, and $\theta_i(v)$ is the underlying random function that follows some probability distributions assigned to the space of square integrable measurable functions. Classical functional data analysis frameworks use basis expansions for representing $\theta_i(\cdot)$ \citep{Morris2015}.
While our approach is generalizable to various types of basis expansions, we focus on wavelet basis functions \citep{ray2006functional, Mallat2008} that has been used extensively for modeling spatial functions such as images \citep{reiss2015wavelet}. Wavelet basis expansion provides us with several advantages, such as providing a natural partition of the coefficient space based on varying resolution levels that are conducive for the proposed multi-scale global-local clustering approach. 


In particular, we represent
\begin{align}\label{eq:wavelet}
    \theta_i(v) = \phi(v)\nTxt{a_{i}} + \sum_{j=0}^\infty\bpsi^\transp_{j}(v)\bbeta_{ij}, v\in \cV,
\end{align}
where $\phi(\cdot)$ is the scaling function with a scalar scaling coefficient \nTxt{$\bm{a}_i$} and $\bpsi_j(\cdot)=[\psi_{j1}(\cdot),\ldots,\psi_{j2^j}(\cdot)]^\transp$ is a vector of wavelet functions that contains $2^j$ orthonormal wavelet bases at resolution level $j$ with the corresponding wavelet coefficients as a vector of $\bbeta_{ij}=[\beta_{ij1},\ldots,\beta_{ij2^j}]^\transp$. Further, the set of finite observed support points $\{v_1,\ldots,v_L\}$ is assumed to be fixed and identical for all units in the data. 
Notably, the sample size is much smaller than the number of observed instances ($n\ll L$) for high-dimensional functional data. 
We assume the number of observed instances to be dyadic of $L=2^{J+1}$ for the ease of wavelet basis expansion \citep{Mallat2008}. If this is not the case, the dimension can be inflated to satisfy this assumption by simply padding zeros around the function \citep{Ma2022DWT}. 
\nTxt{Denote $\by_i=[y_i(v_1),\ldots,y_i(v_L)]^\transp$ and $\btheta_i = [\theta_i(v_1),\ldots,\theta_i(v_L)]^\transp$ as vectors of the observed data and the realizations of the underlying function, respectively. Let $\bphi=[\phi(v_1),\ldots,\phi(v_L)]^\transp$ be scaling function basis at observed locations $(v_1,\ldots,v_L)$ and $\bPsi_j=[\bpsi_{j}(v_1) \ldots \bpsi_{j}(v_L)]^\transp$ be a matrix of wavelet basis functions of dimension $L\times 2^j$ with each row $\bpsi_{j}(v_l)=[\psi_{j1}(v_1),\ldots,\psi_{j2^{j}}(v_l)]^\transp$ as a vector of $2^j$ basis functions for level $j$.}
\nTxt{Given finite observed instances, the expansion representation can be approximated by a finite sum \citep{ray2006functional} with an approximation error of $O(2^{-J})$ \citep{Ma2022DWT} as:
\begin{align}\label{eq:mvn}
    \by_i = \btheta_i + \bepsilon_{i} = \bphi\nTxt{\bm{a}_i} + \sum_{j=0}^J \bPsi_j \bbeta_{ij} + \bepsilon_{i} , \bepsilon_i\overset{\textrm{indep}}{\sim} \bN_L(\mathbf{0}, \bSigma_i), i=1,\ldots,n,
\end{align}
where $\bepsilon_i = [\epsilon_{i{v_1}},\ldots,\epsilon_{i{v_L}}]^\transp$ and $\bSigma_i (v_L\times v_L)$ is a subject-specific covariance matrix.}
Without loss of generality, we assume the zero scaling coefficients \nTxt{$\bm{a}_i$} for the centered data. Of note, the residual covariance matrix may be  unstructured or follow a structured low rank decomposition that facilitates scalability for high-dimensional applications. Specifically, we consider a low-rank decomposition as
\begin{equation}\label{eq:covLowRank}
    \bSigma_i = \bLambda_i \bLambda_i^\transp + \sigma^2_i \bI_L,
\end{equation}
where $\bI_L$ is an identity matrix of size $L$ and $\bLambda_i=\{\lambda_{ilr}\}$ is the factor loading matrix of dimension $L \times K_i$ with $K_i\ll L$ as the number of factors for $i$th unit.

\subsection{Product of Dirichlet Process Prior}
Equation \eqref{eq:mvn} includes two sets of main parameters of interest: wavelet coefficients $\bbeta_{ij} (2^j\times 1)$ for $j=0,\ldots,J$ and covariance $\bSigma_i$. We complete the Bayesian specification by elaborating the priors for these components below.

\nTxt{{\bf Partition introduced by a random measure.} Mixture models are popular tools in Bayesian nonparametrics \citep{Antoniak1974,Escobar1995}. Suppose that probability density can be written as
\begin{align*}
    f_P(y)=\int K(y|\beta)dP(\beta),
\end{align*}
where $K(y|\beta)$ is a specified parametric density on the sample space with mixing parameter $\beta\in B$ and $P$ is a probability measure on $B$. One simple example is a Gaussian kernel $K(y|\beta)=N(y|\beta,\sigma^2)$ with known variance and an unknown mean parameter $\beta$. Under a Bayesian nonparametric framework, we do not restrict $P$ to lie in a fixed, finite-dimensional family (such as a normal or gamma distribution). Instead, we place a prior directly on the entire space $\cF$ of all probability distributions on $B$. A convenient and widely used choice for such a prior is the Dirichlet process (DP). Particularly, the DP prior is defined as $\beta_{ij}\mid  P_j \sim P_j$,  $P_j\sim\textrm{DP}(\alpha,P^*)$, which has a discrete realization almost surely (a.s.) with $P_j=\sum_{h}^\infty w_h\delta_{\beta_h}$, where the atoms $\beta_h\overset{iid}{\sim} P^*$ are sampled from the base measure $P^*$ and independently of the the weights $w_h$. Because $P$ is discrete, when parameters $\beta_i\sim P$ are sampled for different observations, there is a positive probability that two or more of them take exactly the same value. Whenever several $\beta_i$ coincide on the same atom $\beta_h$, those observations are said to belong to the same cluster. This ``tie'' behavior is the key reason the DP induces a random partition of the data: the discreteness of $P$ naturally groups observations into clusters without requiring the number of clusters to be fixed in advance. We refer the readers to \citet{Antoniak1974} and \citet{Muller1996} for additional details.}

\paragraph{Priors on basis coefficients.} We introduce resolution-specific priors on the basis coefficients such that all elements within a resolution are jointly modeled under a particular DP prior, while coefficients across distinct resolutions are modeled under independent DP priors. This structure induces local clustering by independently partitioning coefficients across distinct resolutions, while imposing global clustering for all coefficients within a given resolution level. The proposed prior is naturally justified for wavelet expansions due to the straightforward partition of the coefficient space based on the resolution level, where coefficients at different resolutions operate at distinct scales. Notably, such a specification renders strong advantages in terms of being able to model the $2^{J+1}$ dimensional coefficient vector using $J+1$ independent resolution-specific priors. In particular, this leads to massive gains in computational \nTxt{cost} and enhanced model parsimony due to a reduced number of possible  clustering configurations (see Section \ref{sec:compareModel}), while simultaneously relaxing the restrictive assumptions under global clustering.
In other words, we specify:
\nTxt{
\begin{align}\label{eq:coef_PDP}
    \bbeta_{ij}\mid\mathbf{P}_j&\stackrel{\textrm{indep}}{\sim} \mathbf{P}_j, \mathbf{P}_j\sim \textrm{DP}(\alpha_j, \mathbf{P}^*_j), j=0,\ldots,J,
\end{align}}
where $\alpha_j$ is the concentration parameter and $P^*_j$ is the base measure of the Dirichlet process. This resolution-specific prior implies that coefficients within the same resolution have identical cluster memberships (global clustering), while coefficients in different resolutions are clustered independently (local clustering), which ultimately results in a global-local clustering. This enables the global features corresponding to high-level resolutions (indexed by smaller $j$ ) to be clustered differentially compared to local features corresponding to low-level resolutions (indexed by bigger $j$). Our approach allows accurate clustering of functions that are dominated by a few high-level global features by limiting the contribution of local low-level features via a weighted posthoc criteria when obtaining the consolidated global clusters. Of note, the classical local clustering setting can be recovered as a special case of the proposed approach when each basis coefficient belongs to a distinct partition. Similarly the proposed approach collapses to the classical global clustering method when there are no partitions applied to the basis coefficient vector.

For the DP base measure $\mathbf{P}^*_j$ in (\ref{eq:coef_PDP}), we consider Laplace priors \citep{Park2008BLASSO}. 
In particular, we express the Laplace priors via scale mixtures of normals as 
\begin{align}\label{eq:coefPrior}
    \mathbf{P}^*_j(\bbeta_j)= \bN_{2^j}(0,\bm{\tau}^2_j), \bm{\tau}^2_j=\textrm{diag}(\tau^2_{j1},\ldots,\tau^2_{j2^j}), \tau^2_{jk}\sim \textrm{Exp}(\omega^2), \textrm{ for } j=0,\ldots,J, k=1,\ldots,2^j.
\end{align}


\paragraph{Priors on covariances.} 
The proposed method provides a flexible characterization by considering correlated errors across support points $\{v_1,\ldots,v_L \}$, where the error covariances are subject-specific and clustered across different subjects. 
We impose the low-rank factor model of \eqref{eq:covLowRank} to further improve the scalability with the priors as follows:
\nTxt{\begin{align}\label{eq:LRD_PDP}
(\bLambda_i,\sigma_i)\mid\mathbf{P}_\Sigma &\stackrel{\textrm{indep}}{\sim} \mathbf{P}_\Sigma,  \mbox{ } \mathbf{P}_{\bSigma} \sim \textrm{DP}(\alpha_\Sigma,\mathbf{P}^*_{\bSigma}),
\end{align}}
where $\mathbf{P}^*_{\bSigma} = \mathbf{P}^*_{\bLambda}\times P^*_\sigma$ is a joint measure of the base measures of factor loadings $\mathbf{P}^*_{\bLambda}$ (assignemd a multiplicative Gamma prior \citep{Bhattacharya2011CovFacAnal, IMIFA_Murphy2020}) and variance $P^*_\sigma$ (assigned an inverse Gamma prior). Denote $h_{is}$ as the membership of covariance for $i$th unit. Under the specified base measures and conditional on the clustering allocations, the prior distribution may be written as
\begin{align}
\label{eq:covPrior}
&\lambda_{h_{is}=s,lr}\mid \phi_{slr},\xi_{sr},e_s \sim N(0,\phi^{-1}_{slr}\xi^{-1}_{sr}e_s^{-1}), \thickspace \thickspace\sigma^{-2}_{h_{is}=s}\sim Ga(a,b), \nonumber \\
\phi_{slr}&\sim Ga(3/2,3/2), e_s\sim Ga(a_e,b_e), \mbox{ } \xi_{sr}=\prod_{m=1}^r\delta_{sm}, \thickspace \delta_{s1}\sim Ga(a_1,1), \delta_{sm}\sim Ga(a_2,1), m>1.
\end{align}
The multiplicative variance terms involves local shrinkages $\phi$ that control the shrinkage of individual factor loadings, and global terms $\xi,e$ that control the number of factors and the cluster-specific shrinkage, respectively. A key advantage of the prior specified in 
\eqref{eq:covPrior} is that the number of factors $K_i$ is decided by the data instead of pre-specification \citep{IMIFA_Murphy2020} and can vary among different clusters. 

In addition to the low-rank representation, we alternatively also consider unstructured covariance matrices for theoretical purpose. We assign: 
\nTxt{\begin{align}\label{eq:unStucCov_PDP}
    \bSigma_{i}\mid\mathbf{P}_{\bSigma} \stackrel{\textrm{indep}}{\sim} \mathbf{P}_{\bSigma}, \mathbf{P}_{\bSigma}\sim \textrm{DP}(\alpha_{\Sigma},\mathbf{P}^*_{\bSigma}),
\end{align}}
where $\alpha_{\Sigma}$ is the concentration parameter and $\mathbf{P}^*_{\bSigma}$ is the base measure. 

\subsection{Theoretical Results}\label{sec:thmTmp}

\nTxt{The priors specified above induce a product of Dirichlet process priors on the parameter space that can be denoted as $\Pi^*(\bbeta_i,\bSigma_i)=\prod_{j=0}^J P_j(\bbeta_{ij})\times P_{\bSigma}(\bSigma_i)$. This prior ultimately induces a distribution $\Pi$ on the class of densities on the observed noisy function denoted as $\cF=\{f_P(\by_i)\}$ via the representation $f_P(\by) = \int \phi_{\bSigma}\left(\by-\sum_{j=0}^J\bPsi_j\bbeta_{j}\right)d\Pi^*(\bbeta,\bSigma)$. We denote the resulting class of priors $\Pi$ on $\cF$ as the product of Dirichlet process mixture on functional data (fPDPM). This method generalizes previous work on product mixture priors that was proposed for multivariate time-series data \citep{Kundu2024PDPM} to the case of spatial functions and generic kernel mixture representation in the nonparametric multivariate density estimation literature \citep{Wu2008, Kundu2024PDPM} to the case of functional data analysis relying on suitable basis expansions.}

\nTxt{To assess the validity of the proposed fPDPM framework, we establish several asymptotic properties. Under mild assumptions on the true density that are routinely assumed in multivariate density estimation literature \citep{Wu2008}, we show that the posterior distribution is consistent and contracts toward the true data-generating distribution if the true distribution $f_0\in\cF$. Specifically, we show that the proposed fPDPM is consistent in the weak form of Kullback-Leibler divergence (KL) and the strong form of $L_1$ distance. The proofs follow from the results in \citet{Wu2008} and \citet{Schwartz1965} to show that a positive prior support for arbitrarily small KL neighborhoods of the true density provides the weak consistency of fPDPM. For strong consistency, the basic strategy follows the framework in \citet{Ghosal2017BNP_textbook} by constructing sieves (a compact subset of $\cF$) that have certain required properties and that eventually grow to cover the whole space $\cF$. Additionally, we show that several common priors, such as the Laplace prior on the coefficients and the low-rank representation with the multiplicative Gamma process for the covariance, satisfy the assumptions used in the proof. Formal statements and proofs are provided in Supplementary Material Sections A1 and A2, respectively.}

\subsection{Comparison to Related Local Clustering Methods}\label{sec:compareModel}

\nTxt{
The proposed method is reminiscent of the LPP approach in \citet{LPP_Dunson2009} that specified independent clustering for each basis coefficient in model \eqref{eq:mvn}, with the coefficients being allocated to either a local cluster or a shared global cluster. Specifically, LPP assigns the prior as $\bbeta_{ijk} \sim P_{ijk}, P_{ijk}=z_{ijk}P^{(G)}_{jk} + (1-z_{ijk})P^{(O)}_{jk}$, where $z_{ijk}$ is a binary indicator selecting between the local component $P^{(O)}_{jk}$ and global component $P^{(G)}_{jk}$. Each element of the local component follows an independent DP prior $P^{(O)}_{jk} \sim \textrm{DP}(M_O,{P_{jk}^{(O)}}^*)$, while the global component jointly follows a DP across all resolutions $[P^{(G)}_{jk}]^\transp\sim \textrm{DP}(M_G,{\mathbf{P}^{(G)}}^*)$ for $j=0,\ldots,J,k=1,\ldots,2^j$. The introduction of the global component enables multiple elements of the global component to belong to the same cluster, while the rest elements of the local components are of independent clustering. However, the global cluster can assign common coefficient values for basis coefficients across distinct resolutions that is not ideal for our settings involving wavelet expansions since they resolutions operate on distinct scales and are not expected to have identical values.}

\nTxt{
The proposed approach also has some similarities with another local clustering method proposed by \citet{suarez2016bayesian} (termed SRZ, hereafter). While SRZ models the observed function as the true function subject to an additive normal noise, it adopts a two-stage approach to compute the basis coefficients directly from the observed function via discrete wavelet transform (DWT) and assign a hierarchical Bayesian specification on the means of the computed basis coefficients. Specifically, SRZ applies the DWT on all both observed and true functions to yield corresponding observed and true wavelet coefficients as 
\begin{align}
    \label{eq:DWT_bothSide}\bPsi \by_i = \bPsi\btheta_i + \bPsi\bepsilon_i, \bepsilon\sim\bN(\bf{0},\sigma^2\bI)
\end{align}
where $\bPsi$ of size $L\times L$ is an orthogonal matrix corresponding to the DWT for a certain wavelet family. Due to the rotating invariance property of the multivariate normal distribution \citep{suarez2016bayesian}, $\bPsi\bepsilon_i\overset{d}{=}\bepsilon_i$, which implies that each observed wavelet coefficient can be written as $b_{ijk} \overset{\textrm{indep}}{\sim} N(\beta_{ijk},\sigma^2)$, where $\beta_{ijk}$ is the wavelet coefficient of the true function. SRZ assumes centered functions, leading to zero scaling coefficients \nTxt{$\nTxt{\bm{a}_i}$}, and specifies priors as $\beta_{ijk} \sim P_{jk}, \thickspace P_{jk}\sim\textrm{DP}(M,P_{j}^*), P_{j}^* = \pi_j N(0,\sigma^2\tau_j^2) + (1-\pi_j) \delta_0$, where $\delta_0$ is the point mass at zero.} 

\nTxt{Based on the DWT of \eqref{eq:DWT_bothSide}, instead of using a Bayesian modeling, we can apply a frequentist clustering method on these wavelet coefficients to obtain a two-stage frequentist method. Specifically, we apply a frequentist bi-clustering algorithm of \citet{biClust_Cheng2000} to cluster both subjects and wavelet coefficients at different resolutions. The bi-clustering algorithm generates groups that differentiate one group from the rest of the groups on only a subset of the observed coefficients. By doing so, bi-clustering algorithm performs clustering on sub-domains and avoids using coefficients of all resolutions as the local clustering. We refer to \citet{biClust_Castanho2024} for more details of the bi-clustering algorithm.}

\nTxt{Both SRZ and the frequentist bi-clustering methods are constructed based on a two-stage algorithms of (1) using the DWT to obtain the observed wavelet coefficients, and (2) (locally) clustering on the observed coefficients. SRZ decouples the coefficient estimation and clustering, with the clustering being done on the prior mean of the estimated coefficients instead of the basis coefficients themselves. Moreover, the clustering performance of these two-stage methods is sensitive to the quality of the estimated basis coefficients that might be mis-specified in the presence of low signal-to-noise ratio, spatially correlated errors, and spatial heterogeneity. In contrast, the proposed fPDPM performs basis coefficient estimation and clustering simultaneously, under a wavelet representation of the observed function, which is more robust to underlying noise in the data. Further, the proposed approach results in a more parsimonious modeling by directly clustering the basis coefficients.}

\begin{lemma}\label{lma:cmpComplexity}
    Assume that the maximum number of clusters is $H$ across all basis coefficients. Under the low rank decomposition of covariance matrix in Equation \eqref{eq:covLowRank} with rank  $K\ll L$, the \nTxt{cost} of the proposed model is $O(HJK^2L)+O(nK^2L)+O(HK^3L)$, while the \nTxt{cost} of LPP and SRZ with independent priors on the basis coefficients and independent error terms is $O(HL^2)$.
\end{lemma}

\begin{figure}[!htb]
    \centering
    \includegraphics[width=0.5\linewidth]{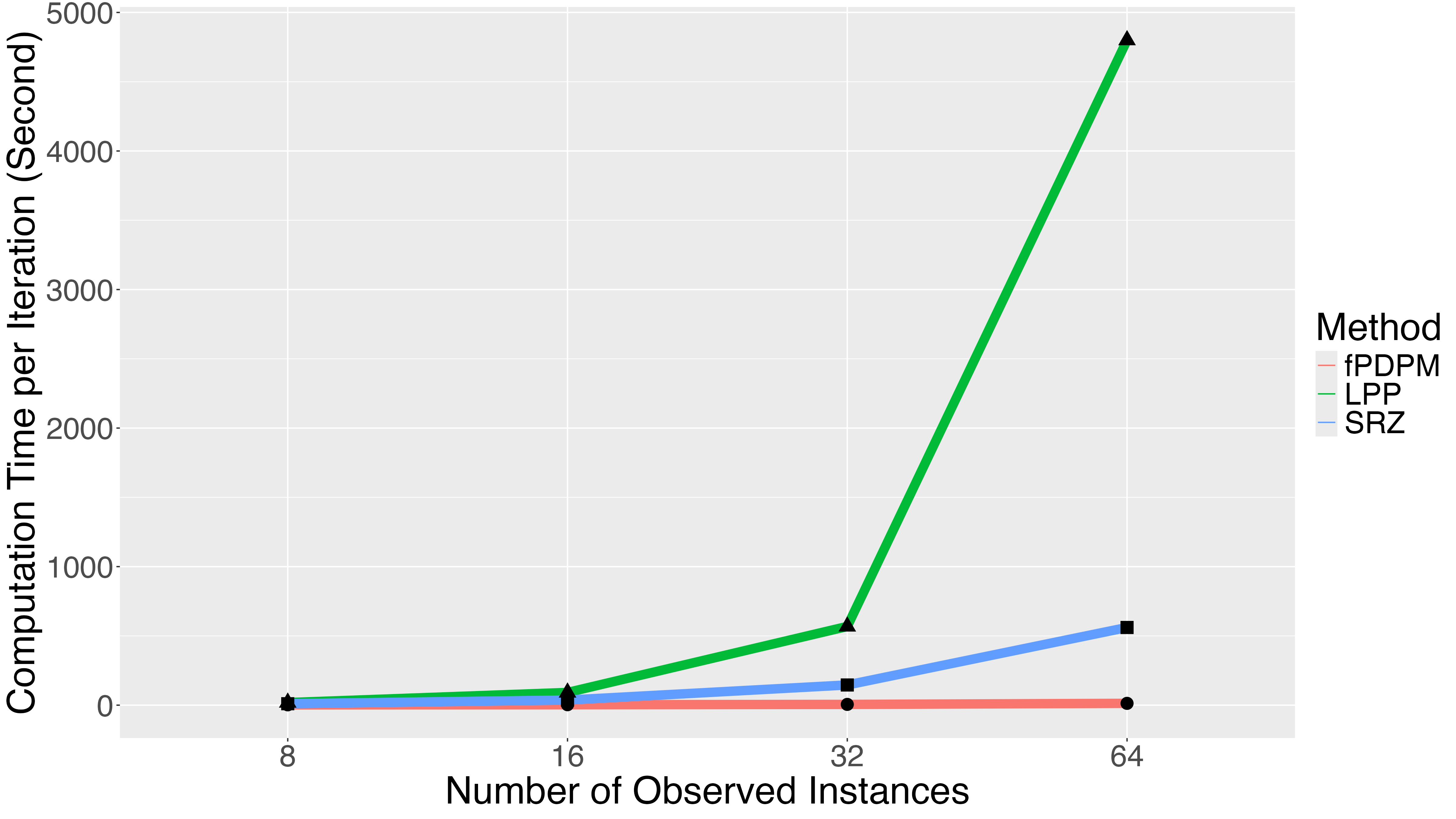}
    \caption{Computation time averaged over $100$ iterations of the proposed method (fPDPM) and the existing local clustering methods (LPP and SRZ). LPP and SRZ are methods proposed by \citet{LPP_Dunson2009} and \citet{suarez2016bayesian}, respectively.}
    \label{fig:timeComplexity}
\end{figure}

\nTxt{A detailed proof is offered in Supplementary Material Section A2.1. The main improvement in computation cost comes from the reduction in the number of membership parameters when comparing to the Bayesian methods of LPP and SRZ. As in our proof in Supplementary Material Section A2, the proposed fPDPM and other local clustering of LPP and SRZ contain three sets of parameters of (1) membership parameters, (2) coefficients, and (3) covariance. For the membership parameters, the proposed fPDPM requires only $J+1$ parameters under wavelet basis functions, while LPP and SRZ assign a random measure $P_{jk}$ to each coefficient and require $L=2^{J+1}$ membership parameter. The result is a straightforward consequence of the Gibbs sampler, and clearly shows the considerable gains in computational cost under the proposed approach for $K \ll L$ especially in high-dimensional settings.}
 Notably, the gain in computational \nTxt{cost} in Lemma \ref{lma:cmpComplexity} holds even under the assumption of correlated errors for the proposed method compared to a more straightforward independent error assumption for the LPP and SRZ. An empirical simulation reporting computation times under various numbers of observed functional instances ($L$) is shown in Figure \ref{fig:timeComplexity}. It is clearly evident that the computational \nTxt{cost} under the proposed method is linear in $L$, wheresa LPP and SRZ have quadratic computational \nTxt{cost} in $L$. This numerical example corroborates the result in Lemma \ref{lma:cmpComplexity}. Note that the Lemma \ref{lma:cmpComplexity} focuses on the computation cost based on the number of the parameters and the corresponding updating operator \citep{cmpComplex_Arora2009} without considering the mixing of different sampler.
 

\section{Posterior Computation and Inference}\label{sec:posteriorComp}
\subsection{Gibbs Sampler}
We implement the fPDPM approach via an efficient Gibbs sampler that proceeds via a slice sampling technique. Using the slice sampler \citep{Walker2007}, the proposed Gibbs sampler facilitates the computation for the stick-breaking representation of the fPDPM (Equation (A1) in the Supplementary Material). The main idea of the slice sampler is to harness the number of possible clusters and reduce an infinite sum into a finite sum by introducing auxiliary parameters. Denote $u_{ij}$ and $u_{is}$ as the auxiliary parameters of the $i$th unit for the coefficients $\bbeta_{ij}$ and the covariance, and $h_{ij}$ and $h_{is}$ as the cluster  membership parameters of the $i$th unit for the coefficients at the $j$th resolution level and the covariance, respectively. Given $u_{ij}$ and $u_{is}$, let $A_j(u_{ij})=\{h_{ij}: w_{j,h_{ij}} > u_{ij}\}$ and $A_\sigma(u_{is})=\{h_{is}: w_{s,h_{is}} > u_{is}\}$ be the sets of the clusters considered for coefficients at resolution level $j$ and the covariance for $i$th unit, respectively. If $u_{ij}\sim U(0,w_{j,h_{ij}})$, the sum over infinite clusters then becomes the sum over a finite set $h_{ij}\in A_j(u_{ij})$. To ensure the set $A_j$ is finite across all units \citep{Walker2007}, we generate enough clusters $h_j^*$ at each iteration such that $\sum_{h=1}^{h_j^*} w_{j,h}>1-u^*_{j}$, where $u^*_{j}=\min_{i}u_{ij}$. The same process is applied to the set $A_\sigma(u_{is})$. Conditional on the membership indicators, the complete data likelihood for the $i$th unit can be easily obtained by
\begin{align*}
    L_i= \phi_{\bSigma_{h_{is}}}\left(\by_i - \sum_{j=0}^J\bPsi_j \bbeta_{h_{ij}j} \right)\prod_{j=0}^J\mathbb{I}(h_{ij}\in A_{j}(u_{ij}))\mathbb{I}(h_{is}\in A_{\sigma}(u_{is})).
\end{align*},
where $\phi_{\bSigma}(\cdot-\bm{\mu})$ is the normal density of mean $\bm{\mu}$ with covariance $\bSigma$ and $\mathbb{I}$ is the indicator function. Due to the low-rank representation for the covariance of \eqref{eq:covLowRank}, the normal likelihood of $\phi_{\bSigma}(\cdot-\bm{\mu})$ can be expressed as a latent factor model \citep{Bhattacharya2011CovFacAnal} with
\begin{align*}
    \by_i = \sum_{j=0}^J\bPsi_j\bbeta_{h_{ij}j} + \bLambda_{h_{is}}\bm{\eta}_{h_{is}} + \tilde{\bepsilon}_{i},
\end{align*}
where $\bm{\eta}_{h_{is}}\sim \bN(0,\bI_{K_{h_{is}}})$ are latent factors with $K_{h_{is}}$ factors, $\bLambda_{h_{is}}=\{\lambda_{h_{is}lr}\}$ is the factor loading matrix, and $\tilde{\bepsilon}_{i}\sim \bN(0,\sigma^2_{h_{is}}\bI)$ is the error with diagonal covariance. 

With the Laplace prior for coefficients of \eqref{eq:coefPrior} and the multiplicative Gamma prior for covariance of \eqref{eq:covPrior}, the conjugacy of the Gibbs sampler is straightforward as shown below.

\begin{algorithm}[!htb]
    \caption{Gibbs sampler for fPDPM}\label{algo1}
    \vspace*{1.5mm}
    \begin{itemize}
    \itemsep 0.2em
        \item[(a)] Update the auxiliary parameters $u_{ij}$ and $u_{is}$ by uniform distributions.
        \item[(b)] For $j=0,\ldots,J$ and $h=1,\ldots,h_j^*$, update the weights of $w_{j,h}$ based on the sticks of $\nu_{j,h}$ by truncated beta distributions. For $h=1,\ldots,h_s^*$, update  $w_{s,h}$ based on $\nu_{s,h}$ by truncated beta distributions.
        \item[(c)] Update membership parameter of $h_{ij}$ with the distribution of $P(h_{ij}=g)\propto \bN\left(\by_i; \sum_{j'\neq j}\bPsi_j \bbeta_{h_{i{j'}}j'} + \bPsi_j\bbeta_{gj} ,\bSigma_i\right) \mathbb{I}(h_{ij}\in A_{j}(u_{ij}))$ and $h_{is}$ with the distribution of $P(h_{is}=s)\propto \bN\left(\by_i; \sum_{j}\bPsi_j \bbeta_{h_{ij}j} + , \bSigma_{s}\right) \mathbb{I}(h_{is}\in A_{\sigma}(u_{is}))$..
        \item[(d)] Update the $g$th cluster of the coefficient with $\bbeta_{gj} \sim \bN_{2^j}\left(\bm{M},\bm{A}^{-1}\right)$, where $\bm{A} = \sum_{i:h_{ij}=g}\frac{\bI_{2^j}}{\sigma_i^2} + (\bm{\tau}^2_{gj})^{-1}, \bm{M} = \bm{A}^{-1} \bPsi_j^\transp \sum_{i:h_{ij}=g}\left(\frac{\by_i-\bLambda_i\bm{\eta}_i}{\sigma_i^2}\right)$ and $\bm{\tau}^2_{gj}=\textrm{diag}(\tau^2_{gj1},\ldots,\tau^2_{gj2^j})$
        \item[(e)] Update the factor matrix $\bm{\eta}_{i}$ with normal distributions.
        \item[(f)] For $s$th cluster of the covariance, update the each row of loading matrix $\bm{\lambda}_{sl}, l=1,\ldots,L$ with normal distributions.
        \item[(g)] For $s$th cluster of the covariance, update variance $\sigma^2_s$ with an inverse Gamma distribution.
        \item[(h)] Update hyperparamters of $\tau^2_{gjk}$ and 
        $\phi_{slr},\xi_{sr}, \delta_{sr}$, and $e_s$.
    \end{itemize}
\end{algorithm}

Details of conditional distributions with the adaptive procedure for the number of factors are provided in the Supplementary Material Section A4. The Gibbs sampler results in posterior samples of coefficients and memberships across all resolution levels and units. The posterior samples are summarized for the following analysis.

\subsection{Posterior Summary}
\paragraph{Local clustering.} The fPDPM results in varying cluster memberships for different resolution levels, leading to local clustering. One can choose to report the optimal clustering allocations for each wavelet resolution, providing granular insights into the clustering mechanisms. 
\paragraph{Global clustering.} Further, one can consolidate these local clustering patterns to inform global clusters as follows. The global clustering coefficients are summarized based on pairwise agreement among the subject-specific coefficients by adapting the method in \cite{suarez2016bayesian}. Denote ${\bm h}^{(r)}_{i}=[h^{(r)}_{i0},\ldots,h^{(r)}_{iJ}]^\transp$ as a vector of members with $h^{(r)}_{ij}$ as the membership at resolution level $j$ of unit $i$ for the $r$th posterior sample with $r=1,\ldots,R$. Define the pairwise distance between units $i$ and $i'$ as a weighted sum as below that is adapted from the criteria proposed in \cite{suarez2016bayesian}
\begin{align}\label{eq:pairDist_mem}
    \bar{d}({\bm h}_i, {\bm h}_{i'})= \sum_{r=1}^R \frac{1}{R} \frac{\sum_{j=0}^J w_j\mathbb{I}(h^{(r)}_{ij}\neq h^{(r)}_{i'j})}{\sum_{j=0}^J w_j}, \mbox{ } \bar{d}({\bm h}_i, {\bm h}_{i'})\in [0,1],
\end{align}
where $w_j$ is the weight for the agreement of coefficients at the resolution level $j$. To focus on the global pattern and reduce the impact of the local features on the overall clustering, we assign a higher weight on the global coefficients and downweight the coefficients corresponding to the remaining resolution levels. We achieve this by setting $w_j=\frac{1}{j+1}$ when $2^j <n$ and $w_j=\frac{1}{2(j+1)}$ if $2^j\geq n$. 
One could also specify alternative weights that downweight coefficients with large variability in clustering or coefficient magnitudes across MCMC iterations, potentially indicating unstable or noisy local features that may not be instrumental in detecting meaningful clustering patterns. Although such alternative weight choices are possible, the proposed weights proved to be robust and provided adequate results in a wide range of numerical experiments.
Based on the pairwise distance $\bar{d}(h_i, h_{i'})$, we build an agglomerative clustering with the complete linkage and decide the memberships by cutting the tree at different levels.

\nTxt{For other existing local clustering methods that perform independent clustering for each basis coefficient, we follow a similar step of building an agglomerative clustering on the posterior pairwise distance matrix. These methods assign a membership parameter for each parameter instead of for each resolution. Denote ${\bm h}^{(r)}_i=[h^{(r)}_{i1},\ldots,h^{(r)}_{iL}]^\transp$ with $h^{(r)}_{il}$ as the membership of $l$th element of coefficient vector $\bbeta_{i}$ for $r$th sample. To obtain the posterior pairwise distance, we revise Equation (11) and sum over all coefficients of $l=1,\ldots,L$. To ensure To ensure a fair comparison, we allocate weights according to resolution levels. For the $l$th element at resolution $j$ with $2^{j}<l<2^{j+1}$, we let $w_l=\frac{1}{j+1}$ when $2^j<n$ and consider $w_l=\frac{1}{2(j+1)}$ if $2^j\geq n$.}

\begin{remark}
    \nTxt{Another possible way to summarize the posterior clustering is by minimizing the loss function, such as the variation of information (VI) \citep{VI_Meila2007} and binder's loss \citep{VI_Binder1978},  based on the quantities $P(h_i=h_{i'}\mid \mathcal{D})=1-\bar{d}(\mathbf{h}_{i},\mathbf{h}_{i'})$ that can be derived based on Equation (11) in the manuscript. For example, the VI loss function can be used for estimating the optimal global clustering by leveraging the approach in \citep{Wade2018}. However, although the optimization problem in \citep{Wade2018} can be theoretically formulated, the practical implementation based on the proposed greedy search algorithm in their paper is not straightforward. In particular, the greedy search approach relies on global clustering allocations from the posterior distribution, which is not readily available under the proposed global-local clustering approach. One potential strategy might be to combine multiple clusterings at local resolutions to obtain a global clustering allocation at each MCMC iteration (in the form of a tuple $(h_{i0},h_{i1},\ldots,h_{iJ})$), and subsequently apply the greedy algorithm in \cite{Wade2018}. However, such an approach is not practical for local clustering in high-dimensions due to a exceedingly large number of candidate clustering configurations possible under local clustering. }
\end{remark}

\paragraph{Function estimation.} 
Denote $R$ posterior samples of coefficients across all resolution levels for unit $i$ as $\{\bbeta^{(r)}_{ij},j=0,\ldots,J,r=1,\ldots,R\}$. Let $\btheta^{(r)}_i=\sum_{j=0}^J\bPsi_j\bbeta^{(r)}_{ij}$ be the estimated function from the $r$th posterior sample of coefficients. We calculate the posterior mean function as $\hat{\btheta}_i=\sum_{r=1}^R\btheta^{(r)}_i/R=[\hat{\theta}_i(v_1),\ldots,\hat{\theta}_i(v_L)]^\transp$. 

\section{Simulation Studies}\label{sec:sim}
A series of simulation studies are conducted to empirically investigate the performance of the proposed fPDPM, while comparing it to competing approaches. We generate functional data that are compatible with spatial transcriptomics data applications and consider two-dimensional ($d=2$) spatial functions of size $L=2^5\times 2^5 = 1,024$  with sample size $n=300$. The spatial functions of the larger size of $L=1,024$ match the size of the real spatial transcriptomics data in Section \ref{sec:realDf}. Given that it is not possible to implement existing local clustering methods for these dimensions due to the infeasible computational burden, we consider a second set of toy simulation studies involving smaller images of dimension $L=2^3 \times 2^3=64$ with sample size $n=20$ to compare the modeling performance with other local clustering methods such as LPP and SRZ. However, it should be noted that our main focus is on larger size images as encapsulated via the first set of simulations with image size $2^5 \times 2^5$. 

\subsection{Experiment Setting}\label{sec:simDfGen}
\begin{figure}[!htb]
    \centering
    \includegraphics[width=\textwidth]{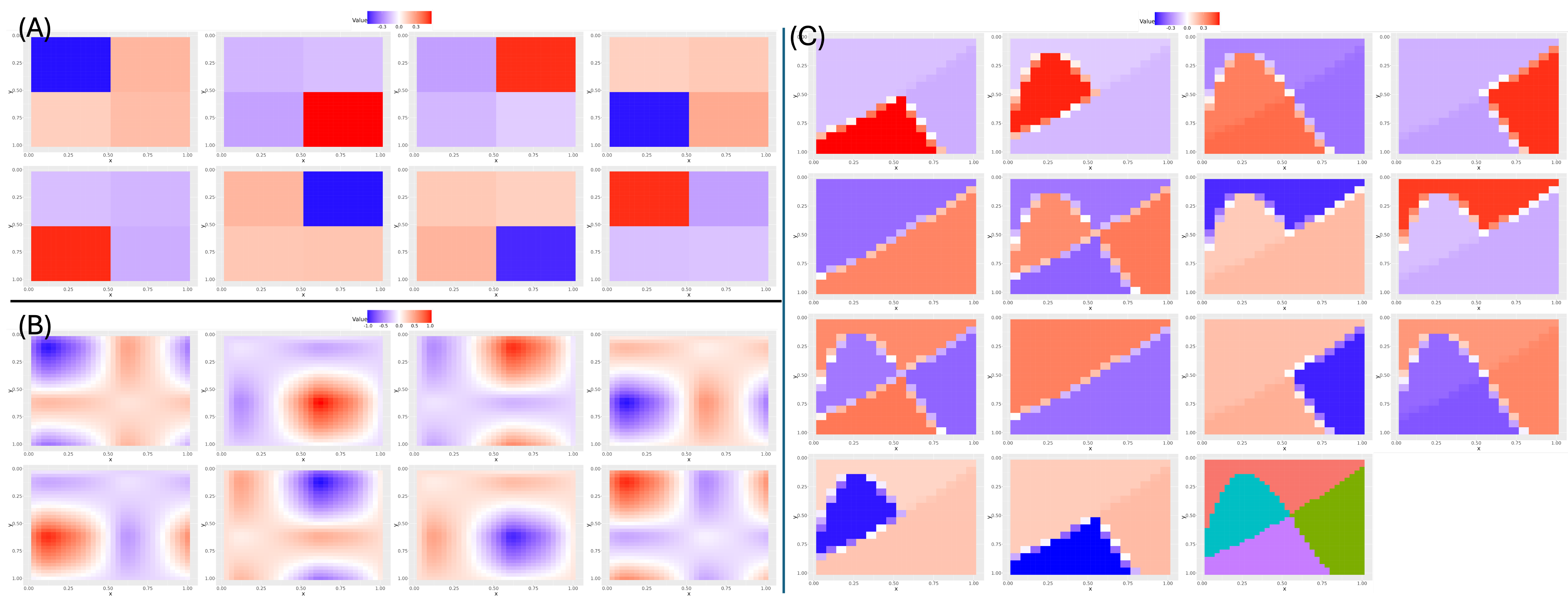}
    \caption{\nTxt{All noise-free simulated images of size $L=1,024$ from scenarios of (A) correct specified model, (B) mis-specified of basis function with Daubechies orthonormal compactly supported wavelet of length eight global clustering, and (C) mis-specified of data generating mechanism for spatial heterogeneity. Panels (A) and (B) share the same wavelet coefficients with eight different values but are generated from different basis functions. Panel (C) generates image directly without basis expansion on four areas shown in the right-most panel of the bottom row. See more details in Section \ref{sec:simDfGen}.}}
    \label{fig:truDf}
\end{figure}

\paragraph{Data-generating mechanism.}
\nTxt{For both sizes of images, we generate functional data under three different scenarios with different clustering structures of (1) correctly specified model, (2) mis-specified basis function, and (3) mis-specified generating mechanism of spatial heterogeneity. The first two scenarios are based on wavelet expansions, while the last one generates the data directly on the image without considering the basis expansion. For the scenario 1, all basis coefficients are zero, $\bbeta_{ij}=\mathbf{0}, j>0$, except for the highest resolution, $\beta_{i0}\neq0$. We consider eight and three different non-zero coefficients for large- and small-size images, respectively. While scenario 2 generated data from the wavelet basis functions with the same basis coefficients as scenario 1 ($\beta_{i0}\neq0, \bbeta_{ij}=\mathbf{0}, j>0$), it considers mis-specified basis functions. Specifically, we generate images by Daubechies \citep{Daubechies1993} orthonormal compactly supported wavelet of length eight and fit the models with the Haar basis functions \citep{Haar1910}. In contrast to scenarios 1 and 2 involving the wavelet expansion, scenario 3 generates images directly and introduces spatial heterogeneity with the clustering patterns varying over the spatial domain. Using a square image (i.e. $v_l\in[0,1]^2$), we assign four non-overlapping areas decided by two functions of $v_2=-7(v_1-0.5)^3+2v_1^2-0.1$ and $v_2=cos(v_1+0.5)$ (see Panel (C) of Figure \ref{fig:truDf}). For each area in large-size images, we consider two potential values of $\theta_i^{(m)}(v_l)\in\{-0.35,0.35\},m=1,3$ and $\theta_i^{(m)}(v_l)\in\{-0.4,0.4\},m=2,4$. We center the data to ensure zero scaling coefficients and exclude two specific patterns that all four areas share the same signs of $\theta_i^{(1)}(v_l)\theta_i^{(3)}(v_l)=(0.35)^2$ and $\theta_i^{(2)}(v_l)\theta_i^{(4)}(v_l)=(0.4)^2$ to ensure a higher signal-to-noise ratio after centering the data, resulting in $14$ different patterns. For small-size images of spatial heterogeneity, due to the small sample size, we consider the same regions with three patterns among $14$ patterns and assign $\lvert\theta_i^{(m)}(v_l)\rvert=0.45, m=1,3$ and $\lvert\theta_i^{(m)}(v_l)\rvert=0.5, m=2,4$}

\begin{figure}[!htb]
    \centering
    \includegraphics[width=\textwidth]{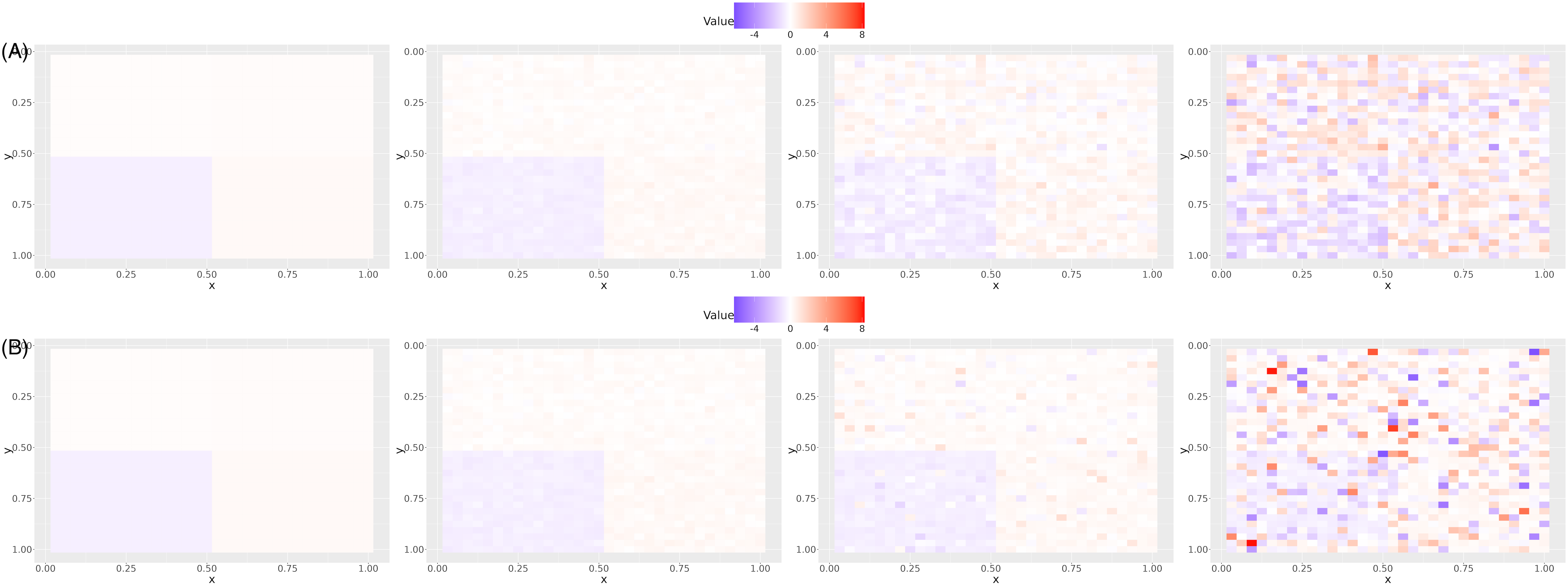}
    \caption{\nTxt{Illustrations of simulated images of size $L=1,024$ from scenarios of correctly-specified model under (A) independent error and (B) correlated error structures. Unobservable noise-free images with three different levels of high, medium, and low SNRs images are shown from the left to right.}}
    \label{fig:ObsImg_sim}
\end{figure}

\nTxt{We consider two different error structures of independent and correlated errors by using the low-rank decomposition for the error term of Equation \eqref{eq:covLowRank} and three levels of signal-to-noise ratio (SNR). We consider a zero loading matrix for independent error. For correlated error, we let loading matrices be non-zero with different numbers of factors. Let $\rho^2_{s}=\sum_{j,k}\beta^2_{jk}$ and $\rho^2_e=\textrm{tr}(\bSigma)$, where $\textrm{tr}$ is the trace function. We calculate the SNR \citep{SNR_Hu2025} as $\frac{\rho^2_s}{\rho^2_s+\rho^2_e}\in[0,1]$. The basis coefficients of scenarios 1 and 2 are directly available from the generating mechanism. For scenario 3, since the underlying images are directly generated without considering basis expansion, we apply the Haar basis expansion on each noise-free image to obtain the corresponding basis coefficients for the calculation of SNRs. We consider SNRs around $0.92, 0.60$, and $0.10$ as high, medium, and low SNRs, respectively. Figure \ref{fig:ObsImg_sim} shows two error structures on three levels of SNRs for one image generated from the correct specified model. Note that under a similar level of the SNR, the independent errors spreads more uniformly over the whole image, while the correlated errors are more concentrated at certain regions of the image. For large-size images, Scenarios 1 and 2 consider both error structures under three SNR levels, while Scenario 3 includes only the high and medium SNRs due to the high degree of model misspecification. For small-size images, all scenarios consider both error structures under high and medium SNRs because of the limited sample size. More details of data-generating mechanism are available in Supplementary Material Section A5.}

\paragraph{Performance metrics and benchmarks.} 
\nTxt{We implemented two variants of the proposed fPDPM model with and without spatially correlated noise terms (denoted as fPDPM and fPDPMi respectively) that perform local clustering. We initialize fPDPM with three different factors of $K\in\{1,3,10\}$ for the low rank covariance and report results for these choices. Due to the intense computation burden, existing Bayesian local clustering methods of LPP and SRZ are infeasible for larger image. For frequentist comparison, to the best of our knowledge, no existing local clustering method is available. Instead, we follow SRZ to obtain the observed coefficients based on Equation \eqref{eq:DWT_bothSide} with Haar functions and apply a bi-clustering algorithm \citep{biClust_Castanho2024} at each resolution on the observed coefficients as a frequentist benchmark. Specifically, we consider the bi-clustering algorithm of \citet{biClust_Cheng2000} to generate groups that differentiate one group from the rest of the groups on only a subset of the observed coefficients, which cluster subjects on subdomains and avoid using coefficients of all resolutions as the local clustering. We also compared the performance with global clustering approaches such as the joint Dirichlet process model (DPM) and a K-means clustering performed on functional principal components (PCA-KM) as proposed in \citep{MFPCA_Happ2018}. For PCA-KM, we select the smallest number of principal components that explain at least $95\%$ variance and decide the number of clusters by maximizing the average silhouette width \citep{Silhouette_Rousseeuw1987}. For the second simulation setting involving smaller images, we also benchmarked against existing local clustering methods, such as a variant of the LPP method proposed by \citet{LPP_Dunson2009} and SRZ proposed by \citet{suarez2016bayesian}. For a fair comparison, we use a Haar wavelet basis expansion in SRZ implementation to obtain the observed Haar coefficients and in the implementation of both fPDPM and LPP. More details of the implementation of these benchmarking methods are presented in Section \ref{sec:compareModel}. The convergence diagnostics of the proposed fPDPM are provided in Supplementary Material Section A5.2.}

We evaluate the clustering performance via the adjusted rand index (ARI) \citep{ARI_Rand1971}, and function estimation error via mean squared error (MSE), noting that the clustering performance is of primary interest. ARI ranges between zero and one with the ARI of one representing that the estimated and the true clustering are perfectly aligned. 
We calculate the ARI based on the memberships for the basis coefficients corresponding to the mean clustering, and ignore the variations due to the covariance clustering. \nTxt{For local clustering methods and the two-stage methods of SRZ and frequentist bi-clustering, we report ARI based on global clustering that are obtained via hierarchical clustering on the pairwise distance. Specifically, for fPDPM, fPDPMi, and two-stage frequentist bi-clustering the pairwise distance of \eqref{eq:pairDist_mem} sums over $J$ resolution, while that from LPP and SRZ sums over $L$ coefficients.}
We cut the tree at the level of the true group number to report the ARI metrics. Due to a large number of the true clusters in Scenario 2 of large-size images, we choose to compute the ARI at each resolution level with non-zero coefficients for $j=0,1$ and $2$. For the global clustering approaches, the ARI from the different resolutions are obtained from comparing the global membership to the true local membership at different resolutions. On the other hand, the function estimation performance (at the observed locations) is obtained by the MSE as $\sum_{i=1}^n(\hat{\btheta}_i - \btheta_i)^2/nL$.
All results are obtained by averaging over $30$ independent replicates for each simulation scenario. 

\subsection{Results of Large-Size Images}
\nTxt{Simulation results of clustering performance for the large-size images are summarized in Table \ref{tab:LargeImg} with the estimation shown in the Supplementary Material Section A5.3. When the data-generating mechanism is correctly specified, the proposed fPDPM variants of $\textrm{fPDPM}_3$ and $\textrm{fPDPM}_{10}$ achieve near-perfect clustering accuracy across all SNR levels. Specifically, both variants consistently yield ARIs above $93$, even under correlated errors, confirming their ability to accurately recover the true clustering structure. In contrast, $\textrm{fPDPM}_1$ and fPDPMi (which assumes independent covariances) show degraded performance under medium and low SNRs, with average ARIs dropping below $54$ under correlated errors. This demonstrates the importance of explicitly modeling residual correlation. The frequentist biclustering method shows intermediate and relatively stable performance with ARI $50-60$, regardless of the error structure, which suggests limited sensitivity to noise correlation but poor alignment with the true clusters. The global clustering methods of Dirichlet process mixture (DPM) and PCA–KM achieve moderate clustering accuracy that is consistently worse than the proposed fPDPM variants, suggesting that conventional global mixture models or dimension-reduction approaches are less effective in recovering multiresolution spatial patterns. Under basis-function misspecification, all proposed fPDPM variants remain highly robust, with ARIs mostly exceeding $98$ and $90$ across all SNR levels under the independent and correlated error, respectively. In contrast, the fPDPMi shows a drop in ARI of $78.4$ when the SNR is low under the correlated error, which implies the importance of incorporating the spatial correlated error. The frequentist biclustering method yields lower ARIs ($<66$) across noise structures, while the global clustering approaches DPM and PCA–KM achieve moderate accuracy ($<85$) but decline under correlated errors with a drastically decrease when the SNR drops. Under the most complex scenario, where the data-generating process is misspecified, both the proposed fPDPM variants and fPDPMi achieve excellent clustering accuracy, with ARI $>90$ across all SNR levels and error structure. The results indicate that the local clustering structure captured by the fPDPM framework remains highly robust even when the assumed model differs from the true generating process. Competing methods show markedly weaker performance, with biclustering yielding ARIs around $<60$ and global clustering methods  achieving only moderate accuracy ($<80$).}

\nTxt{Across all scenarios, the proposed fPDPM framework consistently outperforms existing clustering approaches in recovering the underlying functional partitions. The frequentist benchmark of biclustering method considers the same wavelet basis expansion with the resolution information, but its reliance on noisy observed coefficients and the absence of model-based borrowing lead to consistently lower ARIs. Global methods (DPM and PCA–KM) further degrade under correlated noise or lower SNRs. Overall, fPDPM delivers superior and stable clustering for high-dimensional functional data.}

\begin{table}[!htb]
    \centering
    \caption{Simulation results of large-size images under independent and correlated error structures with various levels of signal-to-noise ratios (SNR). Clustering performance is evaluated in ARI. Results are obtained as the mean with the standard deviation in parentheses over 30 replicates. $\textrm{fPDPM}_{K}$ and fPDPMi, are two variants of the proposed methods with correlated covariance initialized by $K$ factors and independent covariances, respectively. All values have been multiplied by $100$}
    \label{tab:LargeImg}
    \begin{tabular}{crrrrrr}
        & \multicolumn{6}{c}{Scenario 1: correctly specified model}\\
        \cmidrule{2-7}
        & \multicolumn{2}{c}{High SNR} & \multicolumn{2}{c}{Medium SNR} & \multicolumn{2}{c}{Low SNR}\\
        \cmidrule(lr){2-3}  \cmidrule(lr){4-5}  \cmidrule(lr){6-7} 
        & Independent & Correlated & Independent & Correlated & Independent & Correlated\\
        \hline
        $\textrm{fPDPM}_{1}$  &  100 (0)  &  95.3 (10.7)  &  86.2 (17.9)  &  53.0 (14.3)  &  88.7 (11.3)  &  52.2 (30.6)  \\
        $\textrm{fPDPM}_{3}$  &  100 (0)  &  99.9 (0.258)  &  100 (0)  &  100 (0)  &  97.2 (6.00)  &  92.3 (14.0)  \\
        $\textrm{fPDPM}_{10}$  &  100 (0)  &  100 (0)  &  100 (0)  &  100 (0)  &  93.6 (8.00)  &  98.5 (3.80)  \\
        fPDPMi  &  83.8 (7.89)  &  88.7 (11.8)  &  86.4 (11.3)  &  25.6 (9.21)  &  91.8 (9.46)  &  53.3 (10.1)  \\
        Bi-clustering  &  60.1 (2.98)  &  60.1 (3.01)  &  56.3 (1.54)  &  57.0 (1.07)  &  47.4 (13.7)  &  53.2 (3.37)  \\
        DPM  &  83.0 (4.46)  &  85.9 (5.35)  &  84.6 (7.53)  &  37.5 (12.3)  &  60.3 (8.71)  &  1.00 (0.730)  \\
        PCA-KM  &  80.8 (9.21)  &  81.4 (12.2)  &  78.0 (13.28)  &  81.8 (10.5)  &  55.0 (22.7)  &  11.8 (2.00)  \\
    \end{tabular}
    ~
    \begin{tabular}{crrrrrr}
        & \multicolumn{6}{c}{Scenario 2:  mis-specified model in basis function}\\
        \cmidrule{2-7}
        & \multicolumn{2}{c}{High SNR} & \multicolumn{2}{c}{Medium SNR} & \multicolumn{2}{c}{Low SNR}\\
        \cmidrule(lr){2-3}  \cmidrule(lr){4-5}  \cmidrule(lr){6-7} 
        & Independent & Correlated & Independent & Correlated & Independent & Correlated\\
        \hline
        $\textrm{fPDPM}_{1}$  &  100 (0)  &  100 (0)  &  100 (0)  &  100 (0)  &  100 (0)  &  96.6 (7.21)  \\
        $\textrm{fPDPM}_{3}$  &  100 (0)  &  100 (0)  &  100 (0)  &  100 (0)  &  100 (0)  &  100 (0)  \\
        $\textrm{fPDPM}_{10}$  &  100 (0)  &  100 (0)  &  100 (0)  &  90.6 (16.0)  &  98.1 (5.86)  &  99.9 (0.263)  \\
        fPDPMi  &  100 (0)  &  100 (0)  &  98.4 (5.18)  &  93.1 (11.9)  &  98.9 (3.11)  &  78.4 (11.5)  \\
        Bi-clustering  &  65.2 (9.16)  &  65.3 (8.04)  &  51.2 (4.30)  &  52 (2.03)  &  43.2 (6.99)  &  46 (4.68)  \\
        DPM  &  84.4 (6.37)  &  84.4 (6.46)  &  82.3 (9.19)  &  37.6 (14.7)  &  65.8 (11.6)  &  0.900 (0.512)  \\
        PCA-KM  &  78.3 (9.19)  &  66.5 (17.1)  &  70.2 (6.89)  &  78.8 (10.3)  &  50.8 (20.8)  &  13.3 (1.19)  \\
    \end{tabular}
    ~
    \begin{tabular}{crrrr}
    & \multicolumn{4}{c}{Scenario 3: mis-specified model of spatial heterogeneity}\\
        \cmidrule{2-5}
        & \multicolumn{2}{c}{High SNR} & \multicolumn{2}{c}{Medium SNR} \\
        \cmidrule(lr){2-3}  \cmidrule(lr){4-5} 
        & Independent & Correlated & Independent & Correlated \\
        \hline
        $\textrm{fPDPM}_{1}$  &  99 (2.92)  &  99.4 (2.04)  &  98.9 (2.538)  &  97.3 (4.16)\\
        $\textrm{fPDPM}_{3}$  &  100 (0)  &  100 (0)  &  100 (0)  &  100 (0)  \\
        $\textrm{fPDPM}_{10}$  &  89.9 (9.80)  &  95.9 (4.84)  &  99.8 (0.51)  &  93.3 (10.4)\\
        fPDPMi  &  100 (0)  &  100 (0)  &  100 (0)  &  99.9 (0.185)\\
        Bi-clustering  &  59.6 (5.23)  &  61.7 (5.57)  &  44.7 (5.59)  &  45.9 (7.71)\\
        DPM  &  66.4 (7.98)  &  64.9 (7.61)  &  66.0 (5.93)  &  39.9 (7.19) \\
        PCA-KM  &  75.7 (6.33)  &  75.0 (8.11)  &  76.7 (9.11)  &  80.9 (6.79)\\
    \end{tabular}
\end{table}

\subsection{Results of Small-Size Images}
\nTxt{Clustering performance for the small-size images are shown in Table \ref{tab:SmallImg} with the estimation shown in the Supplementary Material Section A5.3. Across all scenarios and error structures, the proposed fPDPM variants achieve excellent clustering accuracy (ARI $>92$) even with limited sample size, except for $\textrm{fPDPM}_1$ under correlated errors at medium SNR (ARI: $80.2$). fPDPMi also performs well overall but shows degradation under correlated noise with medium SNR (scenario 1: $49.3$ and scenario 3: $75.5$), underscoring the benefit of explicitly modeling residual dependence. For Scenario 1, LPP performs much worse than the fPDPM variants (ARI $<78$), while SRZ struggles even more (ARI $<55$). Under Scenario 2, LPP performs well overall (ARI: $100$) but its accuracy drops under correlated errors at medium SNR (ARI: $91.3$), whereas SRZ remains moderate (ARI $<57$). In Scenario 3, LPP achieves satisfactory performance (ARI $81 - 94$) but remains inferior to both fPDPM and fPDPMi, while SRZ performs comparably to fPDPM under high SNR but shows lower accuracy under medium SNR (ARI $\leq97$). Both LPP and SRZ also require substantially greater computation time than the proposed fPDPM framework. The biclustering and global clustering methods (DPM and PCA–KM) perform poorly across all scenarios, particularly under correlated noise. Overall, fPDPM achieves the most accurate and stable clustering for small-size images while maintaining computational efficiency.}

\begin{table}[!htb]
    \centering
    \caption{Simulation results of small-size images under independent and correlated error structures with two levels of signal-to-noise ratios (SNR). Clustering performance is evaluated in ARI. Results are obtained as the mean with the standard deviation in parentheses over 30 replicates. $\textrm{fPDPM}_{K}$ and fPDPMi are two  variants of the proposed methods with correlated covariance initialized by $K$ factors and independent covariances, respectively. All values have been multiplied by $100$}
    \label{tab:SmallImg}
    \begin{tabular}{crrrr}
        & \multicolumn{4}{c}{Scenario 1: correctly specified model}\\
        \cmidrule{2-5}
        & \multicolumn{2}{c}{High SNR} & \multicolumn{2}{c}{Medium SNR}\\
        \cmidrule(lr){2-3}  \cmidrule(lr){4-5}
        & Independent & Correlated & Independent & Correlated\\
        \hline
        $\textrm{fPDPM}_{1}$  &  100 (0)  &  100 (0)  &  100 (0)  &  100 (0)  \\
        $\textrm{fPDPM}_{3}$  &  100 (0)  &  100 (0)  &  95.1 (15.4)  &  93.6 (20.2)\\
        $\textrm{fPDPM}_{10}$  &  100 (0)  &  100 (0)  &  100 (0)  &  93.6 (20.2)\\
        fPDPMi  &  100 (0)  &  100 (0)  &  94.9 (11.9)  &  49.3 (27.9) \\
        LPP  &  60.6 (27.6)  &  77.2 (30.4)  &  74.1 (29.0)  &  57.8 (38.7) \\
        SRZ  &  18.2 (24.7)  &  16.8 (26.2)  &  26.6 (22.4)  &  54.5 (20.2)\\
        Bi-clustering  &  -0.600 (5.19)  &  -0.600 (5.19)  &  -0.600 (5.19)  &  -0.600 (5.19)\\
        DPM  &  100 (0)  &  95.1 (15.6)  &  70.3 (18.1)  &  52.8 (12.6) \\
        PCA-KM  &  90.4 (20.4)  &  80.1 (27.4)  &  76.8 (23.7)  &  62.8 (23.5)\\
    \end{tabular}
    ~
    \begin{tabular}{crrrr}
        & \multicolumn{4}{c}{Scenario 2: mis-specified model in basis function}\\
        \cmidrule{2-5}
        & \multicolumn{2}{c}{High SNR} & \multicolumn{2}{c}{Medium SNR}\\
        \cmidrule(lr){2-3}  \cmidrule(lr){4-5}
        & Independent & Correlated & Independent & Correlated\\
        \hline
        $\textrm{fPDPM}_{1}$  &  100 (0)  &  100 (0)  &  93.7 (20.0)  &  80.2 (32.1)\\
        $\textrm{fPDPM}_{3}$  &  100 (0)  &  100 (0)  &  100 (0)  &  98.7 (4.08)\\
        $\textrm{fPDPM}_{10}$  &  100 (0)  &  100 (0)  &  100 (0)  &  92.7 (19.0)\\
        fPDPMi  &  100 (0)  &  100 (0)  &  100 (0)  &  75.5 (32.1)\\
        LPP  &  100 (0)  &  100 (0)  &  98.6 (4.40)  &  91.3 (23.3)\\
        SRZ  &  55.9 (19.9)  &  56.4 (19.5)  &  55.1 (19.8)  &  56.8 (22.2)\\
        Bi-clustering  &  -0.600 (5.19)  &  -0.600 (5.19)  &  -0.600 (5.19)  &  -0.600 (5.19)\\
        DPM  &  91.5 (19.7)  &  91.5 (19.7)  &  76.4 (12.3)  &  51.2 (7.53)\\
        PCA-KM  &  85.2 (23.1)  &  78.9 (20.7)  &  70.4 (25.0)  &  73.9 (19.9)\\
    \end{tabular}
    ~
    \begin{tabular}{crrrr}
    & \multicolumn{4}{c}{Scenario 3: mis-specified model of spatial heterogeneity}\\
        \cmidrule{2-5}
        & \multicolumn{2}{c}{High SNR} & \multicolumn{2}{c}{Medium SNR} \\
        \cmidrule(lr){2-3}  \cmidrule(lr){4-5} 
        & Independent & Correlated & Independent & Correlated \\
        \hline
        $\textrm{fPDPM}_{1}$  &  100 (0)  &  100 (0)  &  100 (0)  &  100 (0)  \\
        $\textrm{fPDPM}_{3}$  &  100 (0)  &  100 (0)  &  100 (0)  &  100 (0)  \\
        $\textrm{fPDPM}_{10}$  &  100 (0)  &  100 (0)  &  100 (0)  &  100 (0)  \\
        fPDPMi  &  100 (0)  &  100 (0)  &  100 (0)  &  100 (0)\\
        LPP  &  91 (20.8)  &  91.5 (19.9)  &  93.7 (19.9)  &  81.2 (39.7)\\
        SRZ  &  100 (0)  &  100 (0)  &  97 (9.36)  &  94.9 (16.1)\\
        Bi-clustering  &  -0.600 (5.19)  &  -0.600 (5.19)  &  -0.600 (5.19)  &  -0.600 (5.19)\\
        DPM  &  100 (0)  &  100 (0)  &  74.9 (13.3)  &  58.9 (11.8)\\
        PCA-KM  &  83.7 (20.9)  &  79.2 (27.9)  &  93.6 (14.1)  &  84.9 (19.3)\\
    \end{tabular}
\end{table}

\nTxt{Overall, the comparison among Bayesian local clustering frameworks highlights clear distinctions in modeling strategy and performance. LPP assigns a Dirichlet process prior independently to each coefficient, while SRZ places a DP on the observed wavelet coefficients, both of which limit information sharing across resolutions and increase computational demand. In contrast, fPDPM clusters coefficients jointly within each resolution through a hierarchical DP, enabling efficient borrowing of information across scales and yielding both higher accuracy and faster computation. The frequentist biclustering method also utilizes wavelet coefficients at each resolution but fails to encapsulates the basis in the model, leading to the less stable clustering under noise or correlation. These results collectively demonstrate the advantage of the resolution-specific Bayesian construction in fPDPM over existing local and global approaches.}

\section{Spatial Transcriptomics of Breast Cancer}\label{sec:realDf}
 We applied fPDPM to a publicly available spatial transcriptomics dataset of HER2-positive breast tumor from \citet{dataST_Andersson2021}. With the advancement of technologies, measurements of gene expression with spatial localization information have become available, which facilitates the study of spatial patterns of gene expression \citep{SpatialPCA_Shang2022, Moses2022}. These spatial patterns reflect the transcriptional activity of genes \citep{Marx2021} across different tissue locations. Biologically, neighboring spots share a common composite of cell types and have similar expression \citep{SpatialPCA_Shang2022}, which potentially indicates a smooth global pattern. However, local heterogeneity arises due to tissue structure and the interaction among cells \citep{SpatialPCA_Shang2022}. Additionally, spatial transcriptomics data are usually noisy \citep{Rao2021}, and these noises are spatially correlated due to experimental procedures \citep{Du2024}. For example, non-uniform cell permeabilization for spatial transcriptomics chips can lead to higher expression in adjacent spots, while more distant spots exhibit lower expression \citep{Du2024}. Our primary goal is to identify global clusters of genes with similar profiles of spatially varying expression levels, while accounting for local heterogeneity that is well-known in spatial transcriptomics applications. Such local heterogeneity is exemplified via spatial clustering methods in spatial transcriptomics literature that focus on finding spatially coherent regions with homogeneous gene expression profiles in the tissue. The proposed analysis provides a spatially informed method of clustering genes within the tissue that is distinct from scalar gene clustering approaches in literature \citep{Mathur2018}.
 

\subsection{Data Overview and Pre-processing}
We used gene expression data of patient that consists of three consecutive cryosection samples (H1, H2, and H3, henceforth), and compared the clustering reliability of the genes across the tissues, along with the biological interpretation of the clusters. 
\nTxt{Because these three sections are obtained from the same tissue, they provide a natural biological validation setting for assessing the reproducibility of gene clustering across adjacent samples, which we quantify using the ARI.}
Based on the clustering, we construct the co-expression network by Bayesian graphical lasso \citep{BayGLasso} to characterize each cluster. 
\nTxt{Note that the network itself is a postprocessing step to characterize each cluster discovered by the proposed fPDPM model for the biological interpretation.}
We present the results for cluster memberships for the H1 tissue sample in the main manuscript, with the results for the remaining two cryosection samples in Supplementary Material Section A6. We retained genes with more than $20$ non-zero spots and spots with at least $300$ non-zero expression genes \citep{SpatialPCA_Shang2022}. After removing $21$ genes that are identified as technical artifacts \citep{dataST_Andersson2021}, we obtained $10,053$ genes measured on $607$ spots for three samples. For each sample, we normalized the expression data with mean zero and unit variance and selected spatially variable genes \citep{SpatialPCA_Shang2022}, which results in $302$ genes. We padded zeros in all samples to attain the dimensions of $32$ by $32$. 
\nTxt{We consider the Haar basis with the concentration hyperparameter of three in the DP, as the model specified in the Simulation Studies.}
We ran fPDPM of $10,000$ iterations and discarded the burn-in period. \nTxt{The convergence diagnostics are provided in Supplementary Material Section A6.2.}

\subsection{Analysis Results}
We fit the fPDPM model on the pre-processed spatial transcriptomics data to discover resolution-specific clusters. Subsequently, the local clusters under fPDPM are consolidated into global clusters using hierarchical clustering on the distance measure in (\ref{eq:pairDist_mem}). Based on the Silhouette method \citep{Silhouette_Rousseeuw1987}, we cut the estimated hierarchical clustering at $3$ clusters with each cluster contains $36, 142$, and $124$ genes, respectively (see Supplementary Material Table A4 for the full list).
If we focus on common genes among three samples, we obtain a reasonably high clustering reproducibility among three samples, which is measured by ARI of $0.63$ (H1 and H2) and $0.59$ (H1 and H3). 
We identify the first cluster as the innate immune system with $7$ out of $36$ genes in the pathway. For example, CXCL12 activates the leukocyte, a member of the innate immune system, and facilitates tumor cell invasion for breast cancer \citep{HER2_innateImmune}. The second group belongs to the signal transduction pathway and $42$ genes out of $142$ genes are part of the pathway. For example, Akt1 is a member of AKT pathway, which is a common mutated pathway related to the drug resistance in HER2 positive breast cancer \citep{HER2_Akt}. The third group is characterized by the immune system with $50$ of $124$ genes belonging to the immune system pathway. For example, CD4 is a receptor on the surface of many immune cells such as T cells, and the stimulated CD4 improves the treatment efficacy of HER2 positive breast cancer \citep{HER2_CD4}. 



\begin{figure}[!htb]
    \centering
    \includegraphics[width=\linewidth]{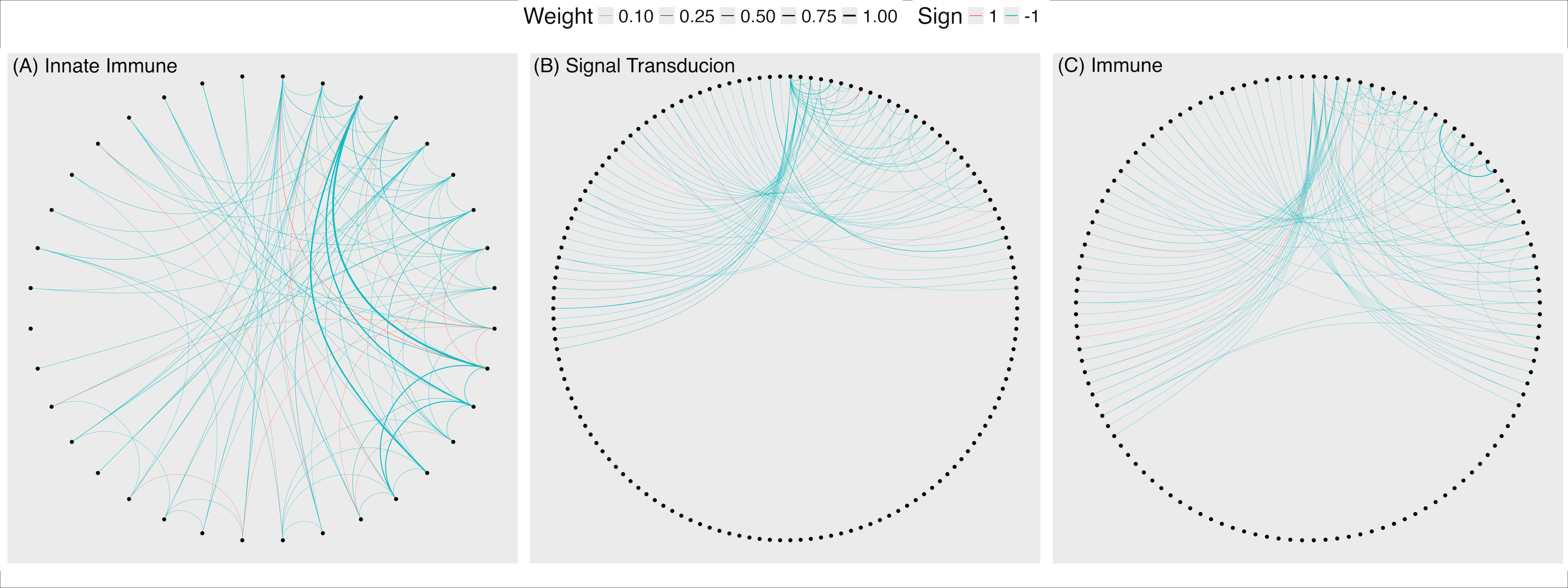}
    \caption{The gene co-expression networks for three clusters identified by the proposed methods with (A) innate immune system, (B) signal transduction, and (C) immune system.}
    \label{fig:cluNetwork}
\end{figure}

In order to understand the interactions between genes lying within a given cluster, we further computed the gene co-expression networks for three clusters using Bayesian graphical lasso \citep{BayGLasso}. The cluster-specific gene networks, which are presented in Figure \ref{fig:cluNetwork}, were computed by treating the spots as independent and ignoring the spatial information. Since the Bayesian graphical lasso cannot generate exact zeros, we construct the networks by including significant edges as inferred via $95\%$ credible intervals. The estimated networks were seen to be sparse and predominantly composed of negative weights. Subsequently, we investigated network-based differences between the clusters through various graph statistics. 

 The network of the first cluster is much denser, whereas the network of the second and third clusters are sparse. This is evident from the average nodal degree for the three clusters that are $(6.28, 1.10, 1.74)$. Moreover, while the network of the second cluster is the most sparse, the edge strengths are slightly higher than those of the first cluster, but considerably lower than those of the third cluster. To assess the central tendency of the networks, we compute the global clustering coefficient by treating the weights of all non-zero edges as ones. The network of the first cluster exhibits the highest global clustering coefficient ($0.27$) that was considerably higher than the second or third clusters (0.077, 0.050). This resulted from the presence of multiple hub genes, such as KRT15, APOD, IGHA, and CFD, (with degrees of $15, 14, 12$, and $12$, respectively), which are known to play an important role in breast cancer. For example, KRT15 is negatively correlated to early stage of HER2-positive tumors \citep{KRT15_Zhang01062023}, APOD is considered carcinogenic for breast cancer \citep{APOD_cancers15235565}, IGHA is associated with improved prognosis \citep{IGHA_Larsson2020}, and CFD is identified as a prognosis in various breast cancer \citep{CFD_01845215-202409000-00038}. The primary hub gene of the second cluster is ERBB2 (degree of $16$ out of $142$ nodes) that is known to be a  the main mutation of HER2 positive breast cancer \citep{ERBB2_Swain2023}. The network of the third cluster identifies TPT1 as the most prominent hub gene (degree of $15$ out of $124$ nodes), which  is known to serve as therapeutic target in HER2 positive breast cancer \citep{TPT1_cdr.2023.21}. In summary, three clusters exhibit distinct network structures that characterize their respective gene co-expression patterns. Moreover, these networks are dominated by distinct hub genes that play a prominent role in driving within network interactions and which are known to be implicated in breast cancer.

In contrast, global clustering under the classical DPM seems produced an overly large number of clusters, with one giant cluster containing the majority of genes and other smaller clusters comprising a small number of genes. Specifically, DPM generated $25$ clusters that contained one major cluster with $229$ out of $302$ genes, five medium-sized clusters with number of genes ranging from $3$ to $18$, and the remaining $19$ clusters with at most two genes. The presence of a non-trivial number of small clusters seems biologically unrealistic and results in lower clustering reproducibility for the set of common genes across tissues, with ARIs of $0.38$ between H1 and H2, and $0.13$ between H1 and H3. In summary, our analysis highlights the challenges of existing global clustering methods for high-dimensional spatial functions with potential local heterogeneity. It simultaneously demonstrates the importance of the proposed global-local clustering approach to infer interpretable and reproducible clusters of spatially varying genes.

\section{Discussion}
In this paper, we introduce fPDPM, a flexible Bayesian nonparametric framework for global-local clustering, specifically designed for high-dimensional functional data. Utilizing a wavelet basis expansion, fPDPM partitions the coefficients into multiple subsets based on resolution levels, applying a product of Dirichlet process priors to enable independent clustering at different resolutions. This approach facilitates global-local clustering, reducing computational \nTxt{cost} relative to traditional local clustering methods while addressing the curse of dimensionality inherent in global clustering techniques. Under mild conditions, we establish the strong consistency of fPDPM and provide examples of common base measures that satisfy these conditions including low-rank decomposition for the residual covariance matrix that enhances scalability. We also propose an efficient Gibbs sampler for posterior inference. Our simulations demonstrate that fPDPM is much improved scalability compared to local clustering methods that encounter tremendous computational burden, and produces more accurate results compared to global clustering methods. We use fPDPM on spatial transcriptomics data of breast cancer and identify replicable clusters comprising genes that corroborate existing biological literature.



Currently, fPDPM constructs the partition of coefficients based on resolution levels, but this partitioning can be adapted for different purposes.  For example, one can use expert knowledge or AI-based deep clustering approaches \citep{Ren2024} to obtain an initial guess of the image partition, and subsequently use this partition as prior knowledge when performing clustering in the spirit of \citet{CPP_Paganin2021, SP_Dahl2023}.  Another potential generalization is to extend fPDPM to longitudinal functional data, with a goal to identify subgroups of functions with homogeneous longitudinal trajectories. Similarly, the fPDPM approach can be extended to cluster images from multiple modalities by considering modality-dependent coefficients and assigning different clustering structures across modalities. Additionally, the current fPDPM framework assumes that the observed function originates from a single underlying true function. However, this assumption may not always hold in spatial transcriptomics data, where a single spatial spot can contain multiple cell types \citep{SpatialPCA_Shang2022}. This issue can be addressed by modeling the true underlying function as a mixture of functions as $y_i(v)=\sum_{b=1}^B \theta_{ib}(v)+\epsilon_{iv}$, where $\theta_{ib}$ represents the $b$th function.
\nTxt{Finally, while the current framework assumes that functions are observed on a common regular grid for the spatial transcriptomics data, an important extension is to accommodate irregular or sparse sampling designs. In such cases, direct application of the discrete wavelet transform becomes suboptimal \citep{Cai1998}. Recent developments provide potential directions \citep{Haris2018} for extending fPDPM to irregularly spaced functional data while preserving theoretical guarantees.}
All these directions are left for future investigations.




\begin{supplement}
\stitle{Supplement Material for ``Bayesian Nonparametric Global-local Clustering of High-dimensional Functions''}
\sdescription{In Supplementary A, theoretical results of the proposed methods are included in Section A1. Proofs of Theorems and additional Lemmas are available in Sections A2 and A3, respectively. In Section A4, we provide details of the conditional distribution and the Gibbs sampler, while Section A5 offers additional results for the simulation studies, including the details of the data generating mechanism for simulation studies (Section A5.1), diagnostics of the proposed method (Section A5.2), results of the image estimation (Section A5.3), and the results of the local clustering (Section A5.4). Section A6 presents more results of the real data analysis on spatial transcriptomics.}
\end{supplement}

\clearpage
\bibliographystyle{ba}
\bibliography{Ref}

\end{document}


\begin{frontmatter}
\title{Supplement Material for ``Bayesian Nonparametric Global-local Clustering of High-dimensional Functions''}
\runtitle{Bayesian Nonparametric Global-local Clustering of High-dimensional Functions}

\begin{aug}
\author[A]{\fnms{Tsung-Hung}~\snm{Yao}\ead[label=e1]{yaot@ufl.edu}\orcid{0000-0001-5342-5430}}
\and
\author[B]{\fnms{Suprateek}~\snm{Kundu}\ead[label=e2]{skundu2@mdanderson.org}}
\address[A]{Department of Biostatistics, University of Florida\printead[presep={,\ }]{e1}}
\address[B]{Department of Biostatistics, The University of Texas MD Anderson Cancer Center\printead[presep={,\ }]{e2}}
\runauthor{T. Yao and S. Kundu}
\end{aug}

\end{frontmatter}

\section{Theoretical Results}\label{supp:thm}
The priors specified in Section \ref{sec:Method} in the Main Paper induce a product of Dirichlet process priors on the parameter space that can be denoted as $\Pi^*(\bbeta_i,\bSigma_i)=\prod_{j=0}^J P_j(\bbeta_{ij})\times P_{\bSigma}(\bSigma_i)$.  
This prior ultimately induces a distribution $\Pi$ on the class of densities on the observed noisy function denoted as $\cF=\{f_P(\by_i)\}$ via the representation $f_P(\by) = \int \phi_{\bSigma}\left(\by-\sum_{j=0}^J\bPsi_j\bbeta_{j}\right)d\Pi^*(\bbeta,\bSigma)$. We denote the resulting class of priors $\Pi$ on $\cF$  as the product of Dirichlet process mixture on functional data (fPDPM). This method generalizes previous work on product mixture priors that was proposed for  multivariate time-series data \citep{Kundu2024PDPM} to the case of spatial functions. By using the stick-breaking representation \citep{Sethuraman1994}, the induced density takes the form:
\begin{align}
    \nonumber f_P(\by) &=\int \ldots \int \phi_{\bSigma}\left(\by-\sum_{j=0}^J\bPsi_j\bbeta_{j}\right)dP_0(\bbeta_{0})\ldots dP_J(\bbeta_{J})dP_{\bSigma}(\bSigma)\\
    \label{eq:stkBreak_pdpm}&=\sum_{h_0=1}^\infty\ldots\sum_{h_J=1}^\infty\sum_{h_s=1}^\infty w_{0,h_0}\ldots w_{J,h_J}w_{s,h_s}{\phi_{\bSigma_{h_s}}}\left(\by-\sum_{j=0}^J\bPsi_j\bbeta_{h_jj}\right),
\end{align}
where $\phi_{\bSigma}(\cdot - \bm{\mu})$ is the normal density function with covariance $\bSigma$ and mean $\mu$, $w_{j,h_j}$ is the weight of $\bbeta_{j}$ in the $h_j$th cluster with $w_{j,h_j}=\nu_{j,h_j}\prod_{e<h_j}(1-\nu_{j,e}), \nu_{j,e} \sim \textrm{Beta}(1,\alpha_j)$ and $\bbeta_{h_jj}\sim \mathbf{P}_j^*$ for $j=0,\ldots,J$, and $w_{s,h_s}$ is the weight of $\bSigma$ in the $h_s$th cluster with $w_{s,h_s}=\nu_{s,h_s}\prod_{e<h_s}(1-\nu_{s,e}), \nu_{s,e}\sim\textrm{Beta}(1,\alpha_{\bSigma})$ and $\bSigma_{h_s}\sim P^*_{\bSigma}$. Equation \eqref{eq:stkBreak_pdpm} generalizes the generic kernel mixture representation in the nonparametric multivariate density estimation literature \citep{Wu2008, Kundu2024PDPM} to the case of functional data analysis relying on suitable basis expansions.


The proposed method defines a prior on the density of the observed functions $y_i(v), v\in\cV$ (as in Equation \eqref{eq:stkBreak_pdpm}) that satisfies several desirable theoretical properties. In our theoretical analysis, particularly for deriving posterior consistency results, we omit the subscript $i$ when appropriate, following standard convention. All theoretical results are constructed under the conditions of $n\rightarrow\infty$ and fixed $L$. For our theoretical treatment, we assume that the number of observed support points for the spatial function is $2^{J+1}$ that produces a square orthonormal basis matrix $\bPsi (L\times L)$.
In general, the theoretical results hold for all orthogonal basis representations with the product of DP prior on basis coefficients, as long as the true density $f_0$ satisfies certain reasonable assumptions. Notably, our posterior consistency results generalized the rich theoretical results for Bayesian non-parametric multivariate density estimation to the case of Bayesian functional data analysis, for which there are limited results. In particular, our focus is on estimating densities of random functions admitting suitable basis expansions.

Denote the Euclidean norm of a vector as $\lVert\cdot\rVert$, the spectral norm of a matrix as $\lVert\cdot\rVert_2$ and the eigenvalues of positive definite matrix $\bSigma (L \times L)$ in decreasing order as $\lambda_1(\bSigma)\geq\ldots\geq\lambda_L(\bSigma)$. Let $KL(f;g)=\int f(\by)\frac{f(\by)}{g(\by)}d\by$ be the Kullback-Leibler (KL) divergence between $f$ and $g$ with $f,g\in\cF$. We first show that the fPDPM prior $\Pi$ on the space of densities $\mathcal{F}$ lead to the posterior consistency of KL neighborhoods with the formal argument below.
\begin{lemma}\label{lma:wkCnsty}
    Let $f_0\in\cF$ and denote $\Pi$ as the prior on $\cF$ induced by the mixing distribution \eqref{eq:stkBreak_pdpm}. Assume that $f_0$ satisfies the following conditions of 
    (i) For some constant $M$ and all $y\in\mathbb{R}^L$, $0<f_0(\by)<M$ and $\lvert\int f_0(\by)\log f_0(\by)d\by<\infty\rvert$; (ii) for some $\delta>0$ and $g_\delta(\by)=\inf_{\lVert \bm{t} - \by \rVert<\delta} f_0(\bm{t})$, $\int f_0(\by)\log \frac{f_0(\by)}{\phi_\delta(\by)}d\by<\infty$; (iii) for some $\eta>0$, $\int \lVert \by\rVert^{2(1+\eta)}f_0(\by)d\by<\infty$.
    Then for prior $\Pi$ defined in \eqref{eq:stkBreak_pdpm}, we have $\Pi(f\in\cF: KL(f_0;f)\leq\eta^*)>0$ for any $\eta^*>0$. 
\end{lemma}

The detailed proof of Lemma \ref{lma:wkCnsty} is available in the Supplementary Material Section \ref{supp:wkCnsty}. Notably, the assumptions  on the true density in Lemma \ref{lma:wkCnsty} are routinely assumed in multivariate density estimation literature \citep{Wu2008}.  Under an orthonormal basis (resulting in a  finite determinant and Frobenius norm of $\bPsi_j$), the proof mainly follows from the results in  \citet{Wu2008}. 
Following the argument of \citet{Schwartz1965}, a positive prior support for arbitrarily small KL neighborhoods of $f_0$ from Lemma \ref{lma:wkCnsty} provides the weak consistency of fPDPM.

While weak consistency is useful, it is of interest to investigate whether fPDPM also results in a more salient strong consistency property. Strong consistency is desirable since it ensures that the posterior concentrates in arbitrarily small $L_1$ neighborhoods of $f_0$, thereby guaranteeing more reasonable numerical estimation. However, given that the space $\cF$ can be non-compact, additional conditions are required to establish strong consistency  \citep{Canale2017}. The basic strategy follows the framework in \citet{Ghosal2017BNP_textbook} by constructing sieves $\cF_n$ (a compact subset of $\cF$) having certain required properties and that eventually grows to cover the whole space $\cF$ as $n\rightarrow\infty$. Next, we consider how to construct such sieves. 

Denote $F_0^n$ as the $n$-product of the measure $F_0$ that is associated with the true density $f_0$ and $\log N(\epsilon,\cG,d)$ as the entropy of the space of densities $\cG\subset\cF$, where $N(\epsilon,\cG,d)$ is the minimum integer for which there exists densities $f_1, \ldots, f_{N} \in \cF$ such that $\cG\subset\cup_{j=1}^N\{f: d(f, f_j) < \epsilon\}$. For the metric $d$ used here, we consider the  Hellinger distance, $d^2(f,g)=\int(\surd{f}-\surd{g})^2$, and $L_1$ distance, $\lVert f-g\rVert_1=\int\lvert f-g\rvert$. We state below the sufficient conditions for the sieves that are required for strong consistency and was originally proposed in \cite{Canale2017}. This result will be used in our theoretical calculations when deriving strong consistency.

\begin{lemma}\label{lma:sumblty} 
Consider sieves $\cF_n\subset \cF$ with $\cF_n\uparrow\cF$ as $n\rightarrow\infty$, where $\cF_n=\cup_{j=-\infty}^\infty\cF_{n,j}$. Suppose the sieves satisfy the following conditions for $\epsilon>0$: (1A) $\Pi(\cF_n^c)\lesssim\exp^{-bn}$ for some $b>0$; and (1B) $\sum_{j=-\infty}^\infty N^{1/2}(2\epsilon,\cF_{n,j},d)\Pi^{1/2}(\cF_{n,j})\exp^{-(4-c)n\epsilon^2}\rightarrow0$ for some $c,\epsilon>0$. 
Then $\Pi(f:d(f_0,f)>8\epsilon\mid y_n)\rightarrow0$ in $F_0^n$-probability for any $f_0$ in the weak support of $\Pi$.
\end{lemma}

Conditions (1A) and (1B) in Lemma \ref{lma:sumblty} regulate the growth of sieve in $\cF_n$  when sample size $n$ increases. Specifically, (1A) suggests the prior probability assigned to $\cF_n^c$ needs to shrink exponentially in $n$, implying that the tails with only exponentially small probability are excluded. Condition (1B) can be considered as a summability condition that stipulates that the weighted sum of the entropy number (weighted by the square root of the prior probability) is required to go toward zero when $n$ increases. To achieve both conditions, we carefully construct sieves (compact sub-spaces) as outlined below. 
\begin{align*}
    \cG=&\left\{f_P \textrm{ with } P=\sum_{h_0\geq1}\ldots\sum_{h_J\geq1}\sum_{h_s\geq1} \left(\prod_{j=0}^J w_{j,h_j}\right)w_{s,h_s}\delta(\bbeta_{h_j,j},\bSigma_{h_s}): \forall j=0,\ldots,J, \sum_{h_j>H}w_{j,h_j}<\epsilon, \right. \\ 
    &\left.  \sum_{h_s>H}w_{s,h_s}<\epsilon  \textrm{ and for } h_j,h_s\leq H, \underline{\sigma}^2\leq\lambda_L(\bSigma_{h_s}), \lambda_1(\bSigma_{h_s})\leq\underline{\sigma}^2(1+\epsilon/\surd{L})^{M},1\leq\frac{\lambda_1(\bSigma_{h_s})}{\lambda_L(\bSigma_{h_s})}\leq u_{h_s},\right.\\
&\left.\underline{a}_{h_j}\leq\lVert\bbeta_{h_j,j}\rVert\leq\bar{a}_{h_j}  \right\}, \mbox{ where } H,M\in\mathbb{N},\underline{\sigma}>0, \mbox{ and } 0\leq\underline{a}_h\leq\bar{a}_h, 1\leq u_h, h=1,\ldots,H.
\end{align*} 

Under the above construction, the entropy of the sieves as given by the following Theorem.

\begin{theorem}\label{thm:entrpBnd}

The entropy number of $\cG$ is given by
\begin{align*}
\cN(\epsilon,\cG,\lVert\cdot\rVert_1)\lesssim\exp&\left[LH\log M + H^{J+2}\log\frac{1}{\epsilon}+ \sum_{h_s<H} \frac{L(L-1)}{2}\log\frac{2Lu_{h_s}}{\epsilon^2} +\right.\\
&\left. \sum_{j=0}^J\sum_{h_j<H}\log\left\{\left(\frac{\bar{a}_{h_j}}{\underline{\sigma}\epsilon/2}+1\right)^{2^j}- \left(\frac{\underline{a}_{h_j}}{\underline{\sigma}\epsilon/2}-1\right)^{2^j}\right\}\right]
\end{align*}
\end{theorem}

Detailed proof is available in the Supplementary Material Section \ref{supp:entrpBnd}. Having established the entropy bounds in Theorem \ref{thm:entrpBnd}, we are now in a position to establish strong consistency results via the Theorem below.

\begin{theorem}\label{thm:strngCnst}
Suppose that $f_0$ satisfies Lemma \ref{lma:wkCnsty} and fPDPM prior  satisfies the following conditions for all sufficiently large $z$ and some positive constants $c_1,c_2,c_3, r$ and $\kappa>L(L-1)$:
\begin{enumerate}
    \item[(2A)] $\mathbf{P}_j^*(\lVert \bbeta_j \rVert>z) < z^{-2(r+1)}$ for $r>(L-1)/2$; (2B) $\mathbf{P}_{\bSigma}^*(\lambda_1(\bSigma^{-1})>z) < \exp(-c_1 z^{c_2})$;
    \item[(2C)] $\mathbf{P}_{\bSigma}^*(\lambda_L(\bSigma^{-1})>1/z) < z^{-c_3}$;
    (2D) $\mathbf{P}_{\bSigma}^*(\lambda_1(\bSigma^{-1})/\lambda_L(\bSigma^{-1})>z) < z^{-\kappa}$.
\end{enumerate} Then, the fPDPM prior  satisfies the sufficient conditions in Lemma \ref{lma:sumblty} and therefore the posterior distribution is strongly consistent at $f_0$ in $F_0^{(n)}$-probability, as $n\to \infty$.
\end{theorem}

Detailed proof of Theorem \ref{thm:strngCnst} is provided in the Supplementary Material Section \ref{supp:strngCnst}. Theorem \ref{thm:strngCnst} provides guidelines for the prior that gives salient strong posterior consistency and a solid theoretical foundation corresponding to a practical implementation of the fPDPM approach. Fortunately, several common priors satisfy conditions (2A) to (2D). 
For example, a Laplace prior on the base measure for basis coefficients satisfies the condition (2A), and the low-rank representation \eqref{eq:LRD_PDP} with the multiplicative Gamma process for the base measure satisfies conditions (2B) to (2D). A formal statement is provided in the Corollary below as: 
\begin{corollary}\label{cor:priorLaLRD}
    Denote $f_0$ as the true density that satisfies conditions in Lemma \ref{lma:wkCnsty}. Consider a prior $\Pi$ on density $\cF$ induced by the $\Pi^*$ based on \eqref{eq:coef_PDP} for coefficients and \eqref{eq:LRD_PDP} for covariances. Let $\mathbf{P}^*_j$ follow a Laplace distribution of \eqref{eq:coefPrior} and $\mathbf{P}^*_{\bSigma}$ be a multiplicative Gamma prior of \eqref{eq:covPrior}. Then conditions (2A) to (2D) are satisfied and the posterior distribution is consistent at $f_0$.
\end{corollary}
Detailed proof is available in the Supplementary Material \ref{supp:priorLaLRD}. Note that the case of independent covariance of $\bSigma_i=\sigma^2_i \bI_L$ can be easily obtain as a special case Corollary \ref{cor:priorLaLRD} with $\bLambda_i=\mathbf{0}$.
Another common prior comes from a multivariate normal prior on the basis coefficients with an inverse-Wishart prior on the unstructured error covariances as the Corollary below.

\begin{corollary}\label{cor:priorNIW}
    Denote $f_0$ as the true density that satisfies conditions in Lemma \ref{lma:wkCnsty}. Consider a prior $\Pi$ on $\cF$ induced by the $\Pi^*(\bbeta,\bSigma)$ of \eqref{eq:coef_PDP} and \eqref{eq:unStucCov_PDP}. Let $\mathbf{P}^*_j(\bbeta)=N(0,\tau^2_j)$ follow a normal distribution and $\mathbf{P}^*_{\bSigma}=IW(\bSigma_0,\nu)$ with. Then conditions (2A) to (2D) are satisfied and the posterior distribution is consistent at $f_0$.
\end{corollary}
The proof is a direct result of Corollary 1 of \citet{Canale2017}.

\section{Proof of Theorems}\label{supp:proof}
\subsection{Proof of Lemma \ref{lma:cmpComplexity}}\label{supp:cmpComplexity}
In this Section, we show the details of the computation cost based on the Gibbs Sampler in Section \ref{supp:GibbsSamp} and compare the results to the cost from existing Baeysian local clustering methods of local partition process (LPP) proposed by \citet{LPP_Dunson2009} and SRZ from \citet{suarez2016bayesian}. Details of LPP and SRZ are included in the Section \ref{sec:compareModel} of the Main Paper. The proposed fPDPM consists of three sets of parameters: (1) membership parameters including $J+1$ levels of resolutions for coefficients and one for covariance, (2) coefficients of dimension $L$, and (3) low-rank covariance of dimension $L$ by $L$. For LPP and SRZ, they include the first two sets of parameters, but consider a larger number of membership parameters with the assignment of one membership parameter to each coefficient. 
In essence, both LPP and SRZ assign a prior on each coefficient $P_{l}$, which requires a large number of $L=2^{J+1}$ membership parameters with an intensive computational bottleneck. 

To calculate computational cost, we assume that the number of clusters is at most $H$ and the maximum rank for covariance is $K\ll L$. The cost for fPDPM comes from the membership parameters and the covariance. For the membership parameter, the main bottleneck is from the calculation of the normal density (Step 3 in Section \ref{supp:GibbsSamp}). For a multivariate normal density with a correlated covariance, the cost of calculating the density via naive Cholesky decomposition is $O(L^3)$. With the Woodbury identity for the low-rank representation of \eqref{eq:covLowRank}, this cost can be reduced to $O(K^2L)$, and the cost to update all membership parameters becomes $O(HJK^2L)$. For updating covariances, the conditional distributions in Section \ref{supp:GibbsSamp} shows that the cost for updating the the factor matrix (Step 5) and loading matrix (Step 6) are $O(nK^2L)$ and $O(HK^3L)$, respectively. In sum, the cost for fPDPM is $O(HJK^2L)+O(nK^2L)+O(HK^3L)$. For LPP and \citet{suarez2016bayesian}, the cost is mainly from updating membership parameters. With independent assumption for normal density, the cost of calculating the density is $O(L)$, and the cost for LPP can be easily obtained as $O(HL^2)$.

\subsection{Proof of Lemma \ref{lma:wkCnsty}}\label{supp:wkCnsty}
Without loss of generality, we express the basis expansion of the underlying function $\btheta$ as a matrix form with $\btheta=\sum_j\bPsi_j\bbeta_j=\bPsi\bbeta$, where $\bPsi$ is the basis matrix of dimension $L\times L$ and $\bbeta=[\bbeta_0, \ldots ,\bbeta_J]^\transp$ is a vector of length $L=2^{J+1}$. Since 
\begin{align*}
    \int f_0(\by) \log\frac{f_0(\by)}{f(\by)}d\by = \int f_0(\by) \log\frac{f_0(\by)}{f_{P_\epsilon}( \by)}d\by + \int f_0(\by) \log\frac{f_{P_\epsilon}( \by)}{f(\by)}d\by,
\end{align*}
we need to construct a distribution $f_{P_\epsilon}( \by)$ so that both terms are smaller than $\epsilon$. To achieve so, construct a mixing distribution $f_m(\by)$ with the defined as follows: 
\begin{align*}
    f_m\left(\by\right)=
    \begin{cases} t_m f_0\left(\by\right) & \textrm{ if } \lVert \by\rVert < m, \\0 & \textrm{ if otherwise} \end{cases},
\end{align*}
where $\lVert\cdot\rVert$ is the Euclidean norm and $t_m^{-1}=\int_{\lVert \bPsi\bbeta\rVert < m}f_0\left( \bPsi\bbeta\right)$ is the normalization constant. Obviously, $t_m\geq1$ and $t_m$ is non-increasing with respect to $m$.  Denote $F_m(\cdot)$ as the probability measure for the mixing distribution $f_m(\cdot)$. For the fPDPM considered in the paper, we can view the measure as $F_m(\cdot)=\prod_{j=0}^J F_{m,j}(\cdot)$ with each $F_{m,j}$ as the probability measure for the coefficients belonging to the jth resolution. For the measure on the covariance matrix, we first focus on the the diagonal matrix of $\bSigma=e^2 \bI$ and generalize our results to the unstructured covariances. We consider the measure $P_m =F_m \times \delta(e_m)$, where $\delta(\cdot)$ is the degenerated distribution. Obviously, $P_m$ is compactly supported since $\lVert\bPsi\bbeta\rVert<m$ implies $\lVert\bbeta\rVert < m\lVert\bPsi\rVert_F$, where $\lVert\cdot\rVert_F$ is the Frobenius norm. By using the variable transformation $\bm{z}=e_m^{-1}(\by-\bPsi\bbeta)$, the distribution $f_{P_m}(\by)$ can be written as
\begin{align*}
    f_{P_m} &= \int \phi_{e_m^2 \bI}\left(\by-\bPsi\bbeta\right)dF_m(\bPsi\bbeta) \\
    &=\int (2\pi)^{-L/2}e_m^{-L}\exp\left(-\frac{\lVert  \by-\bPsi\bbeta \rVert^2}{2e_m^2}\right)f_m(\bPsi\bbeta)\lvert\bPsi\rvert d\bbeta\\
    &= \int_{\lVert\bPsi\bbeta\rVert<m} t_m  (2\pi)^{-L/2}e_m^{-L}\exp\left(-\frac{\lVert \by-\bPsi\bbeta \rVert^2}{2e_m^2}\right)f_0(\bPsi\bbeta)\lvert\bPsi\rvert d\bbeta\\
    &= \int_{\lVert \by -  \bm{z} e_m\rVert<m} t_m  (2\pi)^{-L/2}\exp\left(-\frac{\lVert\bm{z}\rVert^2}{2}\right)f_0(\by-e_m \bm{z})d\bm{z},
\end{align*}
where $\lvert\cdot\rvert$ is the determinant of a matrix. The second last equation holds due to the fact of the Jacobian matrix $\lvert\frac{d\bbeta}{d\bm{z}}\rvert=\lvert\bPsi^{-1}\rvert e_m^L$. Let $e_m=m^{-\eta}$. When $m\rightarrow \infty$, we obtain $e_m\rightarrow0, t_m\rightarrow1$ and $t_mf_0(\by-e_m {\bm{z}})\rightarrow f_0(\by)$. Since $f_0(\by)$ is bounded by assumption, by dominated convergence theorem, 
\begin{align*}
    f_{P_m}(\by)\rightarrow \int (2\pi)^{-L/2} \exp\left(-\frac{\lVert \bm{z} \rVert^2}{2}\right) f_0(\by) d\bm{z} =f_0(\by).
\end{align*}
First, we aim the first term to show that $\int f_0(\by) \log\frac{f_0(\by)}{f_{P_\epsilon}(\by)}d\by\rightarrow 0$ when $m\rightarrow \infty$. To this end, observe that
\begin{align*}
    \int f_0(\by)\log\frac{f_0(\by)}{f_{P_m}(\by)}d\by= \int_{\lVert \by \rVert>m} f_0(\by)\log\frac{f_0(\by)}{f_{P_m}(\by)}d\by + \int_{\lVert \by \rVert\leq m}f_0(\by)\log\frac{f_0(\by)}{f_{P_m}(\by)}d\by.
\end{align*}
For $\lVert \by \rVert>m$, 
\begin{align*}
    f_{P_m}(\by)&=\frac{\lvert\bPsi\rvert}{ (2e_m^2\pi)^{L/2}}\int_{\lVert\bPsi \bbeta\rVert<m} t_m \exp\left(-\frac{\lVert \by-\bPsi\bbeta \rVert^2}{2e_m^2}\right)f_0(\bPsi\bbeta) d\bbeta\\
    &\geq \frac{\lvert\bPsi\rvert}{ (2e_m^2\pi)^{L/2}}\int_{\lVert\bPsi\bbeta\rVert<m} t_m \exp\left(-\frac{\left\lVert \by + m\frac{\by}{\lVert\by\rVert}\right\rVert^2}{2e_m^2}\right)f_0(\bPsi\bbeta) d\bbeta\\
    &= \frac{\lvert\bPsi\rvert}{ (2e_m^2\pi)^{L/2}} \exp\left(-\frac{\left\lVert \by + m\frac{\by}{\lVert\by\rVert}\right\rVert^2}{2e_m^2}\right)\int_{\lVert\bPsi\bbeta\rVert<m} t_m f_0(\bPsi\bbeta) d\bbeta\\
    &= \frac{\lvert\bPsi\rvert}{ (2e_m^2\pi)^{L/2}} \exp\left(-\frac{\left\lVert \by + m\frac{\by}{\lVert\by\rVert}\right\rVert^2}{2e_m^2}\right)\\
    &= \frac{\lvert\bPsi\rvert m^{L\eta}}{ (2\pi)^{L/2}}  \exp\left(-\frac{1}{2}\left\lVert m^\eta\by + m^{\eta+1}\frac{\by}{\lVert\by\rVert}\right\rVert^2\right)\\
    &\geq \frac{\lvert\bPsi\rvert }{ (2\pi)^{L/2}} \lVert\by\rVert^{L\eta} \exp\left(-\frac{1}{2}\left\lVert 2\by\right\rVert^{2\eta+2}\right).
\end{align*}
The last inequality holds is because $r(m)=m^{L\eta} \exp\left(-\frac{1}{2} \lVert m^\eta\by(1+\frac{m}{\lVert\by\rVert})\rVert^2 \right)$ is decreasing in $m$. This can be proved by the Theorem 2 and Theorem 4 from \citet{Wu2008}. 

For $\lVert\by\rVert\leq m$, let $\delta>0$ be fixed and $g_m(\by)=\inf_{\lVert \bm{t}-\by\rVert<\delta e_m}f_0(\bm{t})$. With the convention of $[a,b]=[b,a]$ when $a>b$, the mixing distribution becomes 
\begin{align*}
    f_{P_m}(\by)&=\frac{\lvert\bPsi\rvert}{ (2e_m^2\pi)^{L/2}}\int_{\lVert\bPsi\bbeta\rVert<m} t_m \exp\left(-\frac{\lVert \by-\bPsi\bbeta \rVert^2}{2e_m^2}\right)f_0(\bPsi\bbeta) d\bbeta\\
    &\geq \frac{\lvert\bPsi\rvert}{ (2e_m^2\pi)^{L/2}}\int_{\{\lVert\bPsi\bbeta\rVert<m\} \cap \{\lVert\by-\bPsi\bbeta\}\rVert<\delta e_m} t_m \exp\left(-\frac{\lVert \by-\bPsi\bbeta \rVert^2}{2e_m^2}\right)f_0(\bPsi\bbeta) d\bbeta\\
    &\geq \frac{\lvert\bPsi\rvert}{ (2e_m^2\pi)^{L/2}}g_m(\by)\int_{\{\lVert\bPsi\bbeta\rVert<m\} \cap \{\lVert\by-\bPsi\bbeta\rVert\}<\delta e_m} t_m \exp\left(-\frac{\lVert \by-\bPsi\bbeta \rVert^2}{2e_m^2}\right)d\bbeta\\
    &= \frac{t_m g_m(\by)}{ (2\pi)^{L/2}}\int_{\{\lVert\by -  \bm{z} e_m\rVert<m\} \cap \{\lVert \bm{z} \rVert\}<\delta } \exp\left(-\frac{\lVert  \bm{z}\rVert^2}{2}\right)d\bm{z}\\
    &\geq \frac{t_m g_m(\by)}{ (2\pi)^{L/2}}\int_{\prod_{l=1}^L[0,\textrm{sgn}(y_l)\delta/\surd{L}]} \exp\left(-\frac{\lVert  \bm{z}\rVert^2}{2}\right)d\bm{z}, 
\end{align*}
where $\bm{z}=e_m^{-1}(\by-\bPsi\bbeta)$. The last inequality holds because $\left\{ \bm{z}:  \bm{z}\in\prod_{l=1}^L[0,\textrm{sgn}(y_l)\delta/\surd{L}] \right\} \subset \left\{\bm{z}: \{\lVert \bm{z} - \by/e_m \rVert<m/e_m\} \cap \{\lVert \bm{z} \rVert\}<\delta \right\}$, when $\lVert\by\rVert\leq m$. 

Let 
\begin{align*}
    c=\min_{\by\in\{-\delta/\surd{L},\delta/\surd{L}\}^L}\int_{\prod_{l=1}^L[0,y_l]}\frac{1}{(2\pi)^{L/2}}\exp\left(-\frac{\lVert \bm{z}\rVert^2}{2}\right)d\bm{z}.
\end{align*}
Since $t_m\geq 1$ and $g_m(\by)>g_1(\by), f_{P_m}(\by)\geq c g_1(\by)$ when $\lVert\by\rVert\leq m$.

In sum, for some $0<R<m$
\begin{align*}
    f_{P_m}(\by)\geq 
    \begin{cases}
        \min\left\{c g_1(\by),\frac{\lvert\bPsi\rvert }{ (2\pi)^{L/2}} \lVert\by\rVert^{L\eta} \exp\left(-\frac{1}{2}\left\lVert 2\by\right\rVert^{2\eta+2}\right) \right\} & \textrm{ if } \lVert\by\rVert \geq R \\
        cg_1(\by) & \textrm{ if } \lVert\by\rVert <R
    \end{cases}
\end{align*}
\begin{align}\label{eq:wkCnsty1}
    \log\frac{f_0(\by)}{f_{P_m}(\by)}\leq\xi(\by)\equiv 
    \begin{cases}
        \max\left\{\log\frac{f_0(\by)}{c g_1(\by)},\log\frac{f_0(\by)}{\frac{\lvert\bPsi\rvert }{ (2\pi)^{L/2}} \lVert\by\rVert^{L\eta} \exp\left(-\frac{1}{2}\left\lVert 2\by\right\rVert^{2\eta+2}\right)} \right\} & \textrm{ if } \lVert\by\rVert \geq R \\
        \log \frac{f_0(\by)}{cg_1(\by)} & \textrm{ if } \lVert\by\rVert <R
    \end{cases}
\end{align}
Now, since 
\begin{align*}
    f_{P_m}(\by)&=\frac{\lvert\bPsi\rvert}{ (2e_m^2\pi)^{L/2}}\int_{\lVert\bPsi\bbeta\rVert<m} t_m \exp\left(-\frac{\lVert \by-\bPsi\bbeta \rVert^2}{2e_m^2}\right)f_0(\bPsi\bbeta) d\bbeta \\
    &\leq M\lvert\bPsi\rvert\int_{\lVert\bPsi\bbeta\rVert<m} \frac{t_m}{ (2e_m^2\pi)^{L/2}} \exp\left(-\frac{\lVert \by-\bPsi\bbeta \rVert^2}{2e_m^2}\right) d\bbeta \\
    &< M\lvert\bPsi\rvert t_m < M\lvert\bPsi\rvert t_1
\end{align*}
We can obtain 
\begin{align}\label{eq:wkCnsty2}
    \log\frac{f_0(\by)}{f_{P_m}(\by)}\geq\log\frac{f_0(\by)}{M\lvert\bPsi\rvert t_1}
\end{align}

By combining Equations \ref{eq:wkCnsty1} and \ref{eq:wkCnsty2}, we see that 
\begin{align*}
    \left\lvert\log\frac{f_0(\by)}{f_{P_m}(\by)}\right\rvert \leq \max\left(\xi(\by), \left\lvert\log\frac{f_0(\by)}{M\lvert\bPsi\rvert t_1} \right\rvert\right)
\end{align*}
In this paper, we consider the orthogonal basis with finite $\lvert\bPsi\rvert<\infty$. Following the proof in the Theorem 2 in \citet{Wu2008}, it can be easily shown that $\int f_0(\by) \log\frac{f_0(\by)}{f_{P_\epsilon}(\by)}d\by\rightarrow 0$ when $m\rightarrow \infty$ for the mean function as a basis expansion $\bPsi\bbeta$ with a diagonal covariance. 

Now, we generalize the proof with an unstructured covariance $\bSigma$. Denote $\cS$ as the cone of positive definite matrices of dimension $L$ by $L$ In order to guarantee the measure $P_m$ is compactly supported, define the $P_{m_\epsilon}(D)=1$ with $D=[-m^*,m^*]\times\left\{\bSigma\in\cS:\underline{\sigma}^2\leq\lambda_l(\bSigma)\leq\bar{\sigma}^2, l=1,\ldots,L\right\},$ where $m^* = mK\lVert\bPsi^{-1}\rVert_F$. With the proof from the Lemma 1 from \citet{Canale2017}, $\int f_0(\by) \log\frac{f_0(\by)}{f_{P_\epsilon}(\by)}d\by\rightarrow 0$ when $m\rightarrow \infty$. 

For the second term of $\int f_0(\by) \log\frac{f_{P_\epsilon}(\by)}{f(\by)}d\by\rightarrow 0$, we show that $\int f_0(\by) \log\frac{f_{P_\epsilon}(\by)}{f(\by)}d\by$ satisfies conditions (A7) to (A9) of Lemma 3 in \citet{Wu2008}. Let $D$ be a compact set for $P_m$ as the previous proof and $P_\epsilon$ be chosen to be $P_{m_\epsilon}$. For (A7), we require $\log f_{P_{m_\epsilon}}$ and $\log\inf_{(\bbeta,\bSigma)\in D}\phi_{\bSigma}(\by-\bPsi\bbeta)$ to be $f_0$-integrable. For $\lVert \by\rVert<m^*, \log\inf_{(\bbeta,\bSigma)\in D}\phi_{\bSigma}(\by-\bPsi\bbeta)$ and $\log f_{P_m}$ are bounded. While $\lVert \by\rVert\geq m^*$, we observe that 
\begin{align*}
    \inf_{(\bbeta,\bSigma)\in D}\phi_{\bSigma}(\by-\bPsi\bbeta)&=(2\pi)^{-L/2}\bar{\sigma}^{-L}\exp\left(-\frac{4\lVert\by\rVert^2}{2\underline{\sigma}^2}\right)
\end{align*}
, which is bounded. Hence, for $\lVert \by\rVert\geq m^*$, 
\begin{align*}
    \lvert\log f_{P_m}\rvert \leq \left\lvert\log\left\{ (2\pi)^{-L/2}\bar{\sigma}^{-L}\exp\left(-\frac{4\lVert\by\rVert^2}{2\underline{\sigma}^2}\right)\right\} P_m(D)\right\rvert
\end{align*}
, so that $\log f_{P_m}$ is also $f_0$-integrable. For condition (A8) of Lemma in \citet{Wu2008}, it is clear that the multivariate normal kernel is bounded away zero for $\by$ in a compact set of $\mathbb{R}^L$ and $(\bbeta,\bSigma)\in D$. Last, condition (A9) requires the uniformly equicontinuous as a family of function of $(\bbeta,\bSigma)\in D$ on a given compact set, $\{\phi_{\bSigma}(\by-\bPsi\bbeta), \by\in C\}$. This can be achieved by the same argument used in the proof of Theorem 2 in \citet{Wu2008}.

\subsection{Proof of Theorem \ref{thm:entrpBnd}}\label{supp:entrpBnd}
Consider $f_P\in\cG$ with $P=\sum_{h_0,\ldots,h_J,h_s\geq1}\left(\prod_{j=0}^J w_{j,h_j}\right)w_{s,h_s}\delta(\bbeta_{h_jj},\bSigma_{h_s})$ and $\bSigma_{h_s}=\left(\bm{O}_{h_s}\bLambda_{h_s}\bm{O}_{h_s}\right)^{-1}$ as the eigendecomposition, where $\bLambda_{h_s}=\textrm{diag}\left(\lambda_1(\bSigma_{h_s}),\ldots,\lambda_L(\bSigma_{h_s})\right)$ and $\bm{O}_{h_s}$ is the corresponding orthogonal eigenvectors. From Lemma \ref{lma:diff_Norm} and \ref{lma:boundWeightDiff}, we consider another density $f_{\hat{P}}$ with $\hat{P}=\sum_{h_0,\ldots,h_J,h_s\geq1} \left(\prod_{j=0}^J w_{j,h_j}\right)w_{s,h_s}\delta(\hat{\bbeta}_{h_jj},\hat{\bSigma}_{h_s})$ and $\hat{\bSigma}_{h_s}=\left(\hat{\bm{O}}_{h_s}\hat{\bLambda}_{h_s}\hat{\bm{O}}_{h_s}\right)^{-1}$ within the $\epsilon$-net. To construct such density, choose 
\begin{itemize}
    \item $\hat{\bbeta}_{h_jj}\in\hat{\cR}_{h_jj}$, where $\hat{\cR}_{h_jj}$ is a $\epsilon\underline{\sigma}$-net of $\cR_{h_jj}=\{\bbeta_{h_jj}\in\mathbb{R}^{2^j}: \underline{a}_{h_j}\leq\lVert\bbeta_{h_jj}\rVert\leq\bar{a}_{h_j}\}$ such that $\lVert \bbeta_{h_jj} -\hat{\bbeta}_{h_jj}\rVert<\epsilon\underline{\sigma}$ for all $j=0,\ldots,J$ and $h_j=1,\ldots,H$.
    \item $\left\{\prod_{j=0}^J\hat{w}_{j,h_j}\hat{w}_{s,h_s}: h_0,\ldots,h_J,h_s\leq H\right\}\in\hat{\Delta}$, where $\hat{\Delta}$ is a $\epsilon$-net of a $H^{J+2}$ dimensional probability simplex such that $\sum_{h_0,\ldots,h_J,h_s\leq H}\left\lvert \left(\prod_{j=0}^J \tilde{w}_{j,h_j}\right)\tilde{w}_{s,h_s} - \left(\prod_{j=0}^J \hat{w}_{j,h_j}\right)\hat{w}_{s,h_s}\right\rvert\leq\epsilon$, and $\tilde{w}_{j,h}=\frac{w_{j,h}}{\sum_{c\leq H}w_{j,c}}, h\leq H$.
    \item $\hat{\bm{O}}_{h_s}\in\hat{\cO}_{h_s}$, where $\hat{\cO}_{h_s}$ is a $\delta_{h_s}$-net of the set $\cO_{h_s}$ defined as the set $L\times L$ orthogonal matrices with respect to the spectral norm $\lVert\cdot\rVert_2$ with $\delta_{h_s}=\epsilon^2/(2Lu_{h_s})$ such that $\lVert \bm{O}_{h_s}-\hat{\bm{O}}_{h_s}\rVert\leq L\delta_{h_s}$.
    \item $(m_{h_s,1},\ldots,m_{h_s,L})\in\{1,\ldots,M\}^L, h_s=1,\ldots,H$ such that $\lambda_{l}(\hat{\bSigma}_{h_s})=\underline{\sigma}^2(1+\epsilon L^{1/2})^{m_{h_s,l}-1}$ satisfying $1<\lambda_{l}(\bSigma_{h_s})/\lambda_{l}(\hat{\bSigma}_{h_s})<(1+\epsilon/ L^{1/2})$
\end{itemize}
From Lemma \ref{lma:diff_Norm}, we obtain the norm is bounded by 
\begin{align*}
    \lVert f_P -f_{\hat{P}} \rVert_1 &\leq  \max_{h_0,\ldots,h_J,h_s\leq H} \left[ \left(\frac{2/\pi}{\lambda_L(\hat{\bSigma}_{h_s})}\right)^{1/2}\sum_{j=0}^J\lVert \bPsi_j\rVert_F\left\lVert \bbeta_{h_jj} -\hat{\bbeta}_{k,h_j}\right\rVert
    + \left\{\sum_{l=1}^L \frac{\lambda_l(\bSigma_{h_s})}{\lambda_l(\hat{\bSigma}_{h_s})} - \log \frac{\lambda_l(\bSigma_{h_s})}{\lambda_l(\hat{\bSigma}_{h_s})} -1 \right\}^{1/2}  \right] \\
    & + \sum_{h_0,\ldots,h_J,h_s\leq H}\left(\prod_{j=0}^Jw_{j,h_j}\right)w_{s,h_s} \left\{2L\lVert \bm{O}_{h_s} - \hat{\bm{O}}_{h_s}\rVert_2 \frac{\lambda_1(\bSigma_{h_s})}{\lambda_L(\bSigma_{h_s})} \right\}^{1/2}  \\
    & +  \sum_{h_0,\ldots,h_j,h_s\leq H}\left\lvert \left(\prod_{j=0}^Jw_{j,h_j}\right)w_{s,h_s} - \left(\prod_{j=0}^J\hat{w}_{j,h_j}\right)\hat{w}_{s,h_s}\right\rvert + \sum_{l=1}^{J+2}2\binom{J+2}{l}\epsilon^l\\
    &\leq \max_{h_0,\ldots,h_J,h_s\leq H} \left[ \left(\frac{2/\pi}{\underline{\sigma}^2}\right)^{1/2}\sum_{j=0}^J\lVert \bPsi_j\rVert_F\left\lVert \bbeta_{h_jj} -\hat{\bbeta}_{k,h_j}\right\rVert
    + \left\{\sum_{l=1}^L \left(\frac{\lambda_l(\bSigma_{h_s})}{\lambda_l(\hat{\bSigma}_{h_s})} -1 \right)^2 \right\}^{1/2}  \right] \\
    & + \sum_{h_0,\ldots,h_J,h_s\leq H}\left(\prod_{j=0}^Jw_{j,h_j}\right)w_{s,h_s} \left\{2L\delta_{h_s} \frac{\lambda_1(\bSigma_{h_s})}{\lambda_L(\bSigma_{h_s})} \right\}^{1/2}  + \sum_{l=1}^{J+2}2\binom{J+2}{l}\epsilon^l \\
    & +  \sum_{h_0,\ldots,h_J,h_s\leq H}\left\lvert \prod_{j=0}^J\tilde{w}_{j,h_j}\tilde{w}_{s,h_s} - \prod_{j=0}^J\hat{w}_{j,h_j}\hat{w}_{s,h_s}\right\rvert + \left\lvert 1-(1-\epsilon)^{J+2} \right\rvert\\
    &\leq \left(\frac{2}{\pi}\right)^{1/2}\epsilon \sum_{j=0}^J\lVert\bPsi_j\rVert_F + \left\{\sum_{l=1}^L \left(1+\frac{\epsilon}{ L^{1/2}}-1\right)^2\right\}^{1/2} +
    \sum_{h_0,\ldots,h_J,h_s\leq H}\left(\prod_{j=0}^Jw_{j,h_j}\right)w_{s,h_s} \left\{\frac{\epsilon^2}{u_{h_s}}\frac{\lambda_1(\bSigma_{h_s})}{\lambda_L(\bSigma_{h_s})}\right\}^{1/2}\\
    & + \sum_{h_0,\ldots,h_J,h_s\leq H}\left\lvert \prod_{j=0}^J\tilde{w}_{j,h_j}\tilde{w}_{s,h_s} - \prod_{j=0}^J\hat{w}_{j,h_j}\hat{w}_{s,h_s}\right\rvert + \left\lvert 1-(1-\epsilon)^{J+2} \right\rvert + \sum_{l=1}^{J+2}2\binom{J+2}{l}\epsilon^l\\
    & \leq \epsilon \sum_{j=0}^J\lVert\bPsi_j\rVert_F + 3\epsilon + \left\lvert 1-(1-\epsilon)^{J+2} \right\rvert + \sum_{l=1}^{J+2}2\binom{J+2}{l}\epsilon^l,
\end{align*}
where the second inequality is based on the fact that $x-\log x- 1 \leq (x-1)^2$ for $x\geq 1$ with $x=\lambda_{l}(\bSigma_{h_s})/\lambda_{l}(\hat{\bSigma}_{h_s})$ and results of Lemma \ref{lma:boundWeightDiff}.
When $J$ and $\lVert \bPsi_j\rVert_F$ are finite, $\lVert f_P -f_{\hat{P}} \rVert_1\leq C^*\epsilon$ for some positive constant $C^*$ by some additional algebra. Therefore, a $\epsilon$-net for $\cG$ can be be constructed by $f_{\hat{P}}$ above in $L_1$ distance, and the cardinality of this $\epsilon$-net can be obtained by  cardinality of $\#(\hat{\Delta})\lesssim \epsilon^{-H^{J+2}},\#(\hat{\cO}_{h_s})\lesssim \delta_{h_s}^{-L(L-1)/2}$ and $\#(\hat{\cR}_{h_jj})\lesssim \left(\frac{\bar{a}_{h_j}}{\underline{\sigma}\epsilon/2}+1\right)^{2^j}- \left(\frac{\underline{a}_{h_j}}{\underline{\sigma}\epsilon/2}-1\right)^{2^j}, j=0,\ldots,J$ with 
\begin{align*}
    \lesssim M^{LH}\epsilon^{-H^{J+2}} \prod_{h_0,\ldots,h_J\leq H}\prod_{j=0}^J \left\{\left(\frac{\bar{a}_{h_j}}{\underline{\sigma}\epsilon/2}+1\right)^{2^j}- \left(\frac{\underline{a}_{h_j}}{\underline{\sigma}\epsilon/2}-1\right)^{2^j}\right\}\prod_{h_s\leq H} \left( \frac{2Lu_{h_s}}{\epsilon^2}\right)^{L(L-1)/2}
\end{align*}

\subsection{Proof of Theorem \ref{thm:strngCnst}}\label{supp:strngCnst}
The proof is based on the results of the sieve construction and the corresponding entropy from Theorem \ref{thm:entrpBnd} and aims to show that the conditions (2A) to (2D) satisfy the (1A) and (1B) and the resulting strong posterior consistency. Denote $k=(k_1,\ldots,k_{H_n})$ and $ l=(l_1,\ldots,l_{H_n})$ with $k_h,l_h\in\mathbb{N}$. First, we construct the sieves as follows 
\begin{align*}
    \cF_n =&\left\{ f_p: P=\sum_{h_0,\ldots,h_J,h_s\geq1} \left(\prod_{j=0}^Jw_{j,h_j}\right)w_{s,h_s}\delta(\bbeta_{h_jj},\bSigma_{h_s}): \sum_{h_j>H_n}w_{j,h_j}<\epsilon, \forall j=0,\ldots,J, \sum_{h_s>H_n}w_{s,h_s}<\epsilon \right. \\ 
    &\left.  \textrm{ and for } h_j,h_s\leq H_n, \underline{\sigma}^2_n\leq\lambda_L(\bSigma_{h_s}), \lambda_1(\bSigma_{h_s})\leq\underline{\sigma}^2_n(1+\epsilon/L^{1/2})^{M_n},1\leq\frac{\lambda_1(\bSigma_{h_s})}{\lambda_L(\bSigma_{h_s})}\leq n^{H_n}\right\}\\
    \cF_{n, k, l} =&\left\{ f_p\in\cF_n: \textrm{ for } h_0,\ldots,h_J, h_s<H_n, n^{l_{h_s}-1}\leq\frac{\lambda_1(\bSigma_{h_s})}{\lambda_L(\bSigma_{h_s})}\leq n^{l_{h_s}},  \right. \\
    &\left.  n^{H_n^{J+2}}(k_{h_j}-1)=\underline{a}_{h_j,j}\leq\lVert\bbeta_{h_jj}\rVert\leq\bar{a}_{h_j,j}=n^{H_n^{J+2}}k_{h_j}, \forall j=0,\ldots,J,\right\},
\end{align*}
where $M_n=\underline{\sigma}_n^{-2c_2}=n$ and $H_n=\left\lfloor{\frac{Cn\epsilon^2}{\log n}}\right\rfloor$ for some positive $C$. Clearly, $\cF_n\uparrow\cF$ and $\cF_n\subset\cup_{ k, l}\cF_{n, k, l}$. Following Lemma \ref{lma:nullTest}, $\Pi(\cF_n^c)$ satisfies condition (1A) in Lemma \ref{lma:sumblty}.

Now, we prove the summability condition (1B). By the Theorem \ref{thm:entrpBnd}, we obtain the entropy bound of
\begin{align*}
    \cN(\epsilon,\cF_{n, k, l},\lVert\cdot\rVert_1)\lesssim\exp&\left[LH_n\log M_n + H_n^{J+2}\log\frac{1}{\epsilon}+ \sum_{h_s<H_n} \frac{L(L-1)}{2}\log\frac{2Ln^{l_{h_s}}}{\epsilon^2} +\right.\\
    &\left. \sum_{j=0}^J\sum_{h_j<H_n}\log\left\{\left(\frac{\bar{a}_{h_j}}{\underline{\sigma}_n\epsilon/2}+1\right)^{2^j}- \left(\frac{\underline{a}_{h_j}}{\underline{\sigma}_n\epsilon/2}-1\right)^{2^j}\right\}\right]
\end{align*}
From the proof of Theorem 2 in \citet{Canale2017}, when $n$ and $k_{h_j}$ are larger, we know that 
\begin{align*}
    \left(\frac{\bar{a}_{h_j}}{\underline{\sigma}_n\epsilon/2}+1\right)^{2^j}- \left(\frac{\underline{a}_{h_j}}{\underline{\sigma}_n\epsilon/2}-1\right)^{2^j} &= \left(\frac{n^{H_n^{J+2}}k_{h_j}}{\underline{\sigma}_n\epsilon/2}+1\right)^{2^j}- \left(\frac{n^{H_n^{J+2}}(k_{h_j}-1)}{\underline{\sigma}_n\epsilon/2}-1\right)^{2^j} \\
    &=\left(\frac{2n^{(H_n^{J+2}+\frac{1}{2c_2})}k_{h_j}}{\epsilon}+1\right)^{2^j}- \left(\frac{2n^{(H_n^{J+2}+\frac{1}{2c_2})}(k_{h_j}-1)}{\epsilon}-1\right)^{2^j}\\
    &\lesssim \frac{n^{2^j(H_n^{J+2}+\frac{1}{2c_2})}k_{h_j}^{2^j-1}}{\epsilon^{2^j}}
\end{align*}
Since $\sum_{j=0}^J 2^j=2^{J+1}-1$, we have the entropy bound as
\begin{align*}
    \cN(\epsilon,\cF_{n, k, l},\lVert\cdot\rVert_1)\lesssim\exp&\left[LH_n\log M_n + H_n^{J+2}\log\frac{1}{\epsilon}+ \sum_{h_s<H_n} \frac{L(L-1)}{2}\log\frac{2Ln^{l_{h_s}}}{\epsilon^2} +\right.\\
    &\left. 2^{J+1}(H_n^{J+2}+\frac{1}{2c_2})\log n + 2^{J+1}\log\frac{1}{\epsilon}+\sum_{j=0}^J (2^j-1)\sum_{h_j<H_n}\log k_{h_j}\right]\\
    \approx \exp&\left[(\{H_n^{J+2}+L(L-1)H_n+2^{J+1}\}\log\frac{1}{\epsilon}+\{2^{J+1}(H_n^{J+2}+\frac{1}{2c_2})+LH_n\}\log n\right]\\
    &\times \prod_{h_s<H_n} (2Ln^{l_{h_s}})^{L(L-1)/2} \times \prod_{j=0}^J\prod_{h_j<H_n}k_{h_j}^{2^j-1}\\
\end{align*}

Also, from the tail behavior of condition (2A) and (2D), we have
\begin{align*}
    \Pi(\cF_{n, k, l}) &\leq \prod_{j=0}^J \prod_{h_j<H_n}P^*_j(\lVert\bbeta_{h_jj}\rVert\geq n^{H_n^{J+2}}(k_{h_j}-1))\prod_{h_s<H_n}P^*_{\bSigma}(\lambda_1(\bSigma_{h_s})/\lambda_L(\bSigma_{h_s})>n^{l_{h_s}-1})\\
    &\leq \prod_{j=0}^J \prod_{h_j<H_n} \{n^{H_n^{J+2}}(k_{h_j}-1)\}^{-2(r+1)}\prod_{h_s<H_n}\{n^{l_{h_s}-1})\}^{-\kappa}\\
    &= n^{-2(r+1)(J+1)H_n^{J+3}} \prod_{j=0}^J \prod_{h_j<H_n} (k_{h_j}-1)^{-2(r+1)} \prod_{h_s<H_n}(n^{l_{h_s}-1})^{-\kappa}
\end{align*}

Last, we sum over all $ k$ and $ l$ as $\sum_{ j, l}\cN^{1/2}(\epsilon,\cF_{n, k, l},\lVert\cdot,\rVert_1)\Pi^{1/2}(\cF_{n, k, l})$, and by Lemma \ref{lma:sumCond}, we show that condition (1B) is satisfied.

\subsection{Proof of Corollary \ref{cor:priorLaLRD}}\label{supp:priorLaLRD}
This proof mainly follows the proof of Corollary 2 in \citet{Canale2017}. First, since the Laplace has a exponential tail, the condition (2A) is easily satisfied. For the conditions (2B), we first show that 
\begin{align*}
    \mathbf{P}^*_{\bSigma}(\lambda_1(\bSigma^{-1})>z) \leq  \mathbf{P}^*_{\bSigma}(tr(\bSigma^{-1})>z) \leq \mathbf{P}^*_{\bSigma}(tr(\sigma^{-2} \bI)>z),
\end{align*}
by the Woodbury's identity. Since $\sigma^{-2}$ follows a Gamma distribution, which has a exponential tail, condition (2B) holds. For condition (2C), consider
\begin{align*}
    \mathbf{P}^*_{\bSigma}(\lambda_L(\bSigma^{-1})<1/z)\leq z^{-1}\mathbb{E}_{P^*_{\bSigma}}\{\lambda_1(\bSigma)\}\leq z^{-1}[\mathbb{E}_{P^*_{\bSigma}}\{\lambda_1(\sigma^2 \bI)\} + \mathbb{E}_{P^*_{\bSigma}}\{\lambda_1(\bLambda\bLambda^\transp)\}],
\end{align*}
where the first and the second inequalities hold due to Markov's and Weyl's inequalities, respectively. Obviously, the expectation of $\lambda_1(\sigma^2 \bI)$ is finite since $\sigma^2$ follows an inverse Gamma distribution. Denote $\lambda_{l\cdot s}=[\lambda_{l1s},\lambda_{l2s},\ldots]^\transp, l=1,\ldots,L$. For the second term, by integrating out the local shrinkage of $\phi_{slr}$, we obtain $\lambda_{slr}\mid \xi_{sr},e_s\sim t_3(0,\xi^{-1}_{sr}e^{-1}_s)$ and observe that elements in $\lambda_{l\cdot s}$ are independent given $\xi_{sr}$ and $e_s$. Also, $\lambda_{l\cdot s}$ follow the same distribution over all $l$ and $s$. Since Then
we can obtain
\begin{align*}
    \mathbb{E}_{P^*_{\bSigma}}\left\{\lambda_1(\bLambda\bLambda^\transp)\right\} &\leq \mathbb{E}_{P^*_{\bSigma}}\left\{tr(\bLambda\bLambda^\transp)\right\} = \mathbb{E}_{\xi,e}\left[\mathbb{E}\left\{\sum_{l=1}^L\lambda^\transp_{l\cdot s}\lambda_{l\cdot s}\mid \xi,e\right\}\right]=\mathbb{E}_{\xi,e}\left[\sum_{l=1}^L\mathbb{E}\left\{\lambda^\transp_{l\cdot s}\lambda_{l\cdot s}\mid \xi,e\right\}\right]\\
    &=\mathbb{E}_{\xi,e}\left[\sum_{l=1}^L\sum_{r=1}^\infty 3\xi_{sr}^{-1}e_s^{-1}\right] = 3\sum_{l=1}^L\sum_{r=1}^\infty\mathbb{E}_{\xi,e}\left[ \xi_{sr}^{-1}e_s^{-1}\right]
\end{align*}
Thus $\mathbb{E}_{P^*_{\bSigma}}\left\{\lambda_1(\bLambda\bLambda^\transp)\right\}$ is finite when the expectation of $\mathbb{E}(\delta_1)$ is finite and $\mathbb{E}(\delta_2)<1$ \citep{Bhattacharya2011CovFacAnal}. Both conditions can be achieved by letting $a_1>2$ and $a_2>3$. Finally, for condition (2D), we obtain 
\begin{align}\label{eq:loadingMat}
    \mathbf{P}^*_{\bSigma}\left\{\lambda_1(\bSigma^{-1})/\lambda_L(\bSigma^{-1}\right\}\geq z)=\mathbf{P}^*_{\bSigma}\left\{\left(\lambda_1(\bSigma^{-1})/\lambda_L(\bSigma^{-1})\right)^\kappa\geq z^\kappa\right\} \leq z^{-\kappa}\mathbb{E}_{P^*_{\bSigma}}\left\{\left(\lambda_1(\bSigma)/\lambda_L(\bSigma)\right)^\kappa \right\},
\end{align}
by Markov's inequality, where $\kappa>L(L-1)$ as defined in Theorem \ref{thm:strngCnst}. Again, by using Weyl's inequality, we observe that
\begin{align*}
    \frac{\lambda_1(\bSigma)}{\lambda_L(\bSigma)} \leq \frac{\lambda_1(\sigma^2 I) + \lambda_1(\bLambda\bLambda^\transp)}{\lambda_L(\sigma^2 I) + \lambda_L(\bLambda\bLambda^\transp)} = \frac{\lambda_1(\sigma^2 I) + \lambda_1(\bLambda\bLambda^\transp)}{\lambda_L(\sigma^2 I) } \leq \frac{\lambda_1(\sigma^{-2} I)}{\lambda_L(\sigma^{-2} I)} + \frac{tr(\bLambda\bLambda^\transp)}{\lambda_L(\sigma^2 I)}.
\end{align*}
Since the eigenvalues of the identity matrix are ones, the first term of $\frac{\lambda_1(\sigma^{-2} I)}{\lambda_L(\sigma^{-2} I)}=1$, which is finite after getting $k$th moment with respective to the $P^*_{\bSigma}$. For the second term, we have
\begin{align*}
    \frac{tr(\bLambda\bLambda^\transp)}{\lambda_L(\sigma^2 I)} = \sigma^{-2}tr(\bLambda\bLambda^\transp).
\end{align*}
Based on \eqref{eq:loadingMat}, we observe that $tr(\bLambda\bLambda^\transp)$ has finite $k$th moments, which finished the proof.

\section{Additional Lemmas}\label{supp:AddLemma}
\begin{lemma}\label{lma:diff_Norm}
Denote $\lambda_1(\bSigma)\geq\ldots\geq\lambda_L(\bSigma)$ as eigenvalues of $\bSigma$ and $\bm{O}_{h_s}^{(b)}$ as the matrix of orthonormal eigenvectors from the spectral decomposition of $\bSigma_{h_s}^{(b)}$ for $b=1,2$. The $L_1$ norm of the difference between two densities under the sieve constructed in Theorem \ref{thm:entrpBnd} will be bounded as 
\begin{align*}
\lVert f_{P_1} - f_{P_2}\rVert_1 &\leq \sum_{l=1}^{J+2}2\binom{J+2}{l}\epsilon^l + \sum_{h_0,\ldots,h_J,h_s\leq H}\left(\prod_{j=0}^Jw_{j,h_j}^{(1)}\right)w_{s,h_s}^{(1)} \left[\left\{\sum_{l=1}^L \frac{\lambda_l(\bSigma_{h_s}^{(1)})}{\lambda_l(\bSigma_{h_s}^{(2)})} - \log \frac{\lambda_l(\bSigma_{h_s}^{(1)})}{\lambda_l(\bSigma_{h_s}^{(2)})} -1 \right\}^{1/2} \right.\\
&\left. +\left\{2L\lVert \bm{O}_{h_s}^{(1)} - \bm{O}_{h_s}^{(2)}\rVert_2 \frac{\lambda_1(\bSigma_{h_s}^{(1)})}{\lambda_L(\bSigma_{h_s}^{(1)})} \right\}^{1/2} + \left(\frac{2/\pi}{\lambda_L(\bSigma_{h_s}^{(2)})}\right)^{1/2}\sum_{j=0}^J\lVert \bPsi_j\rVert_F\left\lVert \bbeta_{h_jj}^{(1)} -\bbeta_{h_jj}^{(2)}\right\rVert\right] \\
&+  \sum_{h_1,\ldots,h_j,h_s\leq H}\left\lvert \left(\prod_{j=0}^Jw_{j,h_j}^{(1)}\right)w_{s,h_s}^{(1)} - \left(\prod_{j=0}^Jw_{j,h_j}^{(2)}\right)w_{s,h_s}^{(2)}\right\rvert.
\end{align*}
\end{lemma}

\begin{proof}
Denote 
\begin{align*}
    A_{h_j,h_s}=\left(\prod_{j=0}^J w_{j,h_j}^{(1)}\right)w_{s,h_s}^{(1)}\phi_{\bSigma_{h_s}^{(1)}}\left(\by - \sum_{j=0}^J\bPsi_j\bbeta_{h_jj}^{(1)} \right) - \left(\prod_{j=0}^J w_{j,h_j}^{(2)}\right)w_{s,h_s}^{(2)}\phi_{\bSigma_{h_s}^{(2)}}\left(\by - \sum_{j=0}^J\bPsi_j\bbeta_{h_jj}^{(2)}\right).
\end{align*} 
We can express the norm as
\begin{align*}
\lVert f_{P_1} - f_{P_2}\rVert_1&=\left\lVert \sum_{h_0,\ldots,h_J, h_s\geq1}\left(\prod_{j=0}^J w_{j,h_j}^{(1)}\right)w_{s,h_s}^{(1)}\phi_{\bSigma_{h_s}^{(1)}}\left(\by - \sum_{j=0}^J \bPsi_j\bbeta_{h_jj}^{(1)} \right) \right.\\
&\left. - \sum_{h_0,\ldots, h_J,h_s\geq1}\left(\prod_{j=0}^J w_{j,h_j}^{(2)}\right)w_{s,h_s}^{(2)}\phi_{\bSigma_{h_s}^{(2)}}\left(\by - \sum_{j=0}^J \bPsi_j\bbeta_{h_jj}^{(2)} \right)\right\rVert_1
\leq \sum_{h_0,\ldots,h_J,h_s\geq1} \lVert A_{h_j,h_s}\rVert_1.
\end{align*}
For each term of $\lVert A_{h_j,h_s}\rVert_1$, we dichotomize the membership parameters by the threshold at $H$ for all $h_j, j=0,\ldots,J$ and $h_s$. The summation can be decomposed into a sum of $2^{J+2}$ terms. Given the cut-off $H$, denote $S_l,l=0,\ldots,J+2$ as the sum of $\lVert A_{h_j,h_s}\rVert_1$ that contains $l$ membership parameter bigger than the cut-off $H$. For example, $S_0=\sum_{h_0,\ldots,h_J,h_s\leq H}\lVert A_{h_j,h_s} \rVert_1$ as the term summing over all membership parameters smaller than $H$. Therefore, summation becomes $\sum_{h_0,\ldots,h_J,h_s\geq1} \lVert A_{h_j,h_s}\rVert_1=\sum_{l=0}^{J+2} S_l$. Each term can be upper bounded, and the whole norm can be bounded. When $l\geq1$, we rely on the facts that $\lVert A_{h_j,h_s}\rVert_1 \leq \left\lvert\left(\prod_{j=0}^J w_{j,h_j}^{(1)}\right)w_{s,h_s}^{(1)}+\left(\prod_{j=0}^J w_{j,h_j}^{(2)}\right)w_{s,h_s}^{(2)}\right\rvert$ and $\sum_{h_0,\ldots,h_J,h_s}\prod_{j=0}^J w_{j,h_j}w_{s,h_s}\leq \left(\prod_{j=0}^J \sum_{h_j}w_{j,h_j} \right)\sum_{h_s}w_{s,h_s}$. For example $l=J+2$, we obtain 
\begin{align*}
    S_{J+2}&=\sum_{h_0,\ldots,h_J,h_s> H}\lVert A_{h_j,h_s} \rVert_1\leq \sum_{h_0,\ldots,h_J,h_s> H}\left\lvert\left(\prod_{j=0}^J w_{j,h_j}^{(1)}\right)w_{s,h_s}^{(1)}+\left(\prod_{j=0}^J w_{j,h_j}^{(2)}\right)w_{s,h_s}^{(2)}\right\rvert\\
    &\leq \left(\prod_{j=0}^J \sum_{h_j>H}w_{j,h_j}^{(1)} \right)\sum_{h_s>h_n}w_{s,h_s}^{(1)} + \left(\prod_{j=0}^J \sum_{h_j>H}w_{j,h_j}^{(2)} \right)\sum_{h_s>h_n}w_{s,h_s}^{(2)} \leq  2\epsilon^{J+2}
\end{align*}
By induction, we obtain
\begin{align*}
S_l \leq 2\binom{J+2}{l}\epsilon^l, \forall l\geq1.
\end{align*}

Last, for $l=0$, 
\begin{align*}
\lVert A_{h_j,h_s}\rVert_1 = &\left\lVert \left(\prod_{j=0}^J w_{j,h_j}^{(1)}\right)w_{s,h_s}^{(1)}\phi_{\bSigma_{h_s}^{(1)}}\left(\by - \sum_{j=0}^J\bPsi_j\bbeta_{h_jj}^{(1)} \right) - \left(\prod_{j=0}^Jw_{j,h_j}^{(2)}\right)w_{s,h_s}^{(2)}\phi_{\bSigma_{h_s}^{(2)}}\left(\by - \sum_{j=0}^J\bPsi_j\bbeta_{h_jj}^{(2)}\right)\right.\\
&\left. +\left(\prod_{j=0}^Jw_{j,h_j}^{(1)}\right)w_{s,h_s}^{(1)}\phi_{\bSigma_{h_s}^{(2)}}\left(\by - \sum_{j=0}^J\bPsi_j\bbeta_{h_jj}^{(2)} \right) - \left(\prod_{j=0}^Jw_{j,h_j}^{(1)}\right)w_{s,h_s}^{(1)}\phi_{\bSigma_{h_s}^{(2)}}\left(\by - \sum_{j=0}^J\bPsi_j\bbeta_{h_jj}^{(2)}\right) \right\rVert_1\\
& \leq \left\lVert \left(\prod_{j=0}^Jw_{j,h_j}^{(1)}\right)w_{s,h_s}^{(1)}\left\{ \phi_{\bSigma_{h_s}^{(1)}}\left(\by - \sum_{j=0}^J\bPsi_j\bbeta_{h_jj}^{(1)} \right) - \phi_{\bSigma_{h_s}^{(2)}}\left(\by - \sum_{j=0}^J\bPsi_j\bbeta_{h_jj}^{(2)} \right) \right\}  \right\rVert_1 \\
&+\left\lVert\phi_{\bSigma_{h_s}^{(2)}}\left(\by - \sum_{j=0}^J\bPsi_j\bbeta_{h_jj}^{(2)} \right) \left\{ \left(\prod_{j=0}^Jw_{j,h_j}^{(1)}\right)w_{s,h_s}^{(1)} - \left(\prod_{j=0}^Jw_{j,h_j}^{(2)}\right)w_{s,h_s}^{(2)} \right\}\right\rVert_1\\
& \leq \left(\prod_{j=0}^Jw_{j,h_j}^{(1)}\right)w_{s,h_s}^{(1)} \left\lVert\phi_{\bSigma_{h_s}^{(1)}}\left(\by - \sum_{j=0}^J\bPsi_j\bbeta_{h_jj}^{(1)} \right) - \phi_{\bSigma_{h_s}^{(2)}}\left(\by - \sum_{j=0}^J\bPsi_j\bbeta_{h_jj}^{(2)} \right)  \right\rVert_1 \\
&+\left\lvert\left(\prod_{j=0}^Jw_{j,h_j}^{(1)}\right)w_{s,h_s}^{(1)} - \left(\prod_{j=0}^Jw_{j,h_j}^{(2)}\right)w_{s,h_s}^{(2)} \right\rvert.
\end{align*}
By using the triangle inequality, the norm in the first term in the right hand side can be bounded by
\begin{align}
\nonumber&\left\lVert \phi_{\bSigma_{h_s}^{(1)}}\left(\by - \sum_{j=0}^J\bPsi_j\bbeta_{h_jj}^{(1)} \right) - \phi_{\bSigma_{h_s}^{(2)}}\left(\by - \sum_{j=0}^J\bPsi_j\bbeta_{h_jj}^{(2)} \right)\right\rVert_1 \\
\label{eq:lm2Tmp}&\leq \left\lVert \phi_{\bSigma_{h_s}^{(1)}}\left(\by - \sum_{j=0}^J\bPsi_j\bbeta_{h_jj}^{(1)} \right) - \phi_{\bSigma_{h_s}^{(2)}}\left(\by - \sum_{j=0}^J\bPsi_j\bbeta_{h_jj}^{(1)} \right) \right\rVert_1 + \left\lVert \phi_{\bSigma_{h_s}^{(2)}}\left(\by - \sum_{j=0}^J\bPsi_j\bbeta_{h_jj}^{(1)} \right) - \phi_{\bSigma_{h_s}^{(2)}}\left(\by - \sum_{j=0}^J\bPsi_j\bbeta_{h_jj}^{(2)} \right) \right\rVert_1.
\end{align}
Note that the second term in \eqref{eq:lm2Tmp} can be bounded by 
\begin{align*}
\left(\frac{2/\pi}{\lambda_L(\bSigma_{h_s}^{(2)})}\right)^{1/2}\left\lVert \sum_{j=0}^J\bPsi_j\bbeta_{h_jj}^{(1)} -\sum_{j=0}^J\bPsi_j\bbeta_{h_jj}^{(2)}\right\rVert\leq
\left(\frac{2/\pi}{\lambda_L(\bSigma_{h_s}^{(2)})}\right)^{1/2}\sum_{j=0}^J\lVert \bPsi_j\rVert_F\left\lVert \bbeta_{h_jj}^{(1)} -\bbeta_{h_jj}^{(2)}\right\rVert,
\end{align*}
where $\lVert\cdot\rVert_F$ is the Frobenius norm. 

For the first term in \eqref{eq:lm2Tmp}, by using the triangle inequality, it can be bounded
\begin{align}\label{eq:lma2Tmp2}
\left\lVert \phi_{\tilde{\bSigma}_{h_s}}\left(\by - \sum_{j=0}^J\bPsi_j\bbeta_{h_jj}^{(1)} \right) - \phi_{\bSigma_{h_s}^{(1)}}\left(\by - \sum_{j=0}^J\bPsi_j\bbeta_{h_jj}^{(1)} \right) \right\rVert_1 + 
\left\lVert \phi_{\tilde{\bSigma}_{h_s}}\left(\by - \sum_{j=0}^J\bPsi_j\bbeta_{h_jj}^{(1)} \right) - \phi_{\bSigma_{h_s}^{(2)}}\left(\by - \sum_{j=0}^J\bPsi_j\bbeta_{h_jj}^{(1)} \right) \right\rVert_1,
\end{align}
where $\tilde{\bSigma}_{h_s}=\left(\bm{O}^{(2)}_{h_s}\bLambda^{(1)}_{h_s}(\bm{O}^{(2)}_{h_s})^\transp\right)^{-1}$ and $\bLambda_{h_s}^{(s)}=\textrm{diag}\left(\lambda_1(\bSigma_{h_s}^{(s)}),\ldots,\lambda_L(\bSigma_{h_s}^{(s)})\right)$. Following the proof of Lemma 2 in \citet{Canale2017}, the second term in \eqref{eq:lma2Tmp2} is bounded by 
\begin{align*}
\left\lVert \phi_{\tilde{\bSigma}_{h_s}}\left(\by - \sum_{j=0}^J\bPsi_j\bbeta_{h_jj}^{(1)} \right) - \phi_{\bSigma_{h_s}^{(2)}}\left(\by - \sum_{j=0}^J\bPsi_j\bbeta_{h_jj}^{(1)} \right) \right\rVert_1
\leq \left\{\sum_{l=1}^L \frac{\lambda_l(\bSigma_{h_s}^{(1)})}{\lambda_l(\bSigma_{h_s}^{(2)})} - \log \frac{\lambda_l(\bSigma_{h_s}^{(1)})}{\lambda_l(\bSigma_{h_s}^{(2)})} -1 \right\}^{1/2}.
\end{align*}
Similarly, the first term in \eqref{eq:lma2Tmp2} is bounded by
\begin{align*}
\left\lVert \phi_{\tilde{\bSigma}_{h_s}}\left(\by - \sum_{j=0}^J\bPsi_j\bbeta_{h_jj}^{(1)} \right) - \phi_{\bSigma_{h_s}^{(1)}}\left(\by - \sum_{j=0}^J\bPsi_j\bbeta_{h_jj}^{(1)} \right) \right\rVert_1
\leq \left\{2L\lVert \bm{O}_{h_s}^{(1)} - \bm{O}_{h_s}^{(2)}\rVert_2 \frac{\lambda_1(\bSigma_{h_s}^{(1)})}{\lambda_L(\bSigma_{h_s}^{(1)})} \right\}^{1/2}.
\end{align*}
\end{proof}

\begin{lemma}\label{lma:boundWeightDiff}
The upper bound of the last term in Lemma \ref{lma:diff_Norm} is given by
\begin{align*}
    \sum_{h_0,\ldots,h_J,h_s\leq H}\left\lvert \left(\prod_{j=0}^Jw_{j,h_j}\right)w_{s,h_s} - \left(\prod_{j=0}^J\hat{w}_{j,h_j}\right)\hat{w}_{s,h_s}\right\rvert\leq \sum_{h_0,\ldots,h_J,h_s\leq H}\left\lvert \prod_{j=0}^J\tilde{w}_{j,h_j}\tilde{w}_{s,h_s} - \prod_{j=0}^J\hat{w}_{j,h_j}\hat{w}_{s,h_s}\right\rvert + \left\lvert 1-(1-\epsilon)^{J+2} \right\rvert
\end{align*}, where $\tilde{w}_{j,h}=\frac{w_{j,h}}{1-\sum_{l> H}w_{j,l}}$ and $\tilde{w}_{s,h}=\frac{w_{s,h}}{1-\sum_{l> H}w_{s,l}}$.
\end{lemma}
\begin{proof}
Note that $\sum_{h_0,\ldots,h_J,h_s\leq H}\left\lvert\left(\prod_{j=0}^Jw_{j,h_j}\right)w_{s,h_s} - \left(\prod_{j=0}^J\hat{w}_{j,h_j}\right)\hat{w}_{s,h_s}\right\rvert\leq$
\begin{align*}
 &\sum_{h_0,\ldots,h_J,h_s\leq H}\left\lvert\left(\prod_{j=0}^Jw_{j,h_j}\right)w_{s,h_s} - \prod_{j=0}^J\left(1-\sum_{h_j>H}w_{j,h_j}\right)\left(1-\sum_{h_s>H}w_{s,h_s}\right)\left(\prod_{j=0}^J\hat{w}_{j,h_j}\right)\hat{w}_{s,h_s}\right\rvert \\
 +&\sum_{h_0,\ldots,h_J,h_s\leq H}\left\lvert\prod_{j=0}^J\left(1-\sum_{h_j>H}w_{j,h_j}\right)\left(1-\sum_{h_s>H}w_{s,h_s}\right)\left(\prod_{j=0}^J\hat{w}_{j,h_j}\right)\hat{w}_{s,h_s} - \left(\prod_{j=0}^J\hat{w}_{j,h_j}\right)\hat{w}_{s,h_s}\right\rvert\\
 =&\prod_{j=0}^J\left(1-\sum_{h_j>H}w_{j,h_j}\right)\left(1-\sum_{h_s>H}w_{s,h_s}\right)\sum_{h_0,\ldots,h_J,h_s\leq H}\left\lvert \prod_{j=0}^J\tilde{w}_{j,h_j}\tilde{w}_{s,h_s} - \prod_{j=0}^J\hat{w}_{j,h_j}\hat{w}_{s,h_s}\right\rvert\\
 +&\left\lvert \prod_{j=0}^J\left(1-\sum_{h_j>H}w_{j,h_j}\right)\left(1-\sum_{h_s>H}w_{s,h_s}\right)-1 \right\rvert\sum_{h_0,\ldots,h_j,h_s\leq H}\prod_{j=0}^J\hat{w}_{j,h_j}\hat{w}_{s,h_s}\\
 \leq & \sum_{h_0,\ldots,h_J,h_s\leq H}\left\lvert \prod_{j=0}^J\tilde{w}_{j,h_j}\tilde{w}_{s,h_s} - \prod_{j=0}^J\hat{w}_{j,h_j}\hat{w}_{s,h_s}\right\rvert + \left\lvert 1-(1-\epsilon)^{J+2} \right\rvert
\end{align*}
\end{proof}

\begin{lemma}\label{lma:nullTest}
The prior of \eqref{eq:stkBreak_pdpm} satisfies the conditions (2A) to (2D) in Theorem \ref{thm:strngCnst}. Then, $\Pi(\cF^c_n)\lesssim \exp^{-bn}$.
\end{lemma}
\begin{proof}
Consider the $\cF_n$ constructed in Theorem \ref{thm:strngCnst}. Obviously, since $Pr\{(A\cap B)^c\}=Pr(A^c\cup B^c)\leq Pr(A^c) + Pr(B^c)$, we obtain the upper bound as
\begin{align*}
    \Pi(\cF^c_n) \leq & \sum_{j=0}^J Pr(\sum_{h_j>H_n}w_{j,h_j}\geq\epsilon) + Pr(\sum_{h_s>H_n}w_{s,h_s}\geq\epsilon) \\
    & + H_n\left\{P_{\bSigma}^*(\lambda_L(\bSigma)\leq\underline{\sigma}_n^2)+ P_{\bSigma}^*(\lambda_1(\bSigma)\geq\underline{\sigma}_n^2(1+\epsilon/L^{1/2})^{M_n}) + P_{\bSigma}^*\left(\frac{\lambda_1(\bSigma)}{\lambda_L(\bSigma)}>n^{H_n}\right)\right\}
\end{align*}
By using the stick-breaking representation of DP for the first two terms and the prior on the tail behavior of conditions (2B) to (2D), we use the similar results from the proof of Proposition 2 in \citet{Shen2013} and obtain the upper bound as
\begin{align*}
    \Pi(\cF^c_n) &\lesssim \sum_{m=1,\ldots,J+2}\left\{\frac{e\alpha_m}{H_n}\log\frac{1}{\epsilon} \right\}^{H_n} + H_n\left\{\exp(-c_1\underline{\sigma}_n^{-2c_2})+ \underline{\sigma}_n^{-2c_3}(1+\epsilon/L^{1/2})^{-c_3M_n} + n^{-H_n\kappa} \right\}\\
    &\lesssim 2(H_n)^{-H_n} + \frac{Cn\epsilon^2}{\log n}\{\exp(-c_1n)+n^{c_3/c_2}(1+\epsilon/L^{1/2})^{-c_3n}+n^{-Cn\epsilon^2\kappa/\log n}\}\\
    &\lesssim (J+2)\exp(-H_n\log H_n) + \exp(-c_1n) + \exp\{-c_3n\log(1+\epsilon/L^{1/2})\} +\exp(-Cn\epsilon^2\kappa),
\end{align*}
by using the fact that $M_n=\underline{\sigma}_n^{-2c_2}=n, n^{1/\log n}=e$ and $H_n=\lfloor \frac{Cn\epsilon^2}{\log n}\rfloor$. Note that $-H_n\log H_n <-\frac{Cn\epsilon^2}{\log n}\log\frac{Cn\epsilon^2}{\log n}<-Cn\epsilon^2$ for sufficiently large $n$, the upper bound can be further written as
\begin{align*}
    \Pi(\cF^c_n)\lesssim 2\exp(-Cn\epsilon^2) + \exp(-c_1n) + \exp\{-c_3n\log(1+\epsilon/L^{1/2})\} +\exp(-Cn\epsilon^2\kappa) \lesssim \exp(-bn),
\end{align*}
for $0<b<\min(C\epsilon^2,c_1,c_3\log(1+\epsilon/T^{1/2}),C\epsilon^2\kappa)$.
\end{proof}

\begin{lemma}\label{lma:sumCond}
Under suitable choices of positive constants, we have the summability condition in Theorem \ref{thm:strngCnst} with $\sum_{ j, l}\cN^{1/2}(\epsilon,\cF_{n, k, l},\lVert\cdot,\rVert_1)\Pi^{1/2}(\cF_{n, k, l})\exp^{-(4-c)n\epsilon^2}\rightarrow0$ as $n\rightarrow\infty$ .
\end{lemma}
\begin{proof}
First, we can bound $\cN^{1/2}(\epsilon,\cF_{n, k, l},\lVert\cdot,\rVert_1)\Pi^{1/2}(\cF_{n, k, l})$
\begin{align}
\nonumber\lesssim &\exp\left[\frac{1}{2}(\{H_n^{J+2}+L(L-1)H_n+2^{J+1}\}\log\frac{1}{\epsilon}+\frac{1}{2}\{2^{J+1}(H_n^{J+2}+\frac{1}{2c_2})+LH_n\}\log n\right]\\
\nonumber&\times \prod_{h_s<H_n} (2Ln^{l_{h_s}})^{L(L-1)/4} \times \prod_{j=0}^J\prod_{h_j<H_n}k_{h_j}^{(2^j-1)/2}\\
\nonumber&\times n^{-(r+1)(J+1)H_n^{J+3}} \times \prod_{j=0}^J \prod_{h_j<H_n} (k_{h_j}-1)^{-(r+1)} \times \prod_{h_s<H_n}(n^{l_{h_s}-1})^{-\kappa/2}\\
\nonumber&\approx \exp\left[\log n\{2^{J}(H_n^{J+2}+\frac{1}{2c_2})+\frac{LH_n}{2} -(r+1)(J+1)H_n^{J+3}\}\right]\\
\nonumber&\times \exp\left\{\log\frac{1}{\epsilon}\left(\frac{H_n^{J+2}+L(L-1)H_n}{2}+2^{J} \right)\right\}\\
\label{eq:sumLma_k}&\times \prod_{j=0}^J\prod_{h_j<H_n} k_{h_j}^{(2^j-1)/2} (k_{h_j}-1)^{-(r+1)}\\
\label{eq:sumLma_l}&\times \prod_{h_s<H_n}(n^{l_{h_s}-1})^{-\kappa/2} (2Ln^{l_{h_s}})^{L(L-1)/4}
\end{align}
To obtain the bound of $\sum_{ k, l}\cN^{1/2}(\epsilon,\cF_{n, k, l},\lVert\cdot,\rVert_1)\Pi^{1/2}(\cF_{n, k, l})$, we sum over $ k$ and $ l$ for \eqref{eq:sumLma_k} and \eqref{eq:sumLma_l} and use the fact $\sum_{k}a_kb_k\leq\sum_{k}a_k\sum_{k}b_k$ for $a_k,b_k>0$. Given $j$ and $h_j$, we sum over $ k$ to obtain $K=\sum_{k\geq2}\frac{k_{h_j}^{(2^j-1})/2}{(k_{h_j}-1)^{(r+1)}}$, and $K<\infty$ by the hypothesis of $r>(L-1)/2>(2^j-1)/2$ for all $j=0,\ldots,J$. The term in \eqref{eq:sumLma_k} is bounded by $\prod_{j=0}^J\prod_{h_j<H_n}1+K=(1+K)^{(J+1)H_n}$. For summing over $ l$ for the term \eqref{eq:sumLma_l}, following the proof of Theorem 2 in \citet{Canale2017}, we obtain the bound $\sum_{
 l}(n^{l_{h_s}-1})^{-\kappa/2} (n^{l_{h_s}})^{L(L-1)/4}\leq 2n^{L(L-1)/4}$.

Hence, we combine all terms and get the upper bound
\begin{align*}
  &\exp\left[\log n\left\{2^{J}(H_n^{J+2}+\frac{1}{2c_2})+\frac{LH_n}{2} -(r+1)(J+1)H_n^{J+3}+H_n\frac{L(L-1)}{4}\right\}\right]\\
  &\times \exp\left\{\log\frac{1}{\epsilon}\left(\frac{H_n^{J+2}+L(L-1)H_n}{2}+2^{J} \right)\right\}\\  
  &\times (1+K)^{(J+1)H_n} \times (4L)^{H_nL(L-1)/4}\\
  &\lesssim \exp\left[Cn\epsilon^2\left\{H_n^{J+1}(2^J-(r+1)(J+1)H_n+\frac{L}{2}+\frac{L(L-1)}{4}\right\}\right]\\
  &\lesssim \exp\left[Cn\epsilon^2\left\{\frac{L}{2}+\frac{L(L-1)}{4}\right\}\right],
\end{align*}
where the last inequality holds when $(r+1)(J+1)H_n>2^J$ as $n\rightarrow\infty$. By choosing $C<2(4-c)/\{L+L(L-1)/2\}$, the condition is satisfied and compltes the proof.
\end{proof}

\section{Gibbs Sampler}\label{supp:GibbsSamp}
In this Section, we offer details Gibbs sampler for Algorithm \ref{algo1} with of the conditional distribution. We update the parameters as follows:
\begin{itemize}
    \item[{\bf Step 1}] Update the auxiliary parameters with 
    \begin{align*}
        u_{ij}\sim U(0,w_{h_{ij}}), j=0,\ldots,J; \thickspace u_{is}\sim U(0,w_{h_{is}}).
    \end{align*}
    \item[{\bf Step 2}] For $j=0,\ldots,J$ and $h=1,\ldots,h_j^*$, update the weight of 
    \begin{align*}
        w_{j,h}=\nu_{j,h}\prod_{e<{h}}(1-\nu_{j,e})\nu_{j,h}; \thickspace F(\nu_{j,h}) = \frac{(1-a_{j,h})^{\alpha_j}-(1-\nu_{j,h})^{\alpha_j}}{(1-a_{j,h})^{\alpha_j}-(1-b_{j,h})^{\alpha_j}},
    \end{align*}
    where $a_{j,h}=\max_{h_{ij}=h}\frac{u_{ij}}{\prod_{e<h}(1-\nu_{j,e})}$ and $b_{j,h}=1-\max_{h_{ij}>h}\frac{u_{ij}}{\nu_{j,h_{ij}}\prod_{s<h_{ij},s\neq h}(1-\nu_{j,s})}$. Similar update can be used to $w_{s,h_s}$ with replacing the index $h_j$ by $h_s$. 
    \item[{\bf Step 3}] For $j=0,\ldots,J$, update the membership indicators with
    \begin{align*}
        P(h_{ij}=g)\propto \bN\left(\by_i; \sum_{j'\neq j}\bPsi_j \bbeta_{h_{i{j'}}j'} + \bPsi_j\bbeta_{gj} ,\bSigma_i\right) \mathbb{I}(h_{ij}\in A_{j}(u_{ij})).
    \end{align*}
    Similarly, update the membership indicator for covariance with
    \begin{align*}
        P(h_{is}=s)\propto \bN\left(\by_i; \sum_{j=0}^J\bPsi_j \bbeta_{ij}, \bSigma_{s}\right) \mathbb{I}(h_{is}\in A_{\sigma}(u_{is})).
    \end{align*}
    \item[{\bf Step 4}] For $g$th cluster of the coefficient, update the coefficients with
    \begin{align*}
        \bbeta_{gj} \sim \bN_{2^j}\left(\bm{M},\bm{A}^{-1}\right),
    \end{align*}
    where $\bm{A} = \sum_{i:h_{ij}=g}\frac{\bI_{2^j}}{\sigma_i^2} + (\bm{\tau}^2_{gj})^{-1}, \bm{M} = \bm{A}^{-1} \bPsi_j^\transp \sum_{i:h_{ij}=g}\left(\frac{\by_i-\bLambda_i\bm{\eta}_i}{\sigma_i^2}\right)$ and $\bm{\tau}^2_{gj}=\textrm{diag}(\tau^2_{gj1},\ldots,\tau^2_{gj2^j})$.
    \item[{\bf Step 5}] Given $s$th cluster of the covariance with the number of the factor $K_s$, update the factor matrix with
    \begin{align*}
        \bm{\eta}_i \sim \bN_{K_s} \left\{\left(\frac{1}{\sigma^2_s}\bLambda_s^\transp \bLambda_s+ \bI_{K_s}\right)^{-1}\frac{\bLambda_s^\transp(\by_i-\btheta_i)}{\sigma^2_s},\left(\frac{1}{\sigma^2_s}\bLambda_s^\transp \bLambda_s+ \bI_{K_s}\right)^{-1}\right\},
    \end{align*}
    where $\btheta_i=\sum_{j=0}^J\bPsi_j\bbeta_{ij}$.
    \item[{\bf Step 6}] For $s$th cluster of the covariance, update each row of the loading matrix $\bm{\lambda}_{sl}, l=1,\ldots,L$ with
    \begin{align*}
        \bm{\lambda}_{sl} \sim \bN_{K_s}\left\{\left( \bm{S}_{sl}^{-1}+\sum_{i: h_{is}=s} \sigma^{-2}_s\bm{\eta}_i\bm{\eta}_i^\transp\right)^{-1}\sum_{i: h_{is}=s} \sigma^{-2}_s\bm{\eta}_i(y_{il}-\theta_{il}), \left( \bm{S}_{sl}+\sum_{i: h_{is}=s} \sigma^{-2}_s\bm{\eta}_i\bm{\eta}_i^\transp\right)^{-1}\right\},
    \end{align*}
    where $\bm{S}_{sl}=\textrm{Diag}(\phi_{sl1}\xi_{s1}e_s,\ldots,\phi_{slK_s}\xi_{sK_s}e_s), y_{il}$ and $\theta_{il}$ are the $l$-th element of $y_i$ and $\theta_i$.
    \item[{\bf Step 7}] For $s$th cluster of the covariance, update the variance with 
    \begin{align*}
        \sigma^2_{s} \sim \textrm{InvGa}\left\{a_s+\frac{n_sL}{2}, b_s+\sum_{i:h_{is}=s}\frac{(\by_i-\btheta_i-\bLambda_i\bm{\eta}_i)^2}{2}\right\},
    \end{align*}
    where $n_s=\sum_{i=1}^n1(h_{is}=s)$.
    \item[{\bf Step 8}] Update the hyper-parameters with 
    \begin{align*}
        \tau^2_{gjk} &\sim \textrm{InvGaussian}\left(\mu'_{gjk},\omega'_{gjk}\right)\\
        \phi_{slr}&\sim \textrm{Ga}\left\{2,\left(3+e_s\xi_{sr}\lambda^2_{slr}\right)/2\right\}\\
        \delta_{s1}&\sim\textrm{Ga}\left\{a_1+LK_s/2, 1+\left(e_s\sum_{h=1}^{K_s}\xi_{sh}^{(1)}\sum_{l=1}^L\phi_{slr}\lambda^2_{slr}\right)/2\right\}\\
        \delta_{sm}&\sim\textrm{Ga}\left\{a_2+L(K_s-m+1)/2, 1+\left(e_s\sum_{h=m}^{K_s}\xi_{sh}^{(m)}\sum_{l=1}^L\phi_{slr}\lambda^2_{slr}\right)/2\right\}, m\geq2\\
        e_s&\sim\textrm{Ga}\left\{a_e+LK_s/2,b_e+\left(\sum_{h=m}^{K_s}\xi_{sh}\sum_{l=1}^L\phi_{slr}\lambda^2_{slr}\right)/2\right\}
    \end{align*}
    where $\mu'_{gjk}=\left(\omega^2/\bbeta^2_{gjk}\right)^{1/2}, \omega'_{gjk}=\omega^2, \xi_{sh}^{(m)}=\prod_{t=1}\delta_{st}/\delta_{sm}$, and $\textrm{Ga}(a,b)$ is a gamma distribution with mean $a/b$.
\end{itemize}
To complete the sampler, we specify the hyper-parameters in the priors with $a_1=2.1, a_2=3.1, a_e=3, b_e=2, a_s=2.5, b_s=3$, and $\omega^2=1$.

We use the adaptive Gibbs sampler \citep{IMIFA_Murphy2020} to decide the number of factors. Specifically, at the $r$-th iteration, the probability of changing the number of factors is $p(r)=\exp(-b_0-b_1r)$. With the probability $p(r)$, the adaptive Gibbs sampler discards factors that have a certain proportion, $q$, of elements with absolute value smaller than $\delta$. Otherwise, if all factors have most of elements with their absolute values bigger than $\delta$, adaptive Gibbs sampler add a new factor from the multiplicative Gamma process. We refer the reader to \citet{IMIFA_Murphy2020} for more details. In this paper, we set $b_0=0.1, b_1=0.0005, q=1$ and $\delta=0.1$.



\section{Additional Details of Simulation Studies}\label{supp:AddSim}

\subsection{Details of Data-Generating Mechanism}\label{supp:AddSim_DataGen}
In this Section, we offer details of the data-generating mechanisms used in Main Paper Section \ref{sec:sim} for both the large-size images ($L=1,024$) small-size ($L=64$) under two error structures of correlated and independent errors. For completeness, there is expected overlap between the Main Paper and this section. For both sizes of images, we generate functional data under three different scenarios with different clustering structures of (1) correctly specified model, (2) mis-specified basis function, and (3) mis-specified generating mechanism of spatial heterogeneity. The first two scenarios are based on wavelet expansions, while the last one generates the data directly on the image without considering the basis expansion. For scenarios 1 and 2, all basis coefficients are zero, $\bbeta_{ij}=\mathbf{0}, j>0$, except for the highest resolution, $\beta_{i0}\sim 0.5N(-\mu_\beta,0.25)+0.5N(-\mu_\beta,0.25)$. We consider eight and three different non-zero coefficients for large- and small-size images, respectively, with $\mu_\beta=6$ for large images and $\mu_\beta=1.5$ for small images. Scenario 1 considers correct specified basis of the Haar functions, while Scenario 2 assigns the mis-specified basis functions of Daubechies \citep{Daubechies1993} orthonormal compactly supported wavelet of length eight and fits the models with the Haar basis functions \citep{Haar1910}. In contrast to scenarios 1 and 2 involving the wavelet expansion, scenario 3 generates images directly and introduces spatial heterogeneity with the clustering patterns varying over the spatial domain. Using a square image (i.e. $v_l\in[0,1]^2$), we assign four non-overlapping areas decided by two functions of $v_2=-7(v_1-0.5)^3+2v_1^2-0.1$ and $v_2=cos(v_1+0.5)$ (see Panel (C) of Figure \ref{fig:truDf} in the Main Paper). For each area in large-size images, we consider two potential values of $\theta_i^{(m)}(v_l)\in\{-0.35,0.35\},m=1,3$ and $\theta_i^{(m)}(v_l)\in\{-0.4,0.4\},m=2,4$. We center the data to ensure zero scaling coefficients and exclude two specific patterns that all four areas share the same signs of $\theta_i^{(1)}(v_l)\theta_i^{(3)}(v_l)=(0.35)^2$ and $\theta_i^{(2)}(v_l)\theta_i^{(4)}(v_l)=(0.4)^2$ to ensure a higher signal-to-noise ratio after centering the data, resulting in $14$ different patterns. For small-size images of spatial heterogeneity, due to the small sample size, we consider the same regions with three patterns among $14$ patterns and assign $\lvert\theta_i^{(m)}(v_l)\rvert=0.45, m=1,3$ and $\lvert\theta_i^{(m)}(v_l)\rvert=0.5, m=2,4$

All scenarios share the same mechanism for error terms. Specifically, we consider the low-rank decomposition of $\bSigma=\bLambda\bLambda^\transp+\sigma^2\bI$ with three signal-to-noise ratios (SNRs). For the independent-error setting, we set the loading matrix to zero, $\bLambda=0$, and vary the diagonal variance as $\sigma^2=0.01, 0.1$, and $1$ to represent high, medium, and low SNRs, respectively. For the correlated-error setting, we consider the diagonal variance of $\sigma^2=0.01$ and introduce correlation through a nonzero loading matrix $\bLambda=\{\lambda_{lr}\}$ of size $L\times K$. Each loading is generated as $\lambda_{lr}=\lambda_{lr}^0Z_{lr}$, where $\lambda_{lr}^0 \overset{i.i.d.}{\sim} N(0,s^2)$ and $Z_{lr}$ is a binary indicator sampled independently from a Bernoulli distribution, $Z\overset{i.i.d.}{\sim}\textrm{Ber}(0.1)$. For the hyperparameters, we consider the tuple of $(K,s^2)=(1,0.05^2), (3,0.5^2)$ and $(10,1^2)$ for high-, medium-, and low-SNR, respectively.

\subsection{Diagnostics of Convergence for the Proposed fPDPM}\label{supp:sim_diagGwk}
\begin{figure}[!htb]
    \centering
    \includegraphics[width=\linewidth]{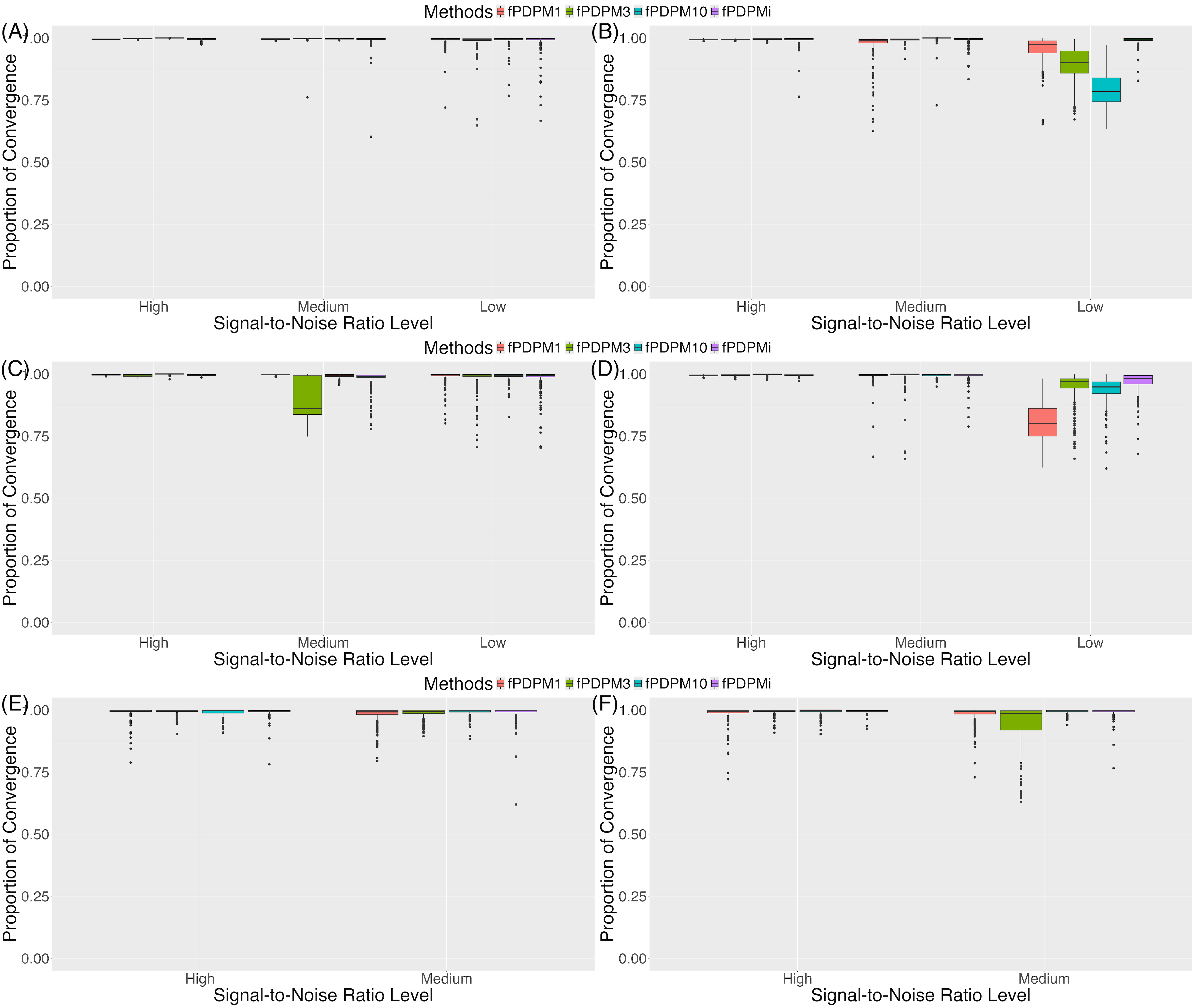}
    \caption{Diagnostics of the proposed method via Geweke statistics fo Scenario 1 (top row, Panels A–B), Scenario 2 (middle row, Panels C–D), and Scenario 3 (bottom row, Panels E–F). In each row, the left panel (A, C, E) shows results for independent errors, and the right panel (B, D, F) shows results for correlated errors. Proportion of images for which the MCMC sampler is assessed to have converged based on the Geweke diagnostic across all simulation settings. Results are shown for the four variants of the proposed method ($\textrm{fPDPM}_K$, and fPDPMi are variants of the proposed methods with correlated covariance initialized by $K$ factors and independent covariances, respectively) under different scenarios, error structures, and SNR levels. For each replicate, Geweke statistics are computed for all $L$ observed instances of each of the $n$ images, and convergence is determined after controlling the false discovery rate using the Benjamini–Hochberg procedure. Higher proportions indicate better overall mixing and convergence behavior of the corresponding sampler.}
    \label{sfig:gwkConv}
\end{figure}

In this section, we provide details of the MCMC diagnostics for the proposed method using the Geweke statistic \citep{geweke1991}, corresponding to the simulation studies in Section \ref{sec:sim} of the Main Paper. Due to the large number of parameters in the model, we focus on the estimated functional values $\hat{\theta}_{i}(v_l)$ and compute the Geweke statistics for these quantities. After discarding the burn-in period, we examine whether the early and late segments of the MCMC samples for $\hat{\theta}_{i}(v_l)$ differ significantly. Under convergence, the Geweke test should fail to reject the null hypothesis of equal means across the two segments. For each scenario, across error structures and SNR levels, we randomly select one simulation replicate and compute Geweke statistics over all $L$ observed instances for each of the $n$ images. Since each image involves $L$ correlated tests, we control the false discovery rate using the Benjamini–Hochberg procedure \citep{BH_Benjamini1995}. We then summarize, for each setting, the proportion of the $n$ images for which the MCMC sampler is deemed convergent (i.e., the null hypothesis is not rejected after FDR correction). These results are reported in Figure \ref{sfig:gwkConv}. Overall, the results indicate that the MCMC chains of the proposed method exhibit satisfactory convergence across all scenarios and error structures.

\subsection{Additional Simulation Results for Image Estimation}\label{supp:simMSE}

\subsubsection{Estimation for Large-Size Images}
Image estimation results for large-size images are summarized in Table \ref{stab:LargeImg_MSE}. Overall, both fPDPM and fPDPMi achieve low reconstruction errors across all scenarios and SNR levels, but with distinct behaviors that reflect the trade-off between pointwise estimation and clustering stability. Under the correctly specified model, fPDPMi attains the smallest MSEs across all SNRs (e.g., $1.3\times10^{-5}$ under high-SNR independent error and $4.3\times10^{-5}$ under correlated error), while fPDPM variants yield moderately larger but still low MSEs ($<0.05$ for most settings). Notably, $\textrm{fPDPM}_{10}$ exhibits occasional instability under correlated noise (MSE $=2.156$ at high SNR), whereas $\textrm{fPDPM}_{1}$ and $\textrm{fPDPM}_{3}$ remain stable. Combined with the ARI results, the extremely low MSEs of fPDPMi alongside weaker clustering performance suggest possible overfitting, where the model captures image-specific noise rather than the shared functional structure. Under basis-function misspecification, the fPDPM variants maintain small MSEs (mostly $<0.09$) and perform comparably to fPDPMi. The independent version continues to achieve the lowest errors, but the marginal difference between fPDPM and fPDPMi diminishes as the signal becomes more complex. This pattern indicates that while correlation modeling has limited impact on overall reconstruction accuracy, it provides smoother and more interpretable estimates that align better with clustering recovery. When the data-generating process differs from the fitted model, both fPDPM and fPDPMi achieve similar MSEs ($<0.06$ across all SNRs and noise structures). The estimation performance remains stable across model specifications, reflecting the robustness of both approaches in capturing dominant functional signals despite structural mismatches. Across all scenarios, fPDPMi consistently yields the lowest MSEs, particularly under high SNR, but this improvement comes at the cost of clustering interpretability and generalization. The fPDPM variants exhibit slightly higher but more stable MSEs, suggesting smoother reconstructions that avoid overfitting. Considering both estimation and clustering performance, fPDPM provides a better overall balance between accurate function recovery and reliable partitioning of functional structures.

\begin{table}[!htb]
    \centering
    \small
    \caption{Image estimation of large-size images under independent and correlated error structures with three levels of signal-to-noise ratios (SNR). Image estimation is measured in mean squared error (MSE). Results are obtained as the mean with the standard deviation in parentheses over 30 replicates. $\textrm{fPDPM}_{K}$ and fPDPMi, variants of the proposed methods with correlated covariance initialized by $K$ factors and independent covariances, respectively. All values have been multiplied by $100$}
    \label{stab:LargeImg_MSE}
    \begin{tabular}{crrrrrr}
        & \multicolumn{6}{c}{Scenario 1: correctly specified model}\\
        \cmidrule{2-7}
        & \multicolumn{2}{c}{High SNR} & \multicolumn{2}{c}{medium SNR} & \multicolumn{2}{c}{Low SNR}\\
        \cmidrule(lr){2-3}  \cmidrule(lr){4-5}  \cmidrule(lr){6-7} 
        & Independent & Correlated & Independent & Correlated & Independent & Correlated\\
        \hline
        $\textrm{fPDPM}_{1}$ & 0.001 (0.001) & 0.001 (0.001) & 0.001 (0.001) & 0.011 (0.001) & 0.002 (0.001) & 0.032 (0.016) \\
        $\textrm{fPDPM}_{3}$ & 0.016 (0.001) & 0.016 (0.001) & 0.016 (0.0002) & 0.017 (0.003) & 0.009 (0.006) & 0.043 (0.010) \\
        $\textrm{fPDPM}_{10}$ & 0.102 (0.211) & 2.156 (5.62) & 0.026 (0.001) & 0.505 (1.51) & 0.003 (0.005) & 0.051 (0.006) \\
        fPDPMi & 1.3$\times 10^{-5}$ (3$\times 10^{-6}$) & 4.3$\times 10^{-5}$ (9$\times 10^{-6}$) & 0.000187 (3.3$\times 10^{-5}$) & 0.008 (0.001) & 0.002 (9.1$\times 10^{-5}$) & 0.053 (0.003) \\
        DPM & 0.003 (0.001) & 0.003 (0.001) & 0.003 (0.001) & 0.021 (0.003) & 0.016 (0.002) & 0.17 (0.012) \\
    \end{tabular}
    ~
    \begin{tabular}{crrrrrr}
        & \multicolumn{6}{c}{Scenario 2 of mis-specified model in basis function}\\
        \cmidrule{2-7}
        & \multicolumn{2}{c}{High SNR} & \multicolumn{2}{c}{medium SNR} & \multicolumn{2}{c}{Low SNR}\\
        \cmidrule(lr){2-3}  \cmidrule(lr){4-5}  \cmidrule(lr){6-7} 
        & Independent & Correlated & Independent & Correlated & Independent & Correlated\\
        \hline
        $\textrm{fPDPM}_{1}$ & 0.005 (0.0004) & 0.005 (0.0005) & 0.005 (0.001) & 0.014 (0.001) & 0.007 (0.0005) & 0.038 (0.011) \\
        $\textrm{fPDPM}_{3}$ & 0.022 (0.0004) & 0.022 (0.0004) & 0.022 (0.0005) & 0.085 (0.131) & 0.005 (0.0001) & 0.046 (0.008) \\
        $\textrm{fPDPM}_{10}$ & 1.68 (4.68) & 0.048 (0.050) & 0.029 (0.001) & 0.033 (0.004) & 0.008 (0.008) & 0.055 (0.007) \\
        fPDPMi & 0.002 (0.0003) & 0.001 (0.0003) & 0.002 (0.0003) & 0.009 (0.001) & 0.005 (0.0001) & 0.056 (0.002) \\
        DPM & 0.003 (0.001) & 0.003 (0.001) & 0.004 (0.002) & 0.020 (0.003) & 0.016 (0.002) & 0.171 (0.012) \\
    \end{tabular}
    ~
    \begin{tabular}{crrrr}
    & \multicolumn{4}{c}{Scenario 3 of mis-specified model in data-generating mechanism}\\
        \cmidrule{2-5}
        & \multicolumn{2}{c}{High SNR} & \multicolumn{2}{c}{medium SNR}\\
        \cmidrule(lr){2-3}  \cmidrule(lr){4-5}
        & Independent & Correlated & Independent & Correlated\\
        \hline
        $\textrm{fPDPM}_{1}$ & 0.005 (0.0004) & 0.005 (0.0004) & 0.004 (0.0003) & 0.012 (0.001)\\
        $\textrm{fPDPM}_{3}$ & 0.576 (1.75) & 0.022 (0.001) & 0.029 (0.023) & 0.025 (0.004)\\
        $\textrm{fPDPM}_{10}$ & 1.52 (4.72) & 0.053 (0.075) & 0.028 (0.001) & 0.035 (0.015)\\
        fPDPMi & 0.004 (0.0004) & 0.004 (0.001) & 0.004 (0.0004) & 0.010 (0.001)\\
        DPM & 0.006 (0.001) & 0.006 (0.002) & 0.008 (0.001) & 0.017 (0.002)\\
    \end{tabular}
\end{table}

\subsubsection{Estimation for Small-Size Images}
Image estimation results for small-size images are summarized in Table \ref{stab:SmallImg_MSE}. Across all scenarios, the MSEs of fPDPM, fPDPMi, LPP, and SRZ are broadly comparable, suggesting that all methods achieve reasonable reconstruction accuracy. Interestingly, the fPDPM variants show remarkably stable MSEs across different SNR levels and error structures, indicating that the proposed resolution-based clustering provides consistent function estimation even under correlated or noisy conditions. In contrast, other methods exhibit more variation as noise levels increase or when errors are correlated. For instance, fPDPMi, LPP, and DPM generally produce low MSEs under independent high-SNR settings but show noticeable increases under correlated or medium-SNR conditions. SRZ demonstrates similar sensitivity and higher variability due to its dependence on noisy observed coefficients. These patterns suggest that while most approaches perform comparably in overall estimation accuracy, fPDPM achieves more stable and reliable reconstruction across varying conditions. Overall, the MSE results reveal minimal differences in reconstruction accuracy across methods but highlight the robustness of fPDPM to changes in SNR and error correlation. The stability of fPDPM’s estimation and  the superior clustering performance underscore its effectiveness in modeling small-size functional data.

\begin{table}[!htb]
    \centering
    \caption{Image estimation of small-size images under independent and correlated error structures with three levels of signal-to-noise ratios (SNR). Image estimation is measured in mean squared error (MSE). Results are obtained as the mean with the standard deviation in parentheses over 30 replicates. $\textrm{fPDPM}_{K}$ and fPDPMi, variants of the proposed methods with correlated covariance initialized by $K$ factors and independent covariances, respectively. All values have been multiplied by $100$}
    \label{stab:SmallImg_MSE}
    \begin{tabular}{crrrr}
        & \multicolumn{4}{c}{Scenario 1: correctly specified model}\\
        \cmidrule{2-5}
        & \multicolumn{2}{c}{High SNR} & \multicolumn{2}{c}{medium SNR}\\
        \cmidrule(lr){2-3}  \cmidrule(lr){4-5}
        & Independent & Correlated & Independent & Correlated\\
        \hline
        $\textrm{fPDPM}_{1}$ & 0.267 (0.025) & 0.273 (0.032) & 0.254 (0.044) & 0.231 (0.066)\\
        $\textrm{fPDPM}_{3}$ & 0.375 (0.039) & 0.357 (0.04) & 0.37 (0.035) & 0.261 (0.043)\\
        $\textrm{fPDPM}_{10}$ & 0.404 (0.027) & 0.410 (0.015) & 0.374 (0.043) & 0.293 (0.037)\\
        fPDPMi & 0.003 (0.0003) & 0.003 (0.0003) & 0.037 (0.007) & 0.189 (0.060)\\
        LPP & 0.005 (0.001) & 0.004 (0.001) & 0.044 (0.011) & 0.129 (0.067)\\
        SRZ & 0.132 (0.099) & 0.195 (0.125) & 0.077 (0.040) & 0.200 (0.055)\\
        DPM & 0.007 (0.001) & 0.026 (0.060) & 0.124 (0.032) & 0.299 (0.083)\\
    \end{tabular}
    ~
    \begin{tabular}{crrrr}
        & \multicolumn{4}{c}{Scenario 2 of mis-specified model in basis function}\\
        \cmidrule{2-5}
        & \multicolumn{2}{c}{High SNR} & \multicolumn{2}{c}{medium SNR}\\
        \cmidrule(lr){2-3}  \cmidrule(lr){4-5}
        & Independent & Correlated & Independent & Correlated\\
        \hline
        $\textrm{fPDPM}_{1}$ & 0.312 (0.049) & 0.296 (0.075) & 0.336 (0.044) & 0.343 (0.060) \\
        $\textrm{fPDPM}_{3}$ & 0.403 (0.025) & 0.437 (0.049) & 0.396 (0.026) & 0.363 (0.045) \\
        $\textrm{fPDPM}_{10}$ & 0.427 (0.03) & 0.419 (0.023) & 0.431 (0.030) & 0.373 (0.049) \\
        fPDPMi & 0.014 (0.006) & 0.026 (0.018) & 0.084 (0.008) & 0.187 (0.079)  \\
        LPP & 0.028 (0.016) & 0.031 (0.022) & 0.090 (0.024) & 0.157 (0.074)\\
        SRZ & 0.200 (0.042) & 0.208 (0.049) & 0.210 (0.020) & 0.278 (0.038)\\
        DPM & 0.041 (0.075) & 0.041 (0.075) & 0.125 (0.019) & 0.300 (0.073)\\
    \end{tabular}
    ~
    \begin{tabular}{crrrr}
    & \multicolumn{4}{c}{Scenario 3 of mis-specified model in data-generating mechanism}\\
        \cmidrule{2-5}
        & \multicolumn{2}{c}{High SNR} & \multicolumn{2}{c}{medium SNR}\\
        \cmidrule(lr){2-3}  \cmidrule(lr){4-5}
        & Independent & Correlated & Independent & Correlated\\
        \hline
        $\textrm{fPDPM}_{1}$ & 0.257 (0.045) & 0.254 (0.045) & 0.278 (0.027) & 0.250 (0.031) \\
        $\textrm{fPDPM}_{3}$ & 0.348 (0.039) & 0.331 (0.049) & 0.341 (0.025) & 0.298 (0.031)\\
        $\textrm{fPDPM}_{10}$ & 0.380 (0.036) & 0.384 (0.031) & 0.354 (0.043) & 0.319 (0.031) \\
        fPDPMi & 0.005 (0.002) & 0.005 (0.003) & 0.067 (0.009) & 0.136 (0.035)\\
        LPP & 0.030 (0.017) & 0.040 (0.029) & 0.086 (0.022) & 0.092 (0.051)\\
        SRZ & 0.169 (0.065) & 0.190 (0.059) & 0.137 (0.020) & 0.158 (0.043) \\
        DPM & 0.007 (0.001) & 0.007 (0.001) & 0.124 (0.029) & 0.177 (0.057)\\
    \end{tabular}
\end{table}

\subsection{Simulation Results for Local Clustering}\label{supp:localClu}
In this section, we conduct an additional simulation scenario that incorporates more complex local patterns for large-size images with $n=300$ and $L=1,024$. Specifically, we consider a larger number of non-zero coefficients to generate the data from the correctly-specified Haar function with independent errors, $\sigma^2=0.01$. In this setting, we allow coefficients at finer resolutions to be non-zero as $\bbeta_{ij}\neq 0, j=0,1,2$. For each resolution level with non-zero coefficients, we include $27$ potential cluster values, $h_{i0},h_{i1},h_{i2}\in\{1,\ldots,27\}$, and the product across resolutions gives $27^3 = 19,683$ possible candidate clustering configurations, which outnumbers the samples size $n=300$. In practice, only a small subset of these configurations occur among the $n=300$ observations. We consider $\bbeta_{h_{ij}jk} = Z_{h_{ij}jk}\bbeta^*_{h_{ij}jk}$ for $k=1,\ldots,2^j$, where $\bbeta^*_{h_{ij}jk}\sim N(\mu_{j},0.25)$ and $Z_{h_{ij}jk}\in\{-1,0,1\}$ is a discrete random variable with probabilities of $P(Z_{h_{ij}jk}=0)=p_{j}$ and $P(Z_{h_{ij}jk}=-1)=P(Z_{h_{ij}jk}=1)=(1-p_j)/2$. Specifically, we assign $(\mu_0,p_{0})=(2,0), (\mu_1,p_{1})=(0.5,0.15)$, and $(\mu_2,p_{2})=(0.15,0.5)$. By doing so, we downweight the local pattern via a smaller mean $\mu_j$ for finer resolutions and further introduce the sparsity with zero indicators of $Z_{h_{ij}jk}=0$. 

In this scenario, we focus on local clustering and evaluate clustering performance separately at each resolution, rather than summarizing clustering across resolutions using the posterior distance. Specifically, for a given resolution $j$, we quantify the agreement between the estimated cluster memberships and the true memberships using the adjusted Rand index, denoted $\textrm{ARI}_j$. Following Section \ref{sec:sim} of the Main Paper, we assess the image estimation accuracy using MSE and compare the two variants of the proposed method (fPDPM with correlated covariance and fPDPMi with independent covariance) to the global clustering methods of DPM and PCA-KM. We ran fPDPM and fPDPM with the hyperparameter $\alpha=3$ for $3,000$ iterations and discarded the first $90\%$ as burn-in.

Simulation results for local clustering are shown in Table \ref{stab:LocalClu}. For data generated under local clustering, fPDPM shows a better ($j=0$) or an equivalent ($j=1,2$) clustering accuracy than fPDPMi. The function  estimation is smaller under fPDPM ($\leq0.31$) and fPDPMi ($0.07$), compared to the global clustering of DPM ($2.55$), although the estimation performance under fPDPMi is much improved, which is not surprising given the data was generated using independent errors. Compared to the proposed fPDPM and fPDPMi, global clustering methods (DPM and PCA-KM) consistently produce considerably poorer clustering performance across resolutions ($<21$). These results suggest that the local clusters in the spatial functions hinders the ability of global clustering approaches to perform accurate clustering, with the performance being adversely affected even further in the presence of correlated measurement errors. In summary, these results highlight the challenges of global clustering approaches in the presence of local heterogeneity and illustrates the advantages of local clustering approaches with correlated noise terms.

\begin{table}[!h]
    \centering
    \caption{Simulation results of local clustering for large-size images under independent. Clustering performance are evaluated under each resolution in $\textrm{ARI}_j, j=0,1,$ and $2$, while the image estimation is measured in mean squared error (MSE). Results are obtained as the mean with the standard deviation in parentheses over 30 replicates. $\textrm{fPDPM}_{K}$ and fPDPMi, variants of the proposed methods with correlated covariance initialized by $K$ factors and independent covariances, respectively. All values have been multiplied by $100$.}~
    \begin{tabular}{crrrr}
	& $\textrm{ARI}_0$ & $\textrm{ARI}_1$ & $\textrm{ARI}_2$ & MSE\\
	$\textrm{fPDPM}_{1}$ & 95.1 (4.5) & 100 (0.006) & 93.9 (8.75) & 0.28 (0.10) \\
        $\textrm{fPDPM}_{3}$ & 95.2 (4.1) & 100 (0.003) & 94.3 (8.84) & 0.27 (0.11) \\
        $\textrm{fPDPM}_{10}$ & 94.9 (4.8) & 100 (0.002) & 97.2 (6.74) & 0.31 (0.26)\\
        fPDPMi & 87.8 (5.6) & 99.0 (1.6) & 95.1 (9.84) & 0.07 (0.13) \\
        DPM & 20.8 (1.4) & 0.92 (0.48) & 1.11 (0.65) & 2.55 (0.06)\\
        PCA-KM & 18.5 (3.6) & 0.17 (0.39) & 0.21 (0.44) & NA
    \end{tabular}
    \label{stab:LocalClu}
\end{table}

\section{Additional Results for Real Data Analysis}\label{supp:ST_Add}
\subsection{Overview for Additional Data}\label{supp:STAddRes}
The pre-processing steps for H2 and H3 tissues were similar to the pre-processing step described in Section 6 in Main Paper. We retained genes with more than $20$ non-zero spots and spots with at least $300$ non-zero expression genes \citep{SpatialPCA_Shang2022}. After removing $21$ genes that are identified as technical artifacts \citep{dataST_Andersson2021}, we obtained $10,053$ genes measured on $579$, and $497$ spots for two samples. For each sample, we normalized the expression data with mean zero and unit variance and selected spatially variable genes \citep{SpatialPCA_Shang2022}, which results in $423$, and $170$ genes for each sample. We padded zeros in all samples to the dimension of $32$ by $32$. We ran fPDPM of $10,000$ iterations and discarded the first $90\%$ of iterations. 

\subsection{Diagnostics of Convergence}\label{supp:STConv}
\begin{figure}[!htb]
    \centering
    \includegraphics[width=0.5\linewidth]{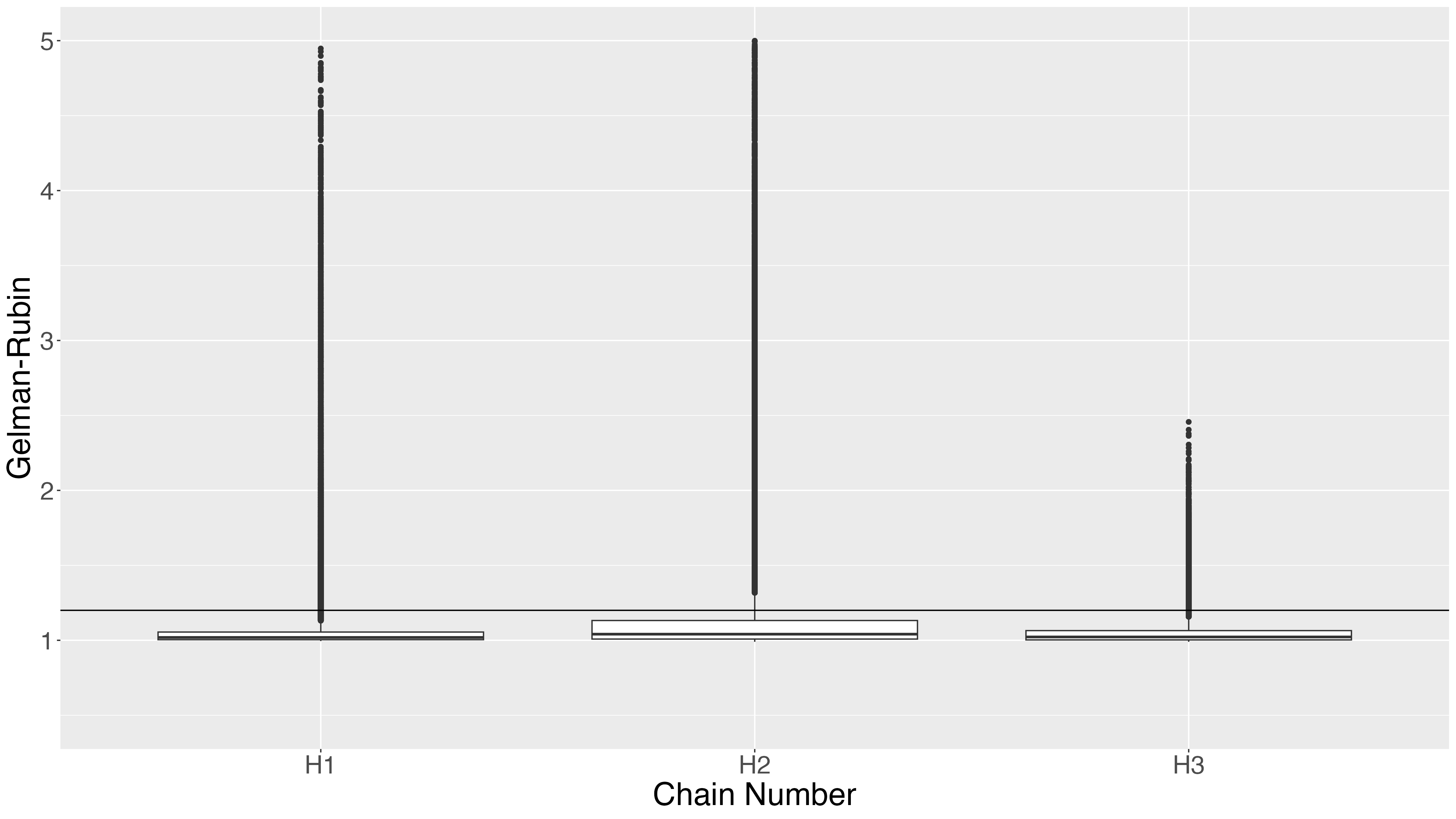}
    \caption{Gelman-Rubin statistics for $\hat{\theta_i}(v_l)$ as the diagnostics of the MCMC for fPDPM with three spatial transcriptomics data (H1, H2, and H3). A threshold of Gelman-Rubin statistics at $1.2$ (black line) are considered with the Gelman-Rubin statistics 
    $\leq1.2$ to indicate convergence.}
    \label{sfig:ST_mcmcDiag_GR}
\end{figure}

In this Section, we provide diagnostics of MCMC for fPDPM using three spatial transcriptomics data (H1, H2, and H3) that are sampled from the same patient. We ran two chains of MCMC with different initial values for $10,000$ iterations and discarded the first $90\%$ iterations. We then estimate the Gelman-Rubin statistics \citep{DiagGR_Gelman1992} for $\hat{\theta}_i(v_l)$. Figure \ref{sfig:ST_mcmcDiag_GR} shows the Gelman-Rubin statistics for three samples with the threshold at $1.2$ (black line). Most Gelman-Rubin statistics for $\hat{\theta}_i(v_l), i=1,\ldots,n, l=1,\ldots,L$ are lower than (H1:$93.3\%$, H2: $82.2\%$, and H3: $93.5\%$) the threshold at $1.2$ indicating the convergence the for MCMC of fPDPM.

\subsection{Gene Co-expression Network}\label{supp:NetExp}
\begin{figure}[!htb]
    \centering
    \includegraphics[width=\linewidth]{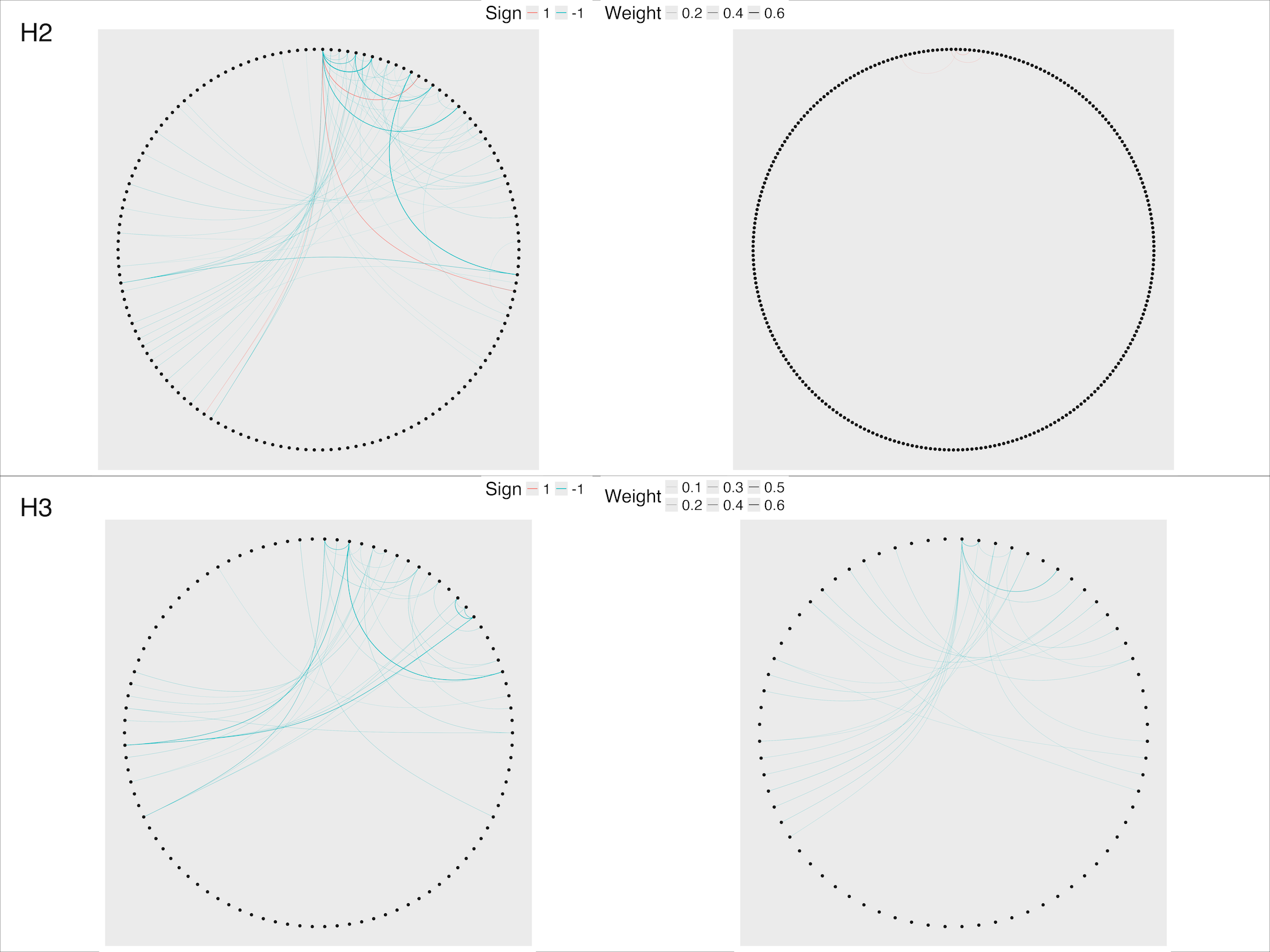}
    \caption{Gene co-expression network of samples of H2 (top row) and H3 (bottom row) with two clusters identified for each sample.}
    \label{sfig:Net_H2H3}
\end{figure}

Aside from the H1 sample, we used fPDPM for H2 and H3 samples. Based on Silhouette methods, we identified two clusters for both samples with the corresponding gene co-expression network shown in Figure \ref{sfig:Net_H2H3}. For H2 sample, the first cluster contains $150$ genes while the second one includes $273$ genes. Obviously, the network of the first cluster is denser and negative dominated (number of positive edges: $19$ and negative edges: $162$), while the network of the second cluster is extremely sparse with only positive edges (numbers of positive edges: $25$ and negative edges: $0$). The first group of the sample H3 contains $98$ genes, and the second one includes $72$ genes. The networks of both clusters are sparse with mostly negative edges (numbers of positive and negative edges of group one: $(98,2)$ and group two: $(45,6)$). For the network of the first cluster, we observe that the cluster coefficient is higher ($0.147$) than that of the second cluster ($0.044$) with the immunoglobulin gene family (IGHA1, IGHA2, IGHG1, IGHG3, IGHG4, IGKC, IGLC2, IGLC3, and IGLC7) that forms a tightly connected sub-network (average weights of the sub-network: $0.247$ and degree: $3.56$ with the weights of the overall network: $0.091$ and the degree: $2.04$). Contrarily, the network of the second group contains a hub gene, NASP, with a degree of $11$. 

\clearpage

\begin{table}
    \centering
    \caption{Full gene list from the sample H1 of three clusters estimated by the proposed model.}
    \begin{tabular}{cl}
        \multirow{13}{*}{\shortstack{Immune}} & A2M, ACTB, ADAM17, ADD1, ADIRF, AHNAK, AKNA, APMAP, AQP1, ARHGAP4, \\ 
        & ARHGAP45, ARHGEF1, ARPC1B, ARPC4, ATG10, ATOX1, ATP5E, B2M, BCAP31, BGN, \\
        & BTG1, C1QA, C1QB, C1QC, C1R, C1S, C3, C6orf62, CALML5, CCDC152, \\
        & CCDC80, CCL19, CCL21, CD2, CD4, CD52, CD6, CD74, CD79B, CDC42,\\
        & COL1A1, COL4A1, COL4A2, CST3, CYTH1, DDX5, DEF6, EIF1, EMP3, FBXW5,\\
        & FN1, FTL, GAPDH, GPX1, GPX3, H2AFJ, HLA-A, HLA-B, HLA-DQB1, HLA-DRA,\\
        & HLA-DRB1, HLA-E, HLA-F, HNRNPA0, IGFBP7, IGHG1, IGHG2, IGHG3, IGHG4, IGHM,\\
        & IL10RA, ITGAL, ITGAX, ITM2B, ITM2C, JAK3, LAPTM5, LCP1, LSM4, LTB,\\
        & LY6E, LYZ, MFAP4, MPEG1, MYL9, MZB1, NACA, NPC2, OGT, PFN1,\\
        & POMP, PTPN6, PTRF, RASAL3, RUNX1, S100A11, S100A6, S100A7, S100A9, SEPP1,\\
        & SERF2, SH3BGRL3, SOD2, SPTBN1, SRSF5, SSR4, TAGLN, TAP1, TCF7, TIMP1,\\
        & TIMP3, TMSB4X, TPI1, TPT1, TRAC, TRBC1, TUBA1B, TXNIP, VIM, VOPP1,\\
        & VWF, WDR34, ZAP70, ZFP36L2\\
        & \\
        \multirow{14}{*}{\shortstack{Signal \\Transduction }} & ABL1, ACTG1, ADGRG1, AGPAT2, AIF1L, AKAP8L, AKT1, AP000769.1, AP3D1, AR,\\
        & ARF1, ATP5A1, ATP6V0B, ATP6V1C2, B4GALT1, B4GALT2, BTF3, C12orf60\\
        & C17orf96, CACFD1, CALM2, CBX3, CCT5, CD164, CD24, CD9, CDH1, CDK12, CHGA, CLDN3,\\
        & CLTC, COL1A2, COX7C, CRABP2, CRACR2B, CSDE1, CSNK1D, CXCL14, DDR1, DDX17,\\
        & DDX3X, DEGS1, DHCR24, DHX15, DUSP4, DYNC1LI2, EFNA1, EHMT1, EPN3, ERBB2,\\
        & FAM134A, FASN, FBRSL1, FXYD3, GIT1, GNAS, GOT2, GRB7, GSE1, GSN,\\
        & GTF3C5, HINT1, HNRNPA2B1, HNRNPAB, HNRNPF, HNRNPH1, HSP90AA1, \\
        & HSP90AB1, HSP90B1, IFI27, IGFBP2, ITGB6, JUN, KLC1, KMT5C, KRT19, KRT7, \\
        & LDHB, LINC00094, MAGED1, MAP3K12, MARCKS, MARCKSL1, MCF2L, MCM7, MDK, MGP, \\
        & MIEN1, MLLT6, MMACHC, MMP14, MT1X, MT2A, MYH9, NASP, NAT14, NEUROD2, \\
        & NFKBIA, NONO, PABPC1, PCGF2, PCSK1N, PDIA6, PEG10, PGK1, PIEZO1, PLEKHG4B,\\
        & PPP1CC, PRKDC, PRRC2B, PRSS8, PTBP1, PTMA, PTMS, PTPRF, RALGDS, RRM2, RXRA, \\
        & S100A8, SBK1, SDC1, SDC4, SDCCAG3, SEC16A, SEMA3C, SF3B2, SKP1, SOX11, SPDEF, \\
        & SSRP1, SURF4, TMED10,  TMEM132A, TMSB10, TPD52, TRMT112, TXNL4A, VAV2, XBP1, \\
        & YWHAQ, YWHAZ, ZNF703 \\
        & \\
        \multirow{4}{*}{\shortstack{Innate \\ Immune}}
        & APOD, AZGP1, C2orf82, CFD, CXCL12, DCN, DERL3, EGR1, EPHB3, EPS8L2,\\
        & FAM3D, IFITM3, IGFBP4, IGHA1, IGHA2, IGKC, IGLC2, IGLC3, IGLC7, JCHAIN,\\
        & KRT15, LGALS3, LRP1, LTF, NDRG2, PTGDS, PYGB, SAA1, SAA2, SCGB3A1,\\
        & SERPING1, SLPI, STAC2, TPSB2, ZG16B\\
    \end{tabular}
    \label{stab:geneCluList}
\end{table}

\clearpage

\bibliographystyle{ba}
\bibliography{Ref}